\documentclass[a4paper,11pt]{article}
\pdfoutput=1 

\usepackage{jheppub} 

\usepackage[T1]{fontenc} 

\usepackage[english]{babel}
\usepackage[utf8]{inputenc}
\usepackage{graphicx}
\usepackage{slashed}
\usepackage{color}
\usepackage{rotating}
\usepackage{ulem}
 \usepackage{float}
 \usepackage{blindtext}
 \usepackage{multirow}

\numberwithin{equation}{section}

\newcommand{\bea}{\begin{eqnarray}}
\newcommand{\eea}{\end{eqnarray}}
\newcommand{\be}{\begin{equation}}
\newcommand{\ee}{\end{equation}}
\newcommand{\nn}{\nonumber}
\newcommand{\ds}{{\sf DarkSUSY}}
\newcommand{\mn}{{\sf MultiNest}}
\newcommand{\likeJ}{\mathcal{L}_{\rm Joint}}
\newcommand{\like}{\mathcal{L}}
\newcommand{\Ohsq}{\Omega_\chi h^2}

\title{Electroweak and Higgs Boson Internal Bremsstrahlung\\
{\large General considerations for Majorana dark matter annihilation and application to MSSM neutralinos}}

\author[a]{Torsten Bringmann,}
\author[b]{Francesca Calore,}
\author[a]{Ahmad Galea}
\author[c]{and Mathias Garny}

\affiliation[a]{Department of Physics, University of Oslo,Box 1048, NO-0371 Oslo, Norway}
\affiliation[b]{LAPTh, CNRS, 9 Chemin de Bellevue, BP-110, Annecy-le-Vieux, 74941, Annecy Cedex, France}
\affiliation[c]{Technical University Munich, James-Franck-Str. 1, D-85748 Garching, Germany}

\emailAdd{torsten.bringmann@fys.uio.no}
\emailAdd{francesca.calore@lapth.cnrs.fr}
\emailAdd{ahmad.galea@fys.uio.no}
\emailAdd{mathias.garny@tum.de}

\abstract{It is well known that the annihilation of Majorana dark matter into
fermions is helicity suppressed. Here, we point out that the underlying
mechanism is a subtle combination of two distinct effects, and we 
present a comprehensive analysis of how the suppression can be 
partially or fully lifted by the internal bremsstrahlung of an 
additional boson in the final state. As a concrete illustration, 
we compute analytically the full amplitudes and annihilation rates 
of supersymmetric neutralinos to final states that contain any 
combination of two standard model fermions, plus one electroweak 
gauge boson or one of the five physical Higgs bosons that appear in the 
minimal supersymmetric standard model. We classify the various 
ways in which these three-body rates can be large compared to 
the two-body rates, identifying cases that have not been 
pointed out before. In our analysis, we put special emphasis on 
how to avoid the double counting of identical kinematic situations
that appear for two-body and three-body final states, in particular on 
how to correctly treat differential rates and the spectrum of 
the resulting stable particles that is relevant for indirect dark 
matter searches. We find that both the total annihilation rates and 
the yields can be significantly enhanced when taking into account 
the corrections computed here, in particular for models with somewhat 
small annihilation rates at tree-level which otherwise
would not be testable with indirect dark matter searches. Even more
importantly, however, we find that the resulting annihilation spectra 
of positrons, neutrinos, gamma-rays and antiprotons differ in general
substantially from the model-independent spectra that are commonly
adopted, for these final states, when constraining particle dark matter 
with indirect detection experiments. }

\begin{document} 
{\large
\flushright TUM-HEP 1083/17 \\
\vspace{-.4cm}\flushright  LAPTH-013/17\\
}
\maketitle
\flushbottom

\section{Introduction}

The prime hypothesis for the cosmologically observed dark matter (DM) \cite{Ade:2015xua} is
a new type of elementary particle \cite{BertoneBook}. Among theoretically well-motivated  
candidates, weakly interacting massive particles (WIMPs) play a prominent role. This is 
because such WIMPs very often appear in theories that attempt to cure
the fine-tuning problems in the Higgs sector of the standard model of particle physics (SM), 
{\it and} because thermal relics with weak masses and cross sections at the electroweak scale 
are typically produced with the correct abundance to account for the DM density today 
\cite{Lee:1977ua,Gondolo:1990dk}.
Another advantage is that the WIMP hypothesis can be tested in multiple ways: at {\it colliders}, 
where the signature consists in missing energy, in {\it direct detection} experiments aiming to 
observe DM particles recoiling off the nuclei of deep underground detectors, and in {\it indirect} 
searches for the debris of DM annihilation in cosmic  regions with large DM densities.  
Direct detection experiments have become extremely competitive in constraining smaller and 
smaller scattering rates \cite{Akerib:2015rjg,Tan:2016zwf}, and collider searches have pushed the 
scale of new physics to TeV energies in many popular 
models \cite{Aaboud:2016wna,Khachatryan:2016nvf}. It is worth stressing, however, that only 
`indirect' searches would eventually allow to test the WIMP DM hypothesis {\it in situ}, 
i.e.~in places that are relevant for the cosmological evidence for DM. Also indirect searches
have become highly competitive during the last decade, now probing the `thermal cross section'
(the one that is needed to produce the observed DM abundance) up to WIMP masses of the order 
of 100\,GeV \cite{Ackermann:2015zua,Bergstrom:2013jra}.

A key quantity for both thermal production of WIMPs and indirect searches is the total annihilation 
cross section. Multiplied by the relative velocity $v$ of the incoming DM particles, it can in the non-
relativistic limit  be expanded as
\be\label{eq:partial}
 \sigma v = a + b v^2 + {\cal O}(v^4)\,.
\ee
It was noted early \cite{Bergstrom:1989jr,Flores:1989ru} that radiative corrections to $\sigma v$ can be 
huge because of 
symmetries of the annihilating DM pair in the $v\to0$ limit. For indirect DM searches, changes in either
the {\it partial} cross section, for a given annihilation channel, or the {\it differential} cross section, 
$d\sigma v/dE$, may be phenomenologically even more important. The reason is that an additional 
photon in the final state can give rise 
to pronounced spectral features in the DM signal in both gamma \cite{Bringmann:2007nk} and 
charged cosmic rays \cite{Bergstrom:2008gr}. 
For electroweak corrections, the situation is in some sense even more interesting because, on top of the 
just mentioned effects, completely new indirect detection channels may open up. In this way, antiproton 
data can for example efficiently constrain DM annihilation to light leptons when considering the 
associated 
emission of $W$ or $Z$ bosons \cite{Kachelriess:2009zy}. In the presence of point-like interactions,
such as described by effective operators, the resulting spectra can be computed in a model-independent 
way by using splitting functions inspired by a parton picture \cite{Ciafaloni:2010ti}. This approach is very 
useful for generic DM phenomenology and is, for example, the one implemented in the `cookbook'
for indirect detection \cite{Cirelli:2010xx}. One of the main results of this article 
(see also \cite{Bringmann:2013oja}) is that the 
resulting cosmic ray spectra from DM annihilation can 
differ substantially from the 
actual spectra, calculated in a fully consistent way from the underlying particle framework.

Here we revisit in detail one of the most often discussed examples where radiative corrections
can be large, namely the case of a Majorana DM particle $\chi$. The tree-level annihilation rate 
into light fermions $f$ is then on general grounds `helicity suppressed', for $v\to0$, as a 
consequence of the conserved quantum numbers of the initial state \cite{Goldberg:1983nd}. 
The resulting suppression 
by a factor of $m_f^2/m_\chi^2$ can be lifted by allowing for an additional 
vector \cite{Bergstrom:1989jr} or scalar \cite{Luo:2013bua} boson in the final state, implying that for 
DM masses at the electroweak scale the radiative `corrections' can be several orders of magnitude 
larger than the result from lowest order in perturbation 
theory\footnote{The lifting of helicity suppression via three-body final states is also relevant for real scalar dark matter
\cite{Toma:2013bka, Giacchino:2013bta, Ibarra:2014qma, Giacchino:2014moa, Giacchino:2015hvk}
and, under certain conditions, for vector dark matter \cite{Bambhaniya:2016cpr}.
The case in which the additional boson is a $Z'$ has been considered in \cite{Bell:2017irk}.}.
 Here, we revisit these arguments and point out that the effect commonly referred to 
as helicity suppression is in fact the culmination of two distinct suppression
mechanisms, in the sense that they can be lifted independently. This 
results, in general, in a rather rich phenomenology of such radiative corrections.

As an application, we consider electroweak corrections to the annihilation cross section of the 
lightest supersymmetric neutralino --
one of the most often discussed DM prototypes \cite{Jungman:1995df} and still a leading candidate despite 
null searches for supersymmetry at ever higher energies and luminosities at the 
LHC \cite{Aaboud:2016wna,Khachatryan:2016nvf} -- though our main findings can be extended in an 
analogous way to other DM candidates that couple to the SM via the electroweak or Higgs sector. 
Concretely, we provide a comprehensive analysis, both analytically and numerically, of all 3-body final states 
from neutralino annihilation
that contain a fermion pair and either an electroweak gauge boson or one of the five Higgs bosons 
contained in the minimal supersymmetric standard model (MSSM), for a neutralino that can be an arbitrary 
admixture of Wino, Bino and Higgsino.\footnote{%
For neutralino annihilation, so far only the cases of photon \cite{Bringmann:2007nk} and
gluon \cite{Bringmann:2015cpa} internal bremsstrahlung (IB) have been considered in full generality. 
Final states with electroweak gauge bosons have been considered for
pure binos in \cite{Ciafaloni:2011sa,Bell:2011if,Bell:2011eu,Garny:2011cj,Shudo:2013lca, Cavasonza:2014xra}, 
for Higgsinos in \cite{Garny:2011ii}, and for pure 
Winos in \cite{Ciafaloni:2012gs}. A first study for a general neutralino has been performed in 
\cite{Bringmann:2013oja}. Finally, final states involving the SM-like Higgs boson have been
considered in \cite{Luo:2013bua} for a toy model encompassing a pure Bino (see also \cite{Kumar:2016mrq}).
}
We find large parameter regions where these 3-body final states significantly enhance the DM annihilation
rate, with  the impact on the {\it shape} of the cosmic-ray spectra relevant for indirect detection being 
even more significant.

One of the technically most involved aspects, apart from the shear number of diagrams to be considered,
is how to avoid `double counting' the on-shell parts of the 3-body amplitudes that are already, implicitly, 
included in the corresponding 2-body results. We provide an in-depth treatment of this 
issue and demonstrate how to accurately treat not only the total cross section but also  
the resulting cosmic-ray spectra. We again find significant effects 
on the latter, indicating the need to correctly adopt this method also for 
other DM candidates. In fact, in order to reliably test the underlying particle models, our 
findings suggest that at least for fermionic final states it is in general not sufficient to use 
the model-independent spectra traditionally provided
by numerical packages. The numerical routines that implement our results for the neutralino case  will be fully
available with the next public release of \ds\ \cite{ds4,ds6}.

This article is organized as follows. We start in Section \ref{sec:neutralinos} 
with a general discussion of Majorana DM annihilating to fermions 
and the relevant symmetries that arise for $v\to0$, revisiting in particular the often invoked `helicity 
suppression' arguments and how this suppression can be lifted fully or partially. 
In Section \ref{sec:enhancement} we then consider the concrete case of  
neutralino DM, and discuss the various possibilities 
of how the presence of an additional final state boson can add sizeable contributions 
to, or even significantly enhance, the 2-body annihilation rates. The double counting 
issues mentioned above are then addressed in detail in a separate Section \ref{sec:doublecount}. 
We scan the parameter space of several MSSM versions and demonstrate the effect of 
these newly implemented corrections to neutralino annihilation in Section \ref{sec:mssm}, 
both for the annihilation rates and the cosmic-ray spectra relevant for indirect DM 
searches, and present our conclusions in Section \ref{sec:conc}. In a more technical 
Appendix, we describe the details of our analytical calculations to obtain the 3-body 
matrix elements for fully general neutralino annihilation in the MSSM 
(Appendix \ref{app:HelicityAmplitudes}), the numerical implementation of these results 
in \ds\ (Appendix \ref{app:num}), and how to correctly treat spin correlations of 
decaying resonances (Appendix \ref{app:corr}).

\section{Majorana dark matter and relevant symmetries}
\label{sec:neutralinos}

For DM annihilation in the Milky Way halo, where DM particles have typical velocities of order 
$10^{-3}$, only the first term in Eq.~(\ref{eq:partial}) gives a sizeable  contribution. 
In the following we therefore neglect $p$ and higher partial wave contributions, and it is 
understood that all (differential) cross sections are effectively evaluated in the zero-velocity limit. For an 
$s$-wave, the relative angular momentum in the initial state is $L=0$. Due to the Majorana nature 
the initial particles are identical, but because we consider fermions  the total wave function 
still needs to be antisymmetric with respect to exchanging the incoming particles.  The orbital 
wave function for $L=0$ is symmetric, so in order to get an anti-symmetric total wave function 
the spins must couple into an antisymmetric state. This is only possible for the singlet state, with
$S=0$, resulting in the following quantum numbers: 
\be
J=0, \qquad C=(-1)^{L+S}=1, \qquad P=(-1)^{L+1}=-1\,.
\ee
Here, the general expressions for $C$ and $P$ apply because we have a system of two {\it fermions}.
Assuming no significant sources of $CP$ violation in the theory, which generally are highly constrained
by measurements of the electric dipole moment and other precision experiments, the symmetry of the 
{\it final} state is 
hence also restricted to be $J_{CP}=0_-$. This implies the well-known `helicity' suppression of the 
annihilation rate into light fermions,  similar to the case of charged pion 
decay.  In the following we first briefly review the origin of this suppression, and then argue that it
can in fact be related to a combination of two rather independent suppression mechanisms.

\subsection{Chiral symmetry, gauge symmetry and helicity suppression}
\label{subsec:chiral}

We want to study  Dirac fermions $f$ as possible final states from the annihilation of Majorana 
particles $\chi$.
Their free Lagrangian is given by $\mathcal{L}_0=\bar f(i\slashed{\partial}-m_f)f$, 
which is invariant under Lorentz transformations, i.e.~invariant under $SU(2)_{L+R}$. In the 
massless limit, $m_f\rightarrow0$, this symmetry is upgraded to a chiral symmetry 
$SU(2)_L\times SU(2)_R$, in which the left and right handed Weyl states transform 
independently of each other, and
helicity and chiral eigenstates unify. 
For a fermion {\it pair} $\bar f f$, the spins can combine to either a singlet ($S=0$), 
\begin{align}
\left(\left|\uparrow\downarrow\right>-\left|\downarrow\uparrow\right>\right)/\sqrt{2},
\label{eq:spinstates1}
\end{align}
or a triplet ($S=1$) spin state,
\begin{align}
\left\{
\left|\downarrow\downarrow\right>, 
\quad\left(\left|\uparrow\downarrow\right>+\left|\downarrow\uparrow\right>\right)/\sqrt{2}, 
\quad\left|\uparrow\uparrow\right>
\right\}\,,
\label{eq:spinstates3}
\end{align}
where the arrows indicate the spin direction along the $z$-axis (the first entry refers to the 
antifermion, the second to the fermion).
If the two fermion momenta are \mbox{(anti-)}parallel -- e.g.~because they 
are emitted back-to-back as the final states of a DM annihilation process -- the $z$-axis can be
chosen to be aligned in the same direction as the momenta, and the above spin projections on the $z$ axis
are directly related to the {\it helicities} of the two particles.
Choosing $\mathbf{p}_f$ ($\mathbf{p}_{\bar f}$) to point along the positive (negative) $z$-axis, the helicity 
configurations $h=S_z p_z/\left| p_z\right|$ of the singlet and triplet state are then given by
\begin{align}
\label{eq:helstates1}
 \frac{1}{\sqrt{2}}\left(\left|-\frac12,-\frac12\right>-\left|+\frac12,+\frac12\right>\right)
 \to \frac{1}{\sqrt{2}}\left(\left|\bar f_R,f_L\right>-\left|\bar f_L,f_R\right>\right)
\end{align}
and
\begin{align}
\label{eq:helstates3}
&\left\{
\left|+\frac12,-\frac12\right>, 
\frac{1}{\sqrt{2}}\left(\left|-\frac12,-\frac12\right>+\left|+\frac12,+\frac12\right>\right), 
\left|-\frac12,+\frac12\right>
\right\} \nonumber\\
\to 
&\left\{
\left|\bar f_L,f_L\right>, 
\frac{1}{\sqrt{2}}\left(\left|\bar f_R,f_L\right>+\left|\bar f_L,f_R\right>\right), 
\left|\bar f_R,f_R\right>
\right\}
\,,
\end{align}
where the arrows indicate the chiral states in the left/right decoupling limit, i.e.~for
$m_f\to0$. 

The momentum configuration thus restricts which helicity states can be associated to the spin 
states. Angular momentum and the assumed $CP$ invariance, on the other hand, restrict 
which spin state can be realized. 
Since $CP=(-1)^{L+S}\times(-1)^{L+1}=(-1)^{S+1}$, for example, only the singlet state with $S=0$ is 
compatible with the odd $CP$ parity of the initial state.
Eq.~(\ref{eq:helstates1}) then tells us that both fermion and antifermion in this momentum
configuration must have the {\it same} helicity. In any chirally symmetric theory, however,
the antifermion must necessarily have the {\it opposite} helicity of the fermion. We note that 
angular momentum conservation alone leads to the same conclusion: since 
$\mathbf{L}=\mathbf{r}\times\mathbf{p}$,  
we must have $L_z=0$ and hence $S_z=J_z=0$.
Eqns.~(\ref{eq:helstates1}, \ref{eq:helstates3}) then imply that fermion and antifermion must, 
independently of the value of $S$, have the same helicity.
The annihillation process $\chi\chi\rightarrow\bar f f$ is therefore only possible if chiral 
symmetry is broken in the Lagrangian, for example through an explicit fermionic mass 
term $m_f\bar f_L f_R$ or through the coupling of the fermion $f$ to a scalar field 
$\lambda\phi\bar f_L f_R$. It follows that the amplitude of the annihilation process must be 
proportional to the chiral symmetry breaking parameters $m_f$ or $\lambda$. 

In addition, it is instructive to consider also the isospin of the involved particles.
Since left-and right-handed SM fermions transform under different representations
of $SU(2)_L$, the final states $\bar f_L f_R$ and $\bar f_R f_L$ have total isospin $I=1/2$.
The initial state $\chi\chi$, on the other hand, has necessarily integer isospin, implying 
$\Delta I\not= 0$. The annihilation rate thus has to vanish for an unbroken $SU(2)_L$,
and therefore has to be proportional to at least one power of the Higgs vacuum expectation 
value (VEV) $v_{EW}$. For heavy DM the ratio $\delta_v\equiv v_{EW}/m_\chi$ becomes small, 
and processes with  $\Delta I\not= 0$ will be suppressed by some 
power of $\delta_v$.

In total, this implies that the amplitude for $\chi\chi\to \bar f f$ has to involve
(at least) one parameter that breaks chiral symmetry, and one power of $v_{EW}$ that 
controls breaking of the $SU(2)_L$ symmetry. For the SM fermions that receive their mass 
from the Higgs mechanism, both of these conditions are fulfilled for the usual helicity 
suppression factor $m_f$. Depending on the model, however, there can be further possibilities, 
as we will discuss in detail for the case of the MSSM below, and it is useful to discriminate 
between the two suppression mechanisms. In the following,
we therefore refer to the suppression related to chiral symmetry breaking as 
{\it Yukawa suppression}, and to the one related to electroweak symmetry breaking as 
{\it isospin suppression}. While isospin suppression is controlled by only one parameter, 
$\delta_v = v_{EW}/m_\chi$, there can in principle be several sources of chiral symmetry
breaking, for example in models with more complicated Higgs sectors. Nevertheless, as we 
discuss in detail in Section \ref{sec:enhancement}, all terms that break chiral invariance 
in the MSSM are accompanied by Yukawa couplings $y_f\propto m_f/v_{EW}$. 
Even though the following discussion of suppression lifting is completely model independent 
we will thus continue to assume, for concreteness, that chiral symmetry breaking is controlled by $y_f$.

\subsection{Lifting of Yukawa and isospin suppression}
\label{sec:liftingModelIndep}

With the above discussion in mind, the only way to avoid  the suppression of non-relativistic 
Majorana DM annihilation is to allow for an additional final state particle. 
Lorentz invariance requires this additional particle to be a boson, such that the
leading process we are interested in is of the form
\be
\label{eq:XX2Bff}
  \chi\chi \to B \bar F f
\ee
where $B$ is a scalar or vector boson, and $F=f$ if $B$ is electrically neutral.
The additional boson can be either a SM particle, in particular a photon (refered to as
electromagnetic IB), a gluon, a weak gauge boson ($W^\pm$, $Z$) or
the Higgs boson $h$, or it can be a new particle beyond the SM (for example a heavy
Higgs boson within the MSSM). For the moment we want to keep the discussion 
model-independent, and therefore focus on the former case. 
A frequently used approximation is to restrict the discussion to $B$ being radiated off a 
fermion line in the final state, as described by soft and/or collinear 
splitting functions \cite{Ciafaloni:2010ti,Cirelli:2010xx}. We emphasize that
this approach does not capture the (partial) lifting of helicity suppression, and therefore is
inadequate for the case of heavy Majorana DM annihilation to fermions.

Taking the gauge restoration limit $v_{EW}\to 0$, it becomes straight-forward to exhibit the 
scaling of a given process with $y_f$ and $v_{EW}$. (We emphasize that we consider this 
limit only in order to discuss the possible mechanisms of Yukawa and isospin suppression 
lifting, while all our numerical results later on take the full dependence on $v_{EW}$ and 
$y_f$ into account). In this limit, the left- and right-handed components of the fermions in 
the final state and in internal lines can not only be considered as gauge interaction eigenstates 
but as independently propagating degrees of freedom. The fermion mass is treated 
perturbatively in the mass insertion approximation, and is associated with a chirality flip 
along with a suppression factor $m_f \propto y_f v_{EW}$. In  addition, longitudinally 
polarized gauge bosons $W_L/Z_L$ can be replaced by the corresponding Goldstone 
bosons $G^\pm, G^0$ by virtue of the Goldstone boson equivalence theorem, 
cf.~Eqs.~(\ref{eq:WardIdentityZ}-\ref{eq:WardIdentityW}) below. All final states thus 
have definite $SU(2)_L$ quantum numbers (i.e. $I=1/2$ for $G^\pm, G^0, h, f_L$, $I=0$ 
for $f_R$, and $I=1$ for $W_T$), except for the $Z_T$, which is a mixture of $I=0, 1$ 
states (even in the gauge restauration limit, we find it convenient to express our results in 
terms of the $Z$ boson instead of the neutral $SU(2)_L$  boson).

\begin{table}
\begin{center}
\begin{tabular}{|c|c|c|c|c|c|c|l|}
\hline
            &  $g$ &  $Z_T/\gamma$ & $W_T$ & $Z_L$ & $W_L$ & $h$  \\
\hline
$\bar F_R f_L$ or $\bar F_L f_R$ & $y_fv_{EW}$ & $y_fv_{EW}$ & $y_fv_{EW}$ & $y_f$ & $y_f$ & $y_f$ \\ 
$\bar F_L f_L$ & $1$ &$1$ & $1$ & $v_{EW}$ & $v_{EW}$ & $v_{EW}$  \\
$\bar F_R f_R$ & $1$ &$1$ & $1$ & $v_{EW}$ & $v_{EW}$ & $v_{EW}$  \\ 
\hline
\end{tabular}
\end{center}
\caption{Summary of Yukawa and isospin suppression(-lifting) in 3-body annihilation 
processes $\chi\chi\to B\bar F f$ for various final state boson $B$ and fermion combinations.
Entries $\propto 1$ correspond to processes that {\it potentially} can lift both Yukawa and isospin 
suppression of the 2-body process. Entries $\propto y_f$ can lift isospin suppression but 
are still suppressed by the Yukawa coupling, while those $\propto v_{EW}$ can 
lift Yukawa suppression but are still suppressed by $\delta_v=v_{EW}/m_\chi$ for large $m_\chi$. 
\label{tab:lifting}
}
\end{table}

The amplitude of the generic 3-body process indicated in Eq.~(\ref{eq:XX2Bff}) can 
be non-zero for $v_{EW}\to 0$ only if $\Delta I=0$, i.e.~if isospin is conserved. Furthermore, the 
amplitude must vanish for $y_f\to 0$ unless both fermions have the same chirality. Note that this is 
possible for 3-body processes because the kinematics does not force the fermions to be emitted 
back-to-back in the center-of-mass (CMS) frame, and therefore the arguments discussed in 
Section \ref{subsec:chiral} do not apply.\footnote{
In the extreme case where both fermions are emitted in the same direction, e.g., one simply has to
exchange $f_R\leftrightarrow f_L$ in Eqs.~(\ref{eq:helstates1}, \ref{eq:helstates3}), which allows equal
chiralities of the fermions in both the singlet and triplet spin state. In this kinematical configuration, it is 
easy to visualize how the fermion momentum can be balanced by the emitted boson $B$, and how their spin 
can combine with $S_B$ and $L$ to the required $J=0$ for both $S_B=0$ and $S_B=1$.
In general, the spin singlet and  triplet states will be linear combinations of {\it all} chiral states, with 
expectation values that depend on the angle between the fermion momenta, thus rendering the above
argument  essentially independent of the specific kinematical configuration. 
Also the requirement of $CP$ conservation is much less restrictive for 3-body than for 2-body final
states. A general discussion is somewhat	complicated by the fact that e.g.~$\bar F f$ is not necessarily 
a $CP$ eigenstate that could be analysed individually, but in principle straight-forward by classifying
all possible effective operators that connect initial and final states (similar in spirit to the analysis of 2-body 
final states presented in Ref.~\cite{Kumar:2013iva}).
}
 These two observations immediately determine which annihilation processes can lift either Yukawa or isospin 
 suppression (or both). In Table~\ref{tab:lifting}, we 
show schematically the required scaling of the amplitude that results from these considerations,
for various combinations of fermion chiralities and final state bosons (where the longitudinal gauge 
bosons represent the corresponding Goldstone bosons).
Both suppression factors can be lifted only in processes where a transverse gauge boson 
($Z_T$, $W_T$, $\gamma$, or gluon $g$) is emitted and the final state fermions are described by spinors of equal chirality 
($\bar F_R f_R$ or $\bar F_L f_L$).  For longitudinal gauge bosons ($Z_L$ or $W_L$) or the Higgs boson 
$h$, only one of the suppression factors can (potentially) be avoided for 3-body final 
states: isospin suppression can be lifted if the fermions are of opposite chirality, and  Yukawa 
suppression can be lifted if the fermions are of equal chirality. 

Let us stress that the symmetry 
arguments presented above simply guarantee that the amplitude must vanish for $y_f\to 0$ and 
$v_{EW}\to 0$, respectively, and the same applies to any {\it gauge invariant sub-sets} of diagrams.
The actual suppression can thus be stronger than indicated by Table~\ref{tab:lifting}, 
i.e.~by additional powers of $v_{EW}$ or $y_f$. 
At the same time, we caution that {\it single} diagrams can scale in a different way, depending on the 
gauge choice, such that the vanishing for $y_f\to 0$ or $v_{EW}\to 0$ is in general not guaranteed.

\subsection{Gauge invariance in IB processes}
\label{subsec:GI}

Following up on the last comment, let us for convenience briefly recall how to verify gauge 
independence and identify gauge invariant subsets of diagrams.
While for photon emission a good test is to check whether a given set of diagrams satisfies the Ward 
identity $\mathcal{M}^\mu(\chi\chi\rightarrow\bar f f\gamma)k_\mu = 0$, where $k_\mu$ is 
the momentum of the photon, this does not work for electroweak 
IB because $SU(2)_L\times U(1)_Y$ has been spontaneously broken. Indeed the 
question of gauge invariance changes in general, as weak hypercharge and isospin are 
no longer conserved in their original form. For the spontaneously broken 
Glashow-Weinberg-Salam theory the correct way to define gauge invariance is in 
terms of the preserved BRST symmetry \cite{Becchi:1974md,Becchi:1975nq}, under 
which SM field transformations involve ghost fields which arise from the electroweak 
gauge fixing procedure. This implies a new set of Ward identities, which in general depend 
on the choice of gauge. Using 
the standard $R_\xi$ class of gauges \cite{Fujikawa:1972fe}, we arrive at the Ward 
identities for electroweak IB as expected from the Goldstone equivalence theorem:
\begin{eqnarray}
\mathcal{M}^\mu(\chi\chi\rightarrow\bar f fZ)k_\mu =  im_Z \mathcal{M}(\chi\chi\rightarrow\bar f fG^0)\,,
\label{eq:WardIdentityZ}
\end{eqnarray}
\begin{eqnarray}
\mathcal{M}^\mu(\chi\chi\rightarrow\bar F fW^{\pm})k_\mu =  m_W \mathcal{M}(\chi\chi\rightarrow\bar F fG^{\pm})\,.
\label{eq:WardIdentityW}
\end{eqnarray}
%
We reiterate that Eqns.~(\ref{eq:WardIdentityZ}) and (\ref{eq:WardIdentityW}) in general apply to 
(subsets of) the full amplitude, 
not  individual diagrams, and are a valuable test for the results outlined in the next section.

\section{Neutralino annihilation to $\bar ff$ and an additional final-state particle}
\label{sec:enhancement}

In this section we apply the general discussion of helicity suppression lifting in Majorana DM
annihilation to the lightest supersymmetric neutralino as DM candidate,
and additional final state bosons charged under $SU(2)_L$. For photon or gluon IB
 we refer to the references listed in the introduction. Concerning the choice of
DM candidate, we note that much of the following discussion is still rather generic and can thus 
be extended in a straight-forward way to any theory with an extended Higgs sector or where the 
DM particles belong to a different electroweak multiplet.
We will introduce the relevant 3-body processes and Feynman diagrams in Section \ref{sec:neutralino-ann}, 
re-visit the discussion of the helicity suppression in light of the specific situation encountered in the MSSM
(Section \ref{subsec:Suppression}) and then demonstrate in detail how these suppressions can be lifted, 
fully or partially, in Section  \ref{sec:lifting}.  In Sections \ref{subsubsec:Mass} and \ref{sec_threshold},
finally, we discuss two mechanisms by which 3-body cross sections can be enhanced which are 
{\it not} related to the helicity suppression of 2-body final states.

\subsection{Full analytic amplitudes and gauge-invariant subsets}
\label{sec:neutralino-ann}

From now on, we thus assume DM to be composed of the lightest neutralino, 
$\chi\equiv \tilde\chi^0_1$, which is a superposition of Wino, Bino and Higgsino states,
\be\label{eq:mixing}
 \chi = N_{11} \tilde B + N_{12} \tilde W^3 + N_{13} \tilde H^0_1 + N_{14} \tilde H^0_2\,,
\ee
obtained by diagonalizing the neutralino mass matrix
\be\label{eq:massmatrix}
  \left(\begin{array}{cccc}
  M_1 & 0 &  \frac{-g'v_1}{\sqrt{2}} & \frac{g'v_2}{\sqrt{2}}\\
 0 & M_2  &  \frac{gv_1}{\sqrt{2}} & \frac{-gv_2}{\sqrt{2}} \\
  \frac{-g'v_1}{\sqrt{2}} & \frac{gv_1}{\sqrt{2}} & 0 & -\mu  \\
  \frac{g'v_2}{\sqrt{2}} & \frac{-gv_2}{\sqrt{2}} & -\mu & 0
  \end{array}\right).
\ee
Here, $M_1$ and $M_2$ are the Bino and Wino mass parameters, respectively, and $\mu$ 
is the Higgsino mass parameter; $v_1$ and $v_2$  are the VEVs of the two Higgs doublets,
with $v_{EW}=\sqrt{v_1^2+v_2^2}$ and $\tan \beta\equiv v_1/v_2$, and $g$ and $g'$ are 
the $SU(2)_L$ and $U(1)_Y$ couplings, 
respectively. We follow the conventions of Ref.~\cite{Edsjo:1997hp}, as implemented in \ds, 
and take all mass eigenvalues to be positive, while the diagonalization matrix $N$ can be complex.

 \begin{figure}[t!]
	\includegraphics[width=1.\columnwidth,clip]{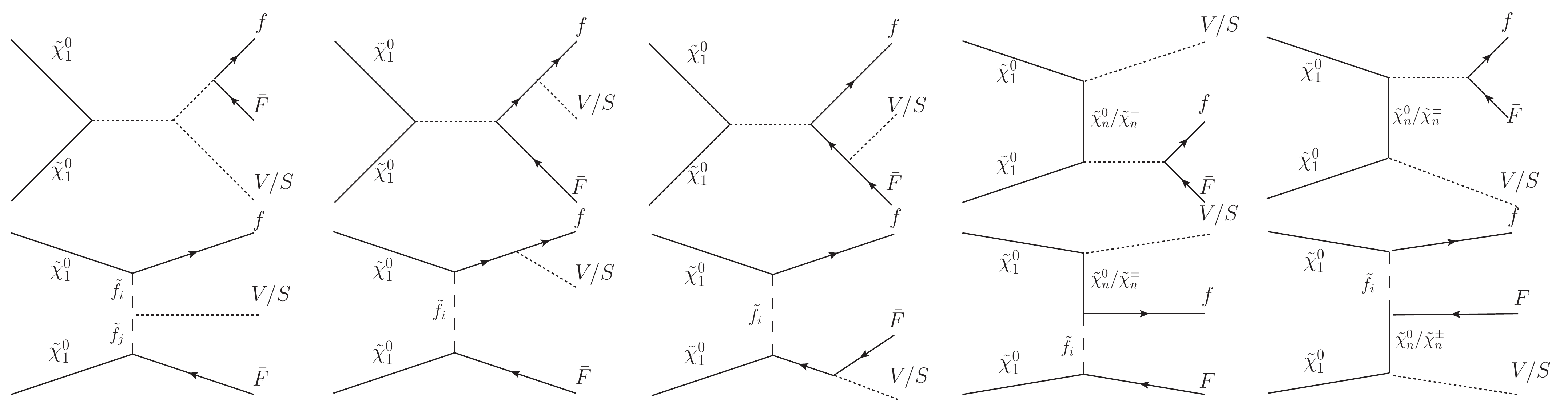}  
 	\caption{Condensed representation of all Feynman diagrams for  neutralino annihilation into 
	$\bar{F} fV$ or $\bar{F} fS$, where dotted lines indicate scalar 
	($S= A, h, H, H^{\pm}$) or vector ($V=Z,W^{\pm}$) mediators, depending on the final state configuration. 
	Fermion final states are identical, $F=f$, for neutral boson emission ($h$, $H$, $A$ or $Z$), while 
	$(f,F)$ constitute the two components of an $SU(2)_{L}$ doublet 
	for charged boson emission ($H^\pm$ or $W^\pm$). See text for more details on how the 
	individual topologies are referred to in this article.
	}
 	\label{fig:Feynman_all}
 \end{figure}

We want to consider here all 3-body final states that contains a fermion pair and 
a boson that is charged under $SU(2)_L$. Assuming $CP$-violating terms to be small,
the full list of processes of interest is thus 
\be
\label{eq:finalstates}
\chi\chi\to W^{+}\bar F f, Z\bar f f, H^{+}\bar F f, A\bar f f, H\bar f f, h\bar f f\,.
\ee
Here, $A$ denotes the $CP$-odd Higgs, $H^{+}$ the charged Higgs, and $H$ and $h$ the heavy
and light $CP$-even Higgs bosons, respectively. For charged boson final states, $f$ denotes any 
fermion doublet component with isospin $+1/2$, and $F$ the corresponding one with isospin $-1/2$; 
for neutral bosons, $f$ can be any SM fermion.

In Fig.~\ref{fig:Feynman_all}, we show all contributing 
Feynman diagrams in a condensed form
(note that some of these diagrams may vanish for specific combinations of internal and external 
particles). For future reference, we follow Ref.~\cite{Bringmann:2013oja} and refer to the top row of 
diagrams as (derived from 2-body) {\it $s$-channel processes},  and to the bottom row of diagrams as 
{\it $t/u$-channel processes} (noting that $t$- and $u$-channel amplitudes are identical in the 
$v\to0$ limit). Likewise, we denote diagrams of the type that appear in the first column as {\it virtual 
internal bremsstrahlung} (VIB), diagrams of the type that appear in the second and third column as {\it final
state radiation} (FSR),\footnote{\label{FSRdef}
We stress that this distinction between VIB and FSR, while useful for the specific 
purpose of our discussion, is {\it not} gauge invariant and exclusively refers 
to the {\it topology} of the involved diagrams. In particular, it should not be confused 
with an often used gauge invariant alternative set of definitions
where FSR refers exclusively to the soft or collinear photons radiated from the final
legs~\cite{Bringmann:2007nk,Ciafaloni:2010ti,Cirelli:2010xx}, while VIB is defined as the 
{\it difference} between the full amplitude squared and the FSR contribution~\cite{Bringmann:2007nk} .
}
and diagrams of the type that appear in the last two columns as {\it initial
state radiation} (ISR).

 \begin{figure}[t!]
 	\includegraphics[width=1.\columnwidth,clip]{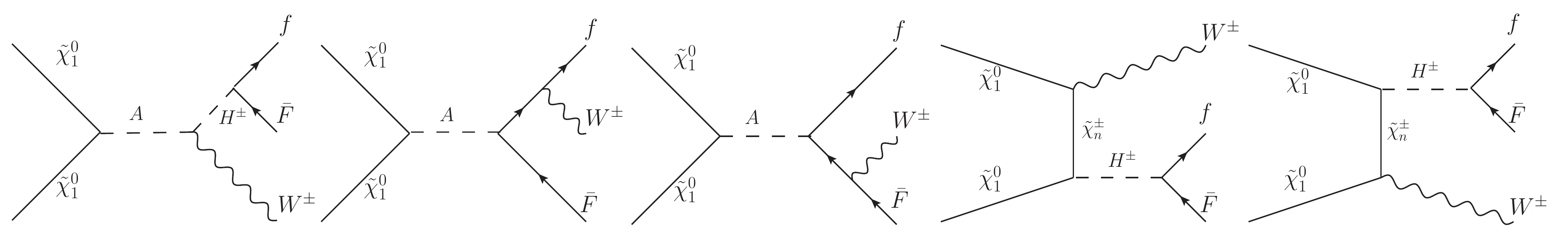}  
 	\caption{Gauge invariant set of amplitudes for neutralino DM annihilation into a fermion pair and 
	a $W$ boson, mediated by $s$-channel bosons with a mass at the scale of the 
	$CP$-odd Higgs $A$.  	
	}
 	\label{fig:Feynman_scalar}
 \end{figure}		

 \begin{figure}[t!]
 	\includegraphics[width=1.\columnwidth,clip]{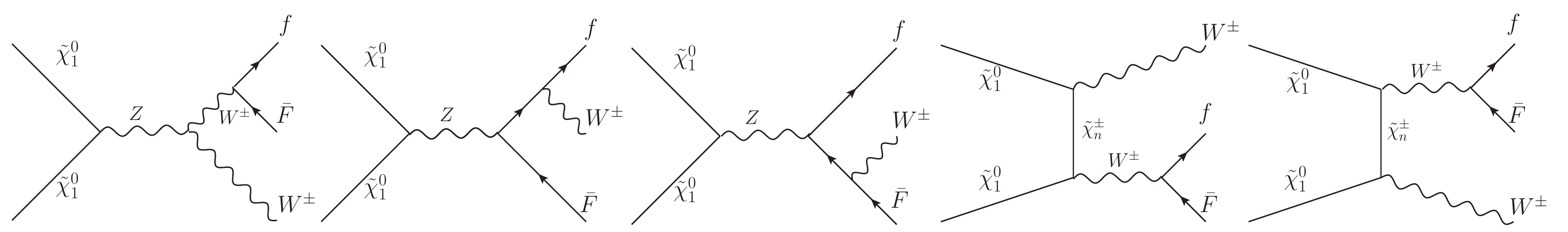}  
 	\caption{Same as Fig.~\ref{fig:Feynman_scalar}, but mediated by $s$-channel bosons with a mass 
	at the electroweak scale.  	
	}
 	\label{fig:Feynman_vector}
 \end{figure}		

 \begin{figure}[t!]
 	\includegraphics[width=1.\columnwidth,clip]{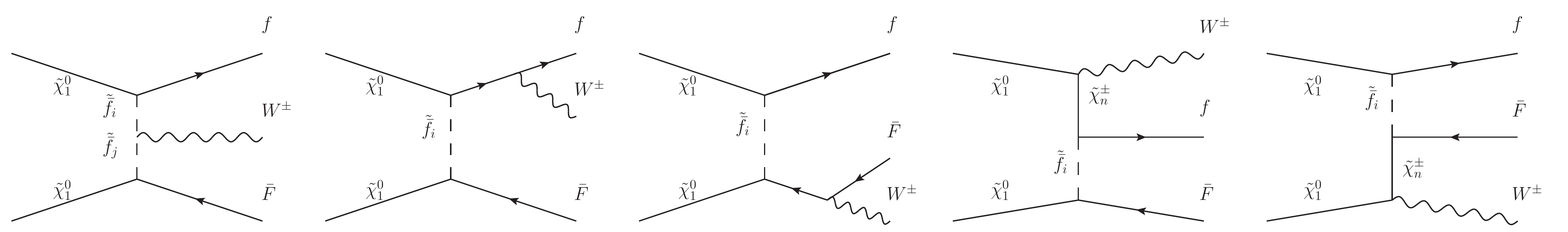}  
 	\caption{Same as Fig.~\ref{fig:Feynman_scalar}, but mediated by $t$-channel sfermions.  	
	}
 	\label{fig:Feynman_tchannel}
 \end{figure}		

We explicitly calculate the full analytical expressions for all these processes in the limit of vanishing 
relative velocity of the annihilating neutralino pair, see Appendix  \ref{app:ExpIntro} for 
technical details. We then use the Ward identities in 
Eqns.~(\ref{eq:WardIdentityZ}) and (\ref{eq:WardIdentityW}) to group diagrams into 
gauge invariant sets for the case of vector boson final states. In general we identified only two of 
such invariant sets: those diagrams that are
derived from 2-body $s$-channel processes and those that are derived from 2-body $t$-channel 
processes. In the limit $m_A\gg m_\chi$ -- which is phenomenologically particularly relevant
because the observed Higgs is very SM like -- the $s$-channel diagrams however split into two 
gauge-invariant subsets.  All diagrams then fall quite neatly into 3 categories: 
{\it heavy Higgs $s$-channel}, which are the set of diagrams with (at least one) mediator at the mass 
scale $M_A$ (see Fig.~\ref{fig:Feynman_scalar}), {\it weak-scale $s$-channel}, 
which are the set of diagrams with $s$-channel mediators at the weak scale 
(see Fig.~\ref{fig:Feynman_vector}), and {\it $t$-channel}, which are the set of diagrams with 
sfermion mediators (see Fig.~\ref{fig:Feynman_tchannel}).\footnote{
We note that for $v\to0$ the two $s$-channel ISR diagrams are actually {\it identical}, but 
for clarity we still include them separately in figures \ref{fig:Feynman_scalar} and 
\ref{fig:Feynman_vector}. For $v\to0$ and $m_F\to0$, also the two $t$-channel ISR 
diagrams are identical; in practice, the difference only matters for final states containing top and bottom
quarks.  
As an important cross-check of our final amplitudes, we confirmed analytically that these identities 
indeed hold.
}
For $Z\bar f f$ and $h\bar f f$ final 
states the three sets of diagrams can be obtained analogously: $t$-channel contributions 
involve at least one sfermion line, while the remaining diagrams belong to the $s$-channel 
category (which can be further split into subsets involving at least one mediator
at scale $M_A$, or none, respectively).

\subsection{Helicity suppression in the MSSM}
\label{subsec:Suppression}

As established in the previous Section, the `helicity suppression' of the 2-body annihilation 
rate by a factor of $m_f^2/m_\chi^2$ is indeed the combination of in principle independent 
Yukawa and isospin suppressions. 
Let us now turn back to this observation and discuss it in more detail in light of the MSSM,
where both mechanisms are still intrinsically linked because of the connection between gauge 
symmetry and chiral structure in the MSSM Lagrangian.

\subsubsection{Yukawa Suppression}
\label{subsubsec:Yukawa}

The chiral symmetry of the MSSM Lagrangian is broken by terms 
proportional to Yukawa couplings (in order to avoid flavour-changing neutral currents, we 
assume as usual that the $A$-terms are proportional to the Yukawa coupling matrices). 
Following the general arguments of Section~\ref{subsec:chiral}, any amplitude 
contributing to $\chi\chi\rightarrow\bar f f$ must therefore be proportional to $y_f$. 
Within the MSSM the values of $y_f$ 
are functions of $\tan\beta$ but, except for the top quark,  in general so small that this 
can lead to a suppression of 
the 2-body amplitudes by many orders of magnitude.
From the point of view of the broken theory, this 
{\it Yukawa suppression} appears to arise from rather different types of contributions to the Lagrangian:
\begin{itemize}
\item[{\it i)}]  fermion mass terms
\item[{\it ii)}]  couplings of any of the five physical Higgs fields to fermions
\item[{\it iii)}] couplings of fermions to sfermion mass eigenstates (which mix 
the left- and right-handed fields). 
\end{itemize}
For example, the first case is relevant for annihilation into fermions via $t$-channel sfermion 
exchange if the sfermion mixing is small (otherwise, the third contribution can dominate 
the amplitude), and the second for annihilation via $s$-channel pseudoscalar mediation.

We note that all three interaction types couple left- and right-handed states and hence 
can `flip' the helicity of one of the final state fermions. The helicity combinations that would 
result in a chirally symmetric theory, $\bar f_{R,L }f_{R,L }$,  can thus be transformed into 
those compatible with the global symmetry requirements outlined in 
Section~\ref{subsec:chiral}, $\bar f_{R,L }f_{L,R }$. 
Traditionally,  the notion of this  helicity flip is sometimes 
taken to refer specifically to the case {\it (i)}, in which it is the (kinematic) fermion mass that breaks chiral 
symmetry in the Lagrangian. Instead, we associate the effect directly with the Yukawa couplings 
in the MSSM Lagrangian (which of course give rise to the SM fermion masses).

\subsubsection{Isospin Suppression}
\label{subsubsec:VEV}

As also discussed in Section \ref{subsec:chiral}, the annihilation process $\chi\chi\to\bar f f$
furthermore violates weak isospin, $\Delta I\not=0$, and therefore its amplitude has to vanish in the
gauge restoration limit $v_{EW}\to 0$. The resulting {\it isospin suppression} by a factor 
$\delta_v\equiv v_{EW}/m_\chi$, for heavy neutralinos,  can arise from different terms in the Lagrangian 
of the broken theory:
\begin{itemize}
\item[{\it a)}] fermion mass terms
\item[{\it b)}] mixing of different gauge multiplets (Bino, Higgsino, Wino) that contribute to the 
lightest neutralino mass eigenstate given by Eq.~(\ref{eq:mixing})
\item[{\it c)}] mixing of left- and right-handed sfermion eigenstates.
\end{itemize}
The structure of the neutralino mass matrix (\ref{eq:massmatrix}) indeed confirms that neutralino mixings vanish
for $v_{EW}\to 0$, as required by $SU(2)_L$ invariance.
Note that case ${\it (a)}$ and ${\it (c)}$ are intrinsically linked
to an accompanying chirality violation, since $m_f\propto y_f v_{EW}$ and the off-diagonal terms in 
the sfermion mass matrix are also proportional to $y_f$ within the MSSM.
Let us consider as an illustration the $t$- and $s$-channel contributions to $\chi\chi\to\bar f f$.
The kinematical helicity suppression due to the fermion mass $m_f$ is relevant for the $t$-channel
(sfermion exchange). In this case Yukawa and isospin suppression simply arise from the two factors
in $m_f\propto y_f v_{EW}$ (case {\it (a)} and {\it (i)}, respectively). In addition, the Yukawa and isospin 
violation can be due to the sfermion mixing (case {\it (c)} and {\it (iii)}). Indeed, due to the mixing, a given 
sfermion mass eigenstate can couple to both left- and right-handed fermions, which then gives rise to the
required chirality flip. 

For $s$-channel annihilation, on the other hand, the situation is more interesting in the sense that 
Yukawa and isospin suppression cannot simply be traced back to the same origin. For a pseudoscalar 
Higgs boson $A$ as mediator, e.g., the Yukawa suppression stems directly from the Yukawa 
coupling $\propto y_f A\bar f f$ (case {\it (ii)}), while the isospin suppression arises from the neutralino 
mixing (case {\it b}): for pure gauge multiplets the coupling $A\bar\chi\chi$ would be forbidden by 
$SU(2)_L$ invariance, and therefore vanishes for $v_{EW}\to 0$. For a $Z$-boson 
in the $s$-channel, the discussion of the limit $v_{EW}\to 0$ is a bit more involved (see Appendix \ref{app:diagrams}),
but is essentially analogous to the case of an $A$ mediator.

\subsection{Yukawa and isospin suppression lifting}
\label{sec:lifting}

In Section \ref{sec:liftingModelIndep}, we discussed which 3-body final states $\chi\chi\to B\bar F f$ can 
potentially lift the Yukawa- and/or isospin suppression of the process $\chi\chi\to \bar f f$, for the case in 
which $B$ is a
SM gauge boson or a Higgs boson. This general discussion based on isospin and chiral symmetry in the 
 limit $v_{EW}\to 0$ can be extended to the MSSM, as shown in Table \ref{tab:liftingMSSM}, 
by noting that all physical Higgs bosons $h, H, A, H^\pm$ have isospin $I=1/2$. Compared to 
Table \ref{tab:lifting}, the amplitudes for $A\bar f f$ scale as expected in the same way as $Z_L\bar f f$, 
noting that in the gauge restoration limit $Z_L$ is given by the Goldstone boson $G^0$ (and hence 
transforms in a similar way as the pseudoscalar $A$). Similar arguments apply to 
the other Higgs bosons.

\begin{table}
\begin{center}
\begin{tabular}{|c|c|c|c|c|c|}
\hline
            &    $Z_T$ & $W_T$ & $Z_L/A$ & $W_L/H^\pm$ & $h/H$  \\
\hline
\multirow{2}{*}{$\bar F_R f_L$ or $\bar F_L f_R$} & $y_fv_{EW}$ & $y_fv_{EW}$ & $y_f$ & $y_f$ & $y_f$ \\ 
&$_{\rm no\ enhancement}$&$_{\rm no\ enhancement}$& $_{t/u,\,s}$ & $_{t/u,\,s}$& $_{t/u,\,s}$\\ \hline
\multirow{2}{*}{$\bar F_L f_L$} & $1$ & $1$ & $v_{EW}$ & $v_{EW}$ & $v_{EW}$  \\
& $_{t/u(\tilde B,\tilde W),\,s(\tilde H)}$ & $_{t/u(\tilde B,\tilde W),\,s(\tilde H,\tilde W)}$ &$_-$& $_{t/u}$& $_{t/u,\,s}$\\ \hline
\multirow{2}{*}{$\bar F_R f_R$} & $1$ & $1$ & $v_{EW}$ & $v_{EW}$ & $v_{EW}$ \\ 
& $_{t/u(\tilde B),\,s(\tilde H)}$ &$_-$&$_-$&$_-$&   $_{t/u,\,s}$\\ \hline
\end{tabular}
\end{center}
\caption{As Table~\ref{tab:lifting}, but applied to weak gauge and Higgs boson final states 
within the MSSM. We also indicated whether the process can be realized with the maximal 
enhancement allowed by chiral and isospin symmetry in $t+u$ and $s$-channel annihilation 
processes, respectively. For the first two columns we also specify for which neutralino 
composition ($\tilde B=$ bino-like, $\tilde W$=wino-like, $\tilde H$=Higgsino-like) the maximal
enhancement occurs. For the last three columns $t+u$-channel processes are possible 
for $\tilde B$- or $\tilde W$-like neutralino as well as mixed $\tilde H/\tilde B$ or 
$\tilde H/\tilde W$, and $s$-channel processes are possible for mixed $\tilde H/\tilde B$ 
or $\tilde H/\tilde W$. Entries with a dash do not contribute to the order we are
working in (see Appendices \ref{app:AmplitudeExpansion} and \ref{app:diagrams} for details).
\label{tab:liftingMSSM}
}
\end{table}

 \begin{table}
\begin{center}
{\scriptsize
\begin{tabular}{|c|c|c|c|c|}
\hline
 & $2\rightarrow 2$ & \multicolumn{3}{|c|}{$2\rightarrow 3$}     \\  \hline
 & & $Z_T$, $W_T$ &  \multicolumn{2}{|c|}{$Z_L$, $W_L$, h} \\  \cline{3-5}
 & & $\bar{F}_L f_L$, $\bar{F}_R f_R$ & $\bar{F}_L f_R$, $\bar{F}_R f_L$ & $\bar{F}_L f_L$, $\bar{F}_R f_R$ \\ 
 \hline
 \hline
 \multirow{15}{*}{$t$-channel I} & \multirow{14}{*}{\includegraphics[width=2.6cm]{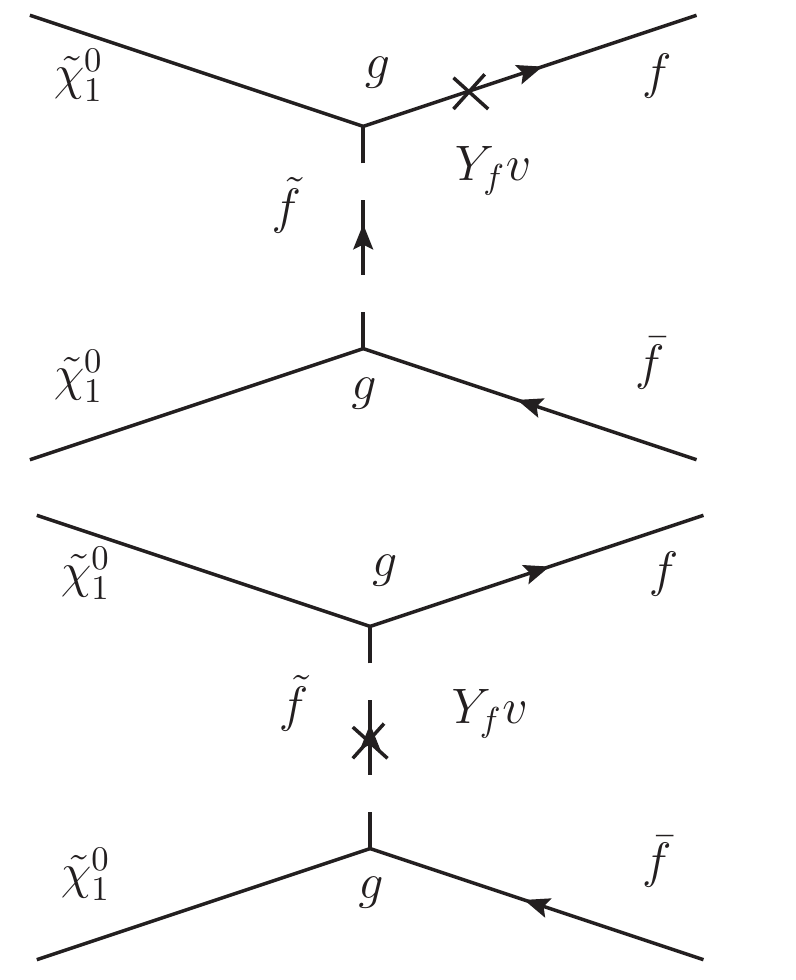}} &  \multirow{14}{*}{\includegraphics[width=2.6cm]{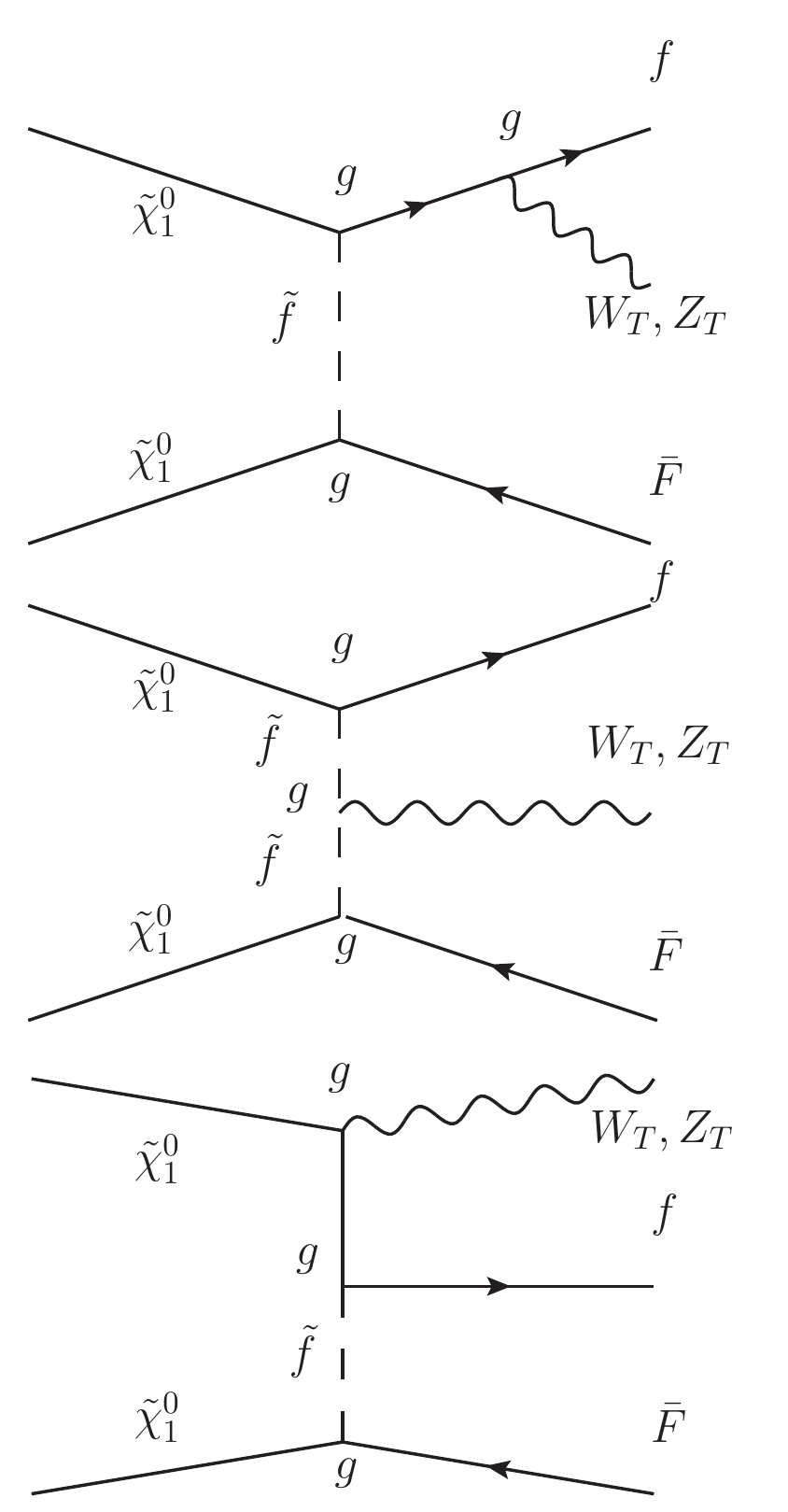}} &  \multirow{14}{*}{\includegraphics[width=2.6cm]{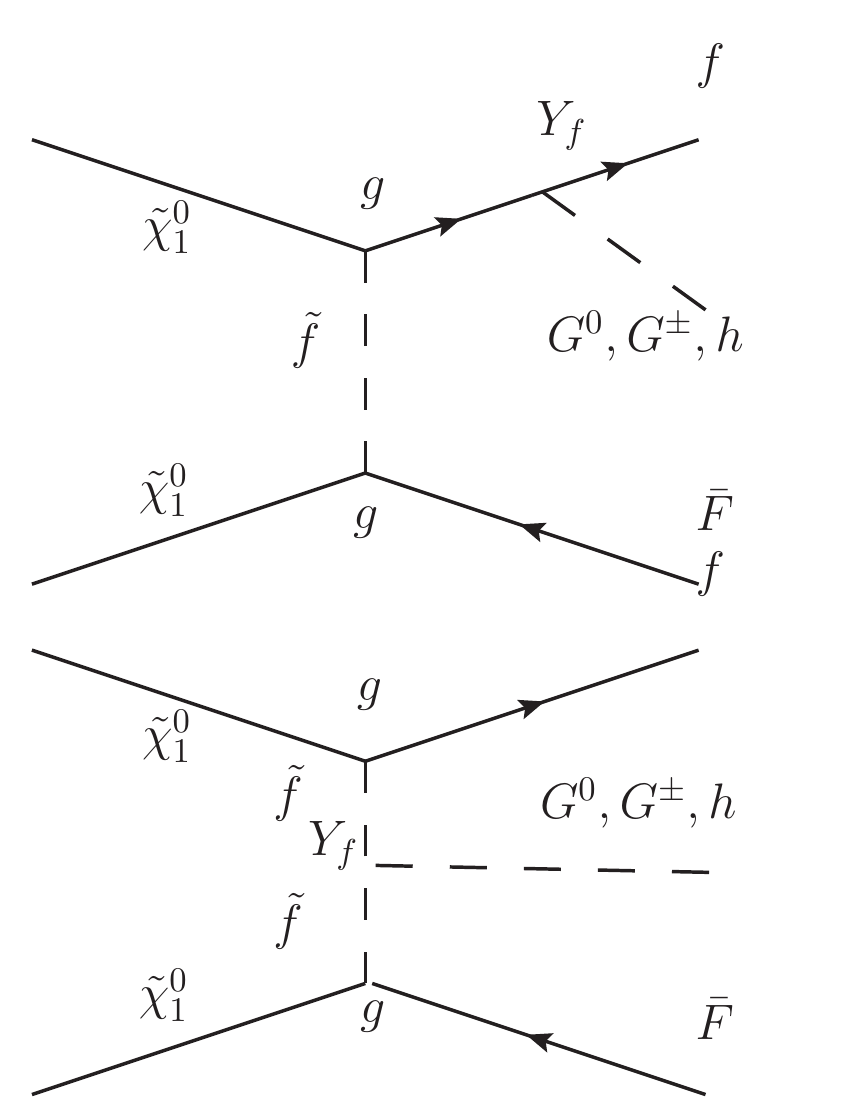}} & \multirow{14}{*}{\includegraphics[width=2.6cm]{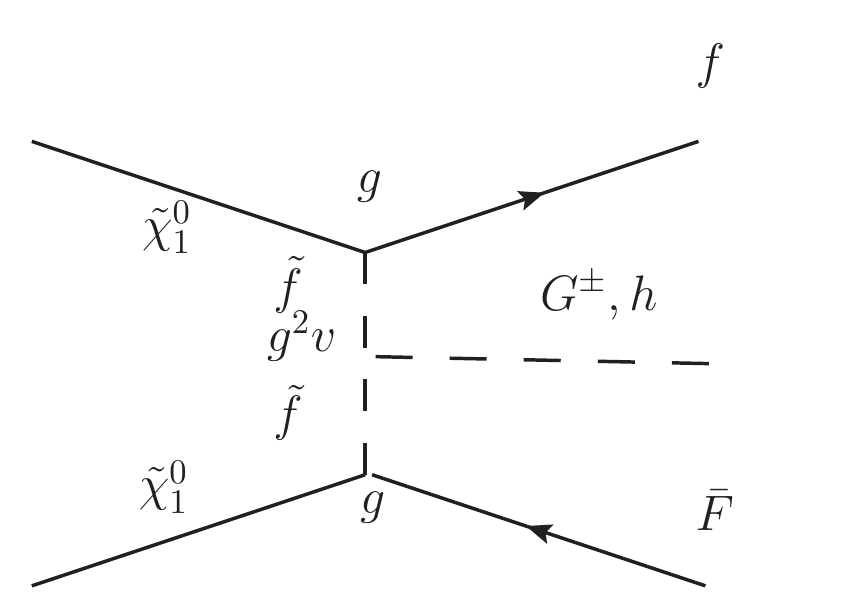}}\\
   & & & &\\
  & & & &\\
  & & & &\\
  & & & & \\
  & & & &\\
  & & & &\\
  & & & &\\
   & & & &\\
  & & & &\\
  & & & & \\
  & & & &\\
  & & & &\\
  & & & &\\
\hline
 \multirow{6}{*}{$t$-channel II} & \multirow{6}{*}{\includegraphics[width=2.6cm]{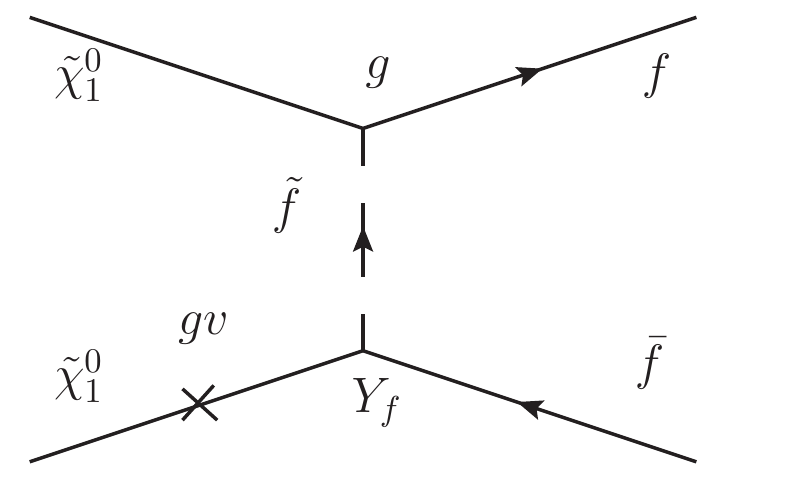}} & & \multirow{6}{*}{\includegraphics[width=2.6cm]{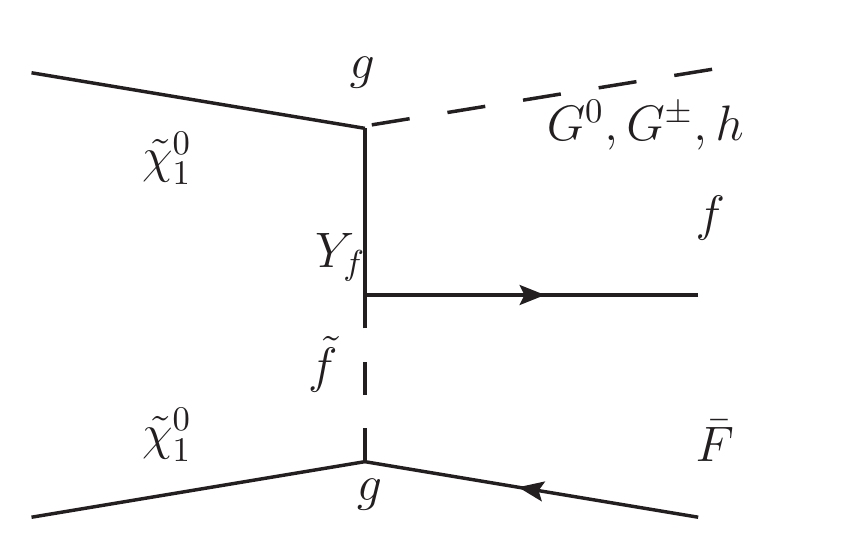}} & \multirow{6}{*} {\includegraphics[width=2.6cm]{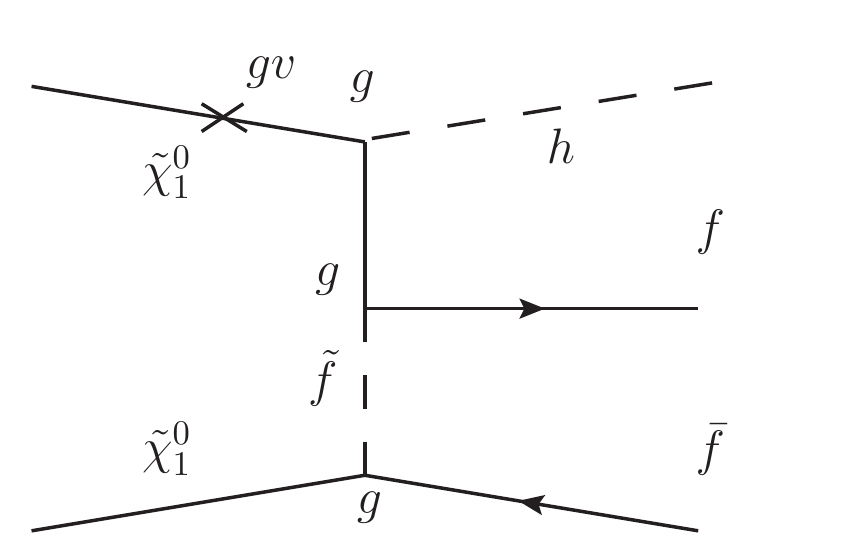}}  \\
   & & & &  \\
   & & & &\\
   & & & &\\
   & & & & \\
   & & & &\\
 \hline
  \multirow{6}{*}{$s$-channel EW} & \multirow{6}{*}{\includegraphics[width=2.6cm]{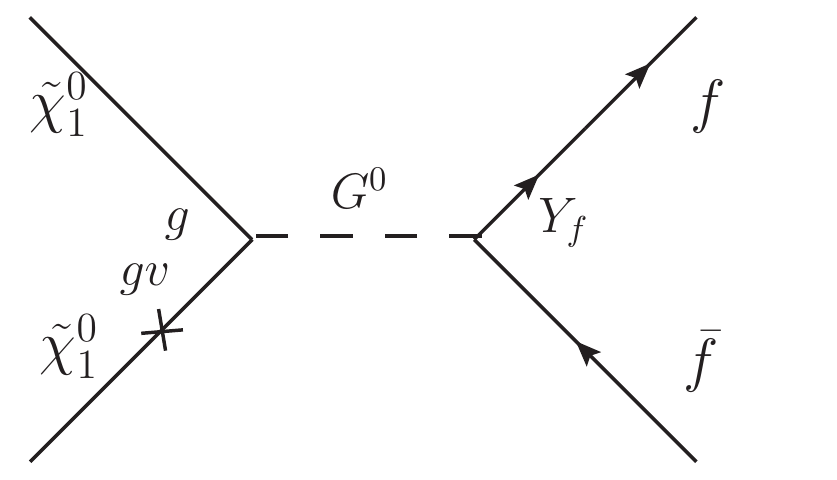}} & \multirow{6}{*}{\includegraphics[width=2.6cm]{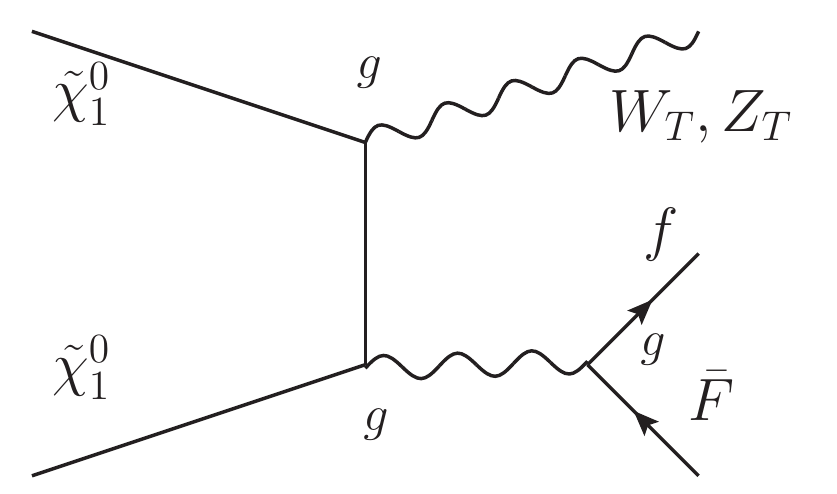}}  & \multirow{6}{*}{\includegraphics[width=2.6cm]{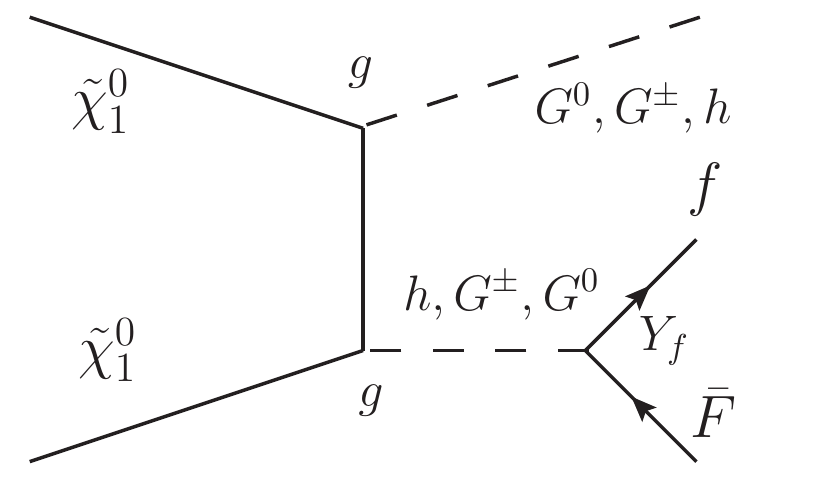}} & \multirow{6}{*}{\includegraphics[width=2.6cm]{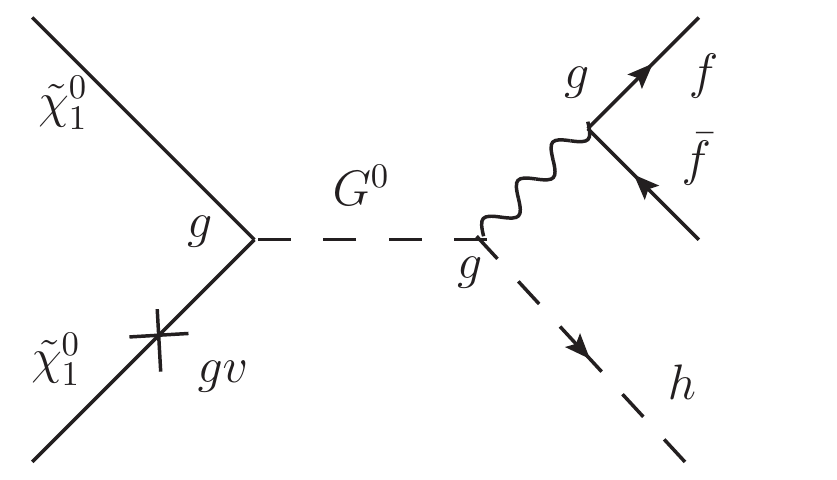}}  \\
   & & & &  \\
   & & & &\\
   & & & &\\
   & & & & \\
   & & & &\\
 \hline
  \multirow{6}{*}{$s$-channel $M_A$} & \multirow{6}{*}{\includegraphics[width=2.6cm]{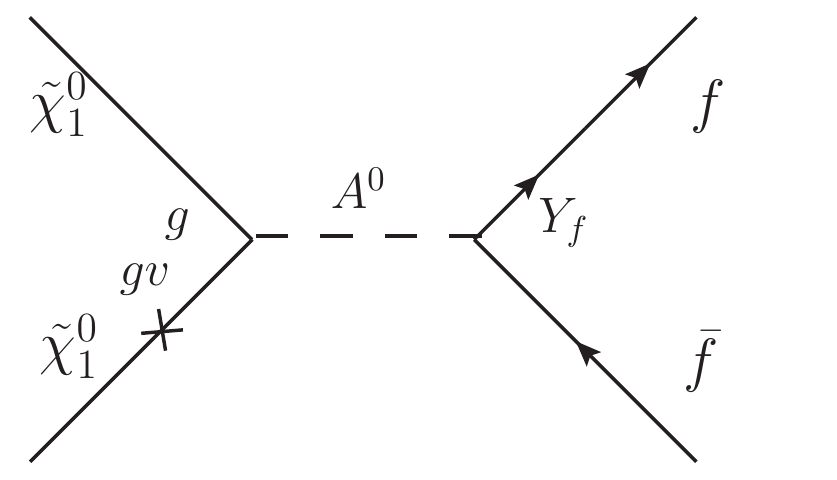}} & & \multirow{6}{*}{\includegraphics[width=2.6cm]{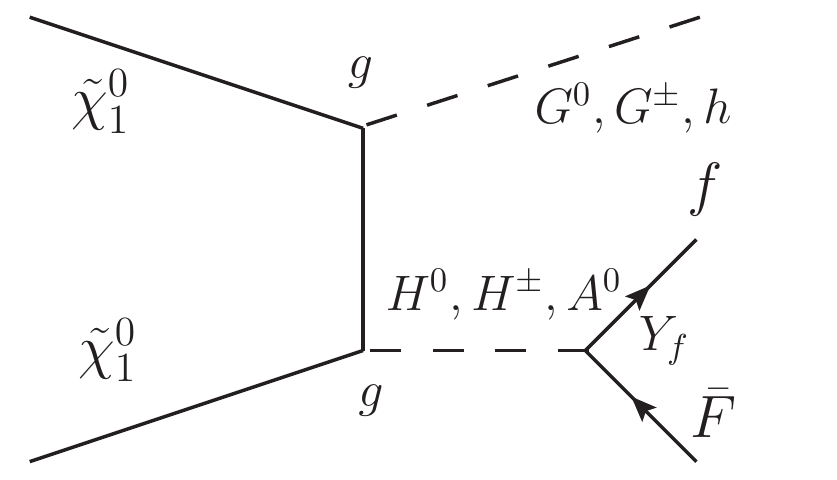}} &   \\
   & & & &  \\
   & & & &\\
   & & & &\\
   & & & & \\
   & & & &\\
\hline
\hline
Lifting of &  & Yukawa + Isospin & Isospin & Yukawa \\
\hline
\end{tabular}
}
\caption{Diagrams for annihilation into fermions $\bar f f$ and $B \bar F f$ (for $B=W,Z,h$) in the gauge restoration 
	limit $v_{EW}\to 0$. The rows correspond to the four gauge-invariant subsets of diagrams that can 
	be discriminated in this limit (see Appendix \ref{app:diagrams} for details).
        The first column corresponds to the 2-body process, and the other 
	columns show various 3-body processes. The diagrams shown in the second column lift both 
	Yukawa and isospin suppression. The diagrams in the third column lift only isospin suppression, 
	and in the fourth column only Yukawa suppression. We show only one representative diagram 
	for each topology (ISR/FSR/VIB) and suppression mechanism. The coupling 
	factors attached to vertices and mass/mixing insertions give the 
	scaling with $y_f$, $v_{EW}$ and $g$ of each diagram (for Bino- or Wino-like neutralinos; 
	modifications for Higgsino-like neutralinos
        are described in Appendix \ref{app:diagrams}).
        Note that contributions with $W_T$ emitted via ISR (second column, first and third row)
        exist for Wino- or Higgsino-like neutralinos; those with $Z_T$ emitted via ISR
        occur only for a Higgsino-like neutralino.
	}
\label{fig:lifting}
\end{center}
\end{table}

In Appendix \ref{app:AmplitudeExpansion}, we consider  the full analytic expressions
for six different mass hierarchies of particular phenomenological interest and determine  
for each of the previously discussed gauge-invariant subsets of diagrams the leading
order in $v_{EW}$ and $y_f$. The result of this exercise is collected in 
 Tables~\ref{tab:EnhancementsW} -- \ref{tab:EnhancementsHiggs},
  where we present the ratio of the leading term for the 3-body amplitude and the corresponding
  2-body amplitude. This allows us, as also indicated in 
 Table~\ref{tab:liftingMSSM},  to identify which contributions to the 3-body amplitudes 
  actually realize the suppression lifting that we can maximally expect on the basis of our general
  symmetry arguments; the `missing' cases, for which we did {not} find a contribution within the 
  MSSM, are marked by a ‘-’. For a detailed technical discussion of the various lifting mechanisms, and 
  how they
  are realized at the level of individual diagrams, we refer to Appendix \ref{app:diagrams}.
We provide  a graphical summary in Table \ref{fig:lifting}, where we show representative diagrams that realize the lifting of
isospin and/or Yukawa suppression, for the sets of gauge invariant classes of diagrams that can be
discriminated in the gauge restoration limit (in addition to the three sets discussed before, the $t$-channel can be
split into contributions that remain non-zero in the limit of pure neutralino states (I), and those that require neutralino
mixing (II)). Isospin suppression can be lifted in all cases by the emission of longitudinal gauge bosons (here represented by the
Goldstone bosons) or a Higgs boson. Lifting of Yukawa suppression, as well as lifting of both suppression factors, is more restricted.
This can be traced back to basic properties of the unbroken MSSM Lagrangian and the conservation of
$J_{CP}=0_-$ (see Appendix \ref{app:diagrams} for details), explaining the 
  `missing' entries in Table~\ref{tab:liftingMSSM}. Let us also highlight that the classification procedure revealed ways to 
lift the 2-body suppression that have not been pointed out for the MSSM before (in particular 
Higgsstrahlung via $t$-channel ISR and a specific $s$-channel VIB process, shown in the last column and second/third row
in Table \ref{fig:lifting}, respectively).

\subsection{Heavy propagator suppression}
\label{subsubsec:Mass}

An additional form of suppression, unrelated to the discussion so far, arises in diagrams that 
rely on mixing between neutralinos or contain heavy propagators. 
This {\it mass suppression} takes the form $\delta_\textrm{X}\equiv m_\chi/M_\textrm{X}$, where X is 
the heavy state in question. 
In particular, both $s$-channel contributions to $\chi\chi\rightarrow\bar f f$ and a subset of $t$-channel 
contributions -- those of type $(II)$, see Appendix \ref{app:AmplitudeExpansion} -- rely on 
mixing the Bino/Wino with the Higgsino. For example, for a Bino- or Wino-like neutralino, 
the 2-body amplitude in the $s$-channel is suppressed by a factor
$\delta\mu^2=m_\chi^2/\mu^2$ if $|\mu|\gg m_\chi$. For a Higgsino-like neutralino, on the other hand, 
it is suppressed by $\delta_{M_i}=m_\chi/M_i$ for $M_i\gg m_\chi$, where $M_i={\rm min}(M_1, M_2)$ 
(see Table \ref{tab:2to2}).

These suppression factors of the $s$-channel annihilation can be lifted for the case of a 
Wino- or Higgsino-like neutralino by the emission of a (transverse) $W$ or $Z$ from one of 
the initial neutralino lines (ISR). (The corresponding diagram is illustrated in the third row, 
second column of Table \ref{fig:lifting}.) Additionally, this 3-body process simultaneously
lifts both isospin- and Yukawa suppression. It is particularly relevant if the 2-body final states 
$WW$ and $ZZ$ are kinematically forbidden, such that the internal gauge boson is off-shell. 
This is a special case of the threshold effects that we turn to next.

\subsection{Threshold effects}
\label{sec_threshold}
A given 2-body channel $\chi\chi\to AB$ is strongly phase-space suppressed 
if the CMS energy is close to the mass of the final-state particles, and for $2m_\chi\leq m_A+m_B$
the corresponding partial cross section vanishes completely in the $v\to0$ limit. If either $A$ or $B$ 
are off-shell and decay 
into much lighter states, however, the phase-space opens up again and thereby
potentially increases even the total 2-body annihilation rate significantly. 
For the MSSM, this is particularly relevant for the $W^+W^-$ and 
$\bar tt$ channels,  which has previously been studied for specific neutralino compositions 
\cite{Chen:1998dp,Yaguna:2010hn} (for an approximate numerical implementation in the
context of relic density calculations, see  \cite{Belanger:2013oya}). 
For the processes we are interested in here, 
threshold effects can in general appear for {\it any} two-boson final states (or $\bar tt$).

For a more detailed discussion of this effect, it is useful to rewrite the 
3-body cross section as (see e.g. \cite{Agashe:2014kda,Uhlemann:2008pm})
\bea\label{eq:2to3forResonance}
  \sigma v_{2\to 3} &=& \frac{S}{4E_{\chi_1}E_{\chi_2}}\int \overline{|{\cal M}_{2\to3}|^2} \; d\Phi_3(P;p_1,p_2,p_3) \nn\\
  &=& \frac{S}{4E_{\chi_1}E_{\chi_2}}\int \overline{|{\cal M}_{2\to3}|^2} \; d\Phi_{2}(P;p_1,q) \times \frac{dq^2}{2\pi} \times d\Phi_{2}(q;p_2,p_3)\,,
\eea
where $d\Phi_n(P;p_1,\dots,p_n) = (2\pi)^4 \delta^{(4)}(P-\sum p_i) \prod_i \frac{d^3p_i}{(2\pi)^32E_i}$ 
is the $n$-body phase space element, $P=p_{\chi_1}+p_{\chi_2}$ the sum of the $4$-momenta of 
the annihilating neutralinos and $E_{\chi_i}$ their energy; the $p_i$ denote the final-state momenta.  
Since $q^2=(p_2+p_3)^2$ is time-like, we will in the following often use the notation 
$q^2\equiv m_{23}^2$ instead.
For the processes considered here, c.f.~Eq.~(\ref{eq:finalstates}),
the symmetry factor $S$ is always $1$.\footnote{\label{foot_Sfactors}
In general, if some of the final state particles are of identical type,
configurations that differ only by exchanging these particles should be counted only once 
in the phase space integration. Since this will be convenient later on, we thus use a convention 
where one integrates over all of the phase space as if all particles were distinct, and then correct 
for the corresponding over-counting by a symmetry factor $S$. It is $S=1$ if all final-state particles 
are distinct, and $S=1/2$ ($S=1/6$) if two (all three) of them are identical. 
} 
Furthermore, $\overline{|{\cal M}|^2}$ denotes the usual squared matrix element, 
averaged over initial spins and summed over final spins/helicities.

We now assume that the amplitude is dominated by a resonant, almost on-shell internal 
propagator that decays into particles 2 and 3, and hence carries 
momentum $q$. For a resonance $R$ with mass $M$, width $\Gamma$, 
and spin $1$, $1/2$, or $0$, respectively, we then have
\be\label{eq:Mdecomposition}
  {\cal M}_{2\to3} = \frac{1}{m_{23}^2-M^2+iM\Gamma} \times \left\{\begin{array}{cl} 
    {\cal M}_{2\to2}^{(q)\,\mu}(-g_{\mu\nu}+q_\mu q_\nu/M^2){\cal M}_{1\to2}^{(q)\,\nu} & {\rm vector} \\
    {\cal M}_{2\to2}^{(q)}(\slashed{q}+M){\cal M}_{1\to2}^{(q)} & {\rm fermion} \\
    {\cal M}_{2\to2}{\cal M}_{1\to2} & {\rm scalar} 
  \end{array}\right.
\ee
where ${\cal M}_{2\to2}$ (${\cal M}_{1\to2}$) is the matrix element for  
$\chi\chi\to p_1 q$ ($R^*\to p_2p_3$), up to polarization vectors or spinors for the `external' 
particle $R$ (as indicated by the superscript $q$).

The decisive observation is now that $\int \left|{\cal M}_{1\to2}\right|^2 d\Phi_{2}(q;p_2,p_3)$ 
must be independent of the polarization state of $R$ once all the final state polarizations 
are summed over. This is familiar from on-shell momenta $q$  -- the total 
(but not differential) decay rate of a particle is independent of its polarization state -- but 
holds more generally for time-like initial momenta $q$  \cite{Uhlemann:2008pm}.
{\it As long as the full phase-space integral is performed} (see Section \ref{sec:NWAdiff} for 
how to treat differential cross sections), one may thus conveniently replace the
correlated polarization or spin structure of Eq.~(\ref{eq:Mdecomposition})  with an unpolarized sum:
\begin{eqnarray}\label{decorr}
  \overline{\left|{\cal M}_{2\to2}^{(q)\,\mu}(-g_{\mu\nu}+q^\mu q^\nu/M^2){\cal M}_{1\to2}^{(q)\,\nu}\right|^2} &=& \Big|\sum_\lambda{\cal M}_{2\to2}^{(q)\,\mu} \epsilon_\mu^{*\lambda}\epsilon_\nu^{\lambda}{\cal M}_{1\to2}^{(q)\,\nu}\Big|^2 \nonumber\\
& \to & \frac13 \sum_{\lambda_1,\lambda_2} \left|{\cal M}_{2\to2}^{(q)\,\mu}\epsilon_\mu^{*\lambda_1}\right|^2 \left|\epsilon_\nu^{\lambda_2}{\cal M}_{1\to2}^{(q)\,\nu}\right|^2 \;, \\
  \overline{\left|{\cal M}^{(q)}_{2\to2}(\slashed{q}+M){\cal M}^{(q)}_{1\to2}\right|^2} &=& \Big|\sum_s{\cal M}^{(q)}_{2\to2} u_s \bar u_s {\cal M}^{(q)}_{1\to2}\Big|^2 \nonumber\\
& \to & \frac12 \sum_{s_1,s_2} |{\cal M}^{(q)}_{2\to2} u_{s_1}|^2 |\bar u_{s_2} {\cal M}^{(q)}_{1\to2}|^2 \,,
\end{eqnarray}
In this way, we can {\it independently of the spin} of $R$ 
replace
\be
 \overline{|{\cal M}_{2\to3}|^2} \to 
 \frac{ \overline{|{\cal M}_{2\to2}|^2} ~\overline{|{\cal M}_{1\to2}|^2}}{(m_{23}^2-M^2)^2+M^2\Gamma^2}
\ee
in Eq.~(\ref{eq:2to3forResonance}) which, for $v\to0$,  leads to
\be
\label{sigmares_offshell}
\sigma v_{2\to 3}^{\mathrm{res}}=\mathcal{S}\int^{(2m_\chi-m_1)^2}_{(m_2+m_3)^2} 
\frac{dm_{23}^2}{\pi} \frac{m_{23}}{(m_{23}^2-M^2)^2+M^2\Gamma^2}\,
\tilde \Gamma_{R\to23}\, \widetilde{\sigma v}_{\chi\chi\to1R}\,.
\ee
Here, the decay rate of the off-shell resonance in the frame where $q=(m_{23},\mathbf{0})$ is given 
by 
\be
\tilde \Gamma_{R\to23}\equiv\frac{S_{23}}{2m_{23}}\int \overline{\left|{\cal M}_{1\to2}\right|}^2 d\Phi_{2}(q;p_2,p_3)
= \frac{S_{23}}{16\pi}\frac{\lambda^\frac12(m_{23}^2,m_2^2,m_3^2)}{m_{23}^3} 
\overline{\left|{\cal M}_{1\to2}\right|}^2\,,
\ee
and the cross section for the annihilation into an off-shell resonance is given by
\be
\label{deftildesv}
\widetilde{\sigma v}_{\chi\chi\to1R}\equiv\frac{S_{1R}}{P^2}\int \overline{\left|{\cal M}_{2\to2}\right|}^2 d\Phi_{2}(P;p_1,q)
= \frac{S_{1R}}{128\pi}\frac{\lambda^\frac12(4m_\chi^2,m_1^2,m_{23}^2)}{m_\chi^4} \overline{\left|{\cal M}_{2\to2}\right|}^2\,.
\ee
In the last step we performed the phase-space integral explicitly by using the fact that for 
$v\to0$ the annihilation process is kinematically the same as a pseudo-scalar decay, implying that 
$\overline{\left|{\cal M}\right|}^2$ cannot have any angular dependence.\footnote{
For this reason, the result takes the same form as for off-shell 
decays~\cite{Chakrabortty:2016idh,Chakrabortty:2017oje}, 
suggesting a straight-forward generalization 
to 4-body final states dominated by the annihilation into two off-shell particles:
\be
\sigma v_{2\to 4}^{\mathrm{res}}=\mathcal{S}\int
\frac{dm_{12}^2}{\pi}\frac{dm_{34}^2}{\pi} 
\frac{m_{12}}{(m_{12}^2-M_{R_1}^2)^2+M_{R_1}^2\Gamma_{R_1}^2}\,
\frac{m_{23}}{(m_{23}^2-M_{R_2}^2)^2+M_{R_2}^2\Gamma_{R_2}^2}\,
\tilde \Gamma_{R_1\to12}\,\tilde \Gamma_{R_2\to34}\, \widetilde{\sigma v}_{\chi\chi\to R_1R_2}\,.
\ee
} 
Eq.~(\ref{sigmares_offshell}) will thus continue to hold for general $s$-wave annihilation, provided one 
replaces $4m_\chi\to s$ in Eq.~(\ref{deftildesv}). 
The squared matrix elements are here again summed (averaged) over final (initial) spins/helicities,
leading to an overall symmetry factor of ${\cal S}=S/(S_{1R}S_{23})$ (with $S_{1R}, S_{23}$ defined in 
accordance with footnote \ref{foot_Sfactors}).

We note that Eq.~(\ref{sigmares_offshell}) can be significantly simplified by a few well-motivated
assumptions. Concretely, let us assume the off-shell particle to decay to massless final states, 
$m_2=m_3=0$, and $\overline{\left|{\cal M}_{1\to2}\right|}^2\propto M^2$ close to the threshold; 
this implies $\tilde{\Gamma}_{R\to 23}=({m_{23}}/{M}){\Gamma}_{R\to 23}$. 
We also introduce a {\it reduced cross section}
\be
 (\sigma v)_{\rm red} \equiv (\sigma v)_{\chi\chi\to1R} / \lambda^{n+1/2}(1,\mu_1,\mu_R)\,,
\ee
with $\mu_R\equiv m_{23}^2/s$ and $\mu_1\equiv m_1^2/s$, allowing for the 2-body
cross section close to threshold to be suppressed not only by a phase-space factor ($n=0$), but 
by an additional such factor from the matrix element itself (as e.g.~in the
example of Higgsino annihilation below, for which we have $n=1$). By definition, $(\sigma v)_{\rm red}$
thus remains finite both above and below the threshold. Assuming $ (\sigma v)_{\rm red}$ to be independent
of $m_{23}$ close to threshold, Eq.~(\ref{sigmares_offshell}) simplifies to 
\be
\label{threshold_gen}
\sigma v_{\chi\chi\to 1R^*}\simeq 
\mathcal{S} (\sigma v)_{\rm red} \int_0^{\mu_{\rm max}}
\frac{d\mu}{\pi}\frac{\gamma \mu}{(\mu-1)^2+\gamma^2} \lambda^{n+1/2}(1,\mu_1,\mu \mu_R)\,,
\ee
where $\mu_{\rm max}={{(\sqrt{s}-m_1)^2}/{m_R^2}}$ and $\gamma\equiv \Gamma_R/M$. This
expression is model-independent in the sense that the threshold correction can be directly 
estimated for any given 2-body cross section (i.e.~without first having to compute $\tilde{\sigma v}$
or $\tilde \Gamma$).

\begin{figure}[t!]
	\centering
		\includegraphics[width=0.7\columnwidth]{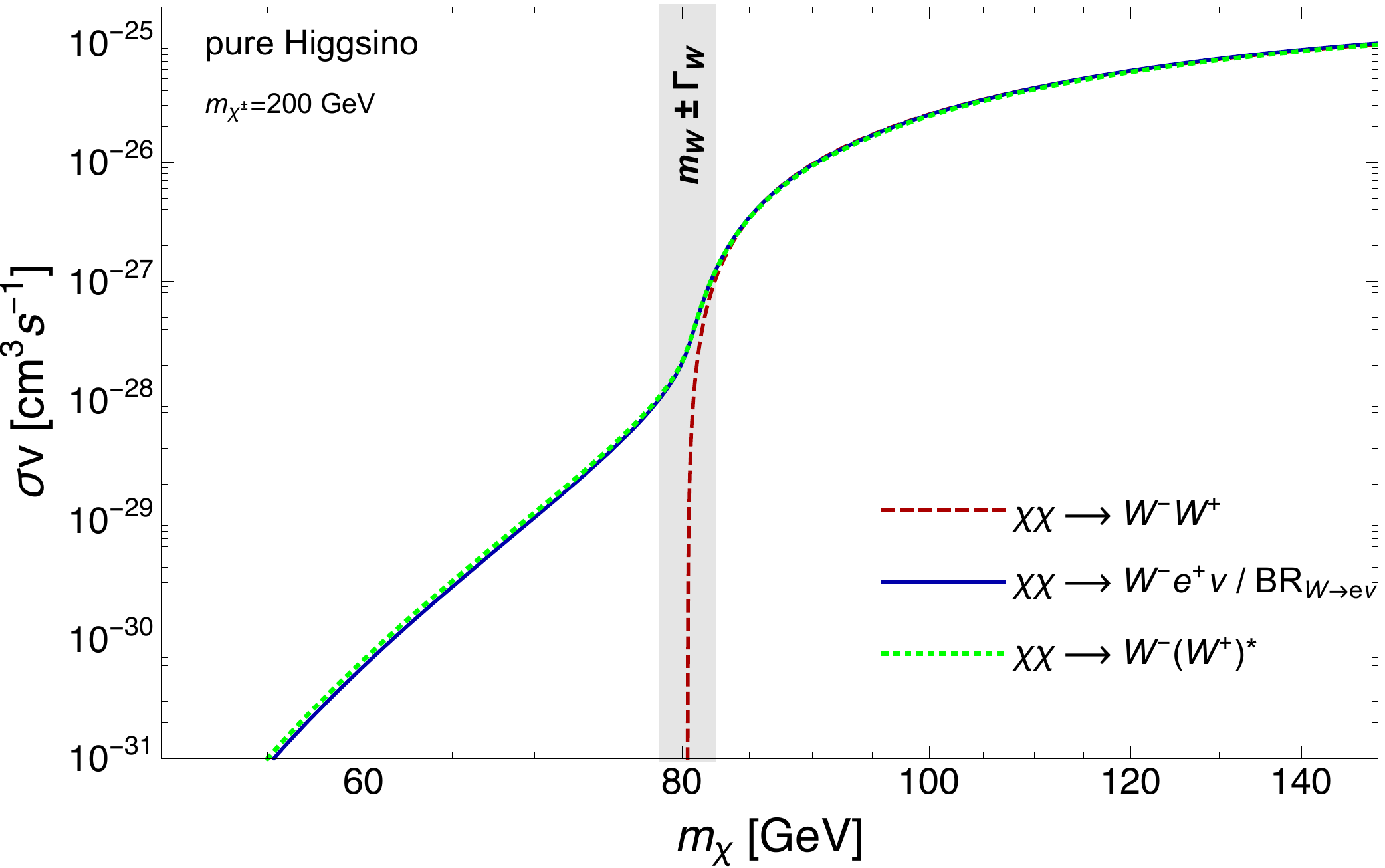}
	\caption{Cross section for pure Higgsinos annihilating to $W$ bosons, $\chi\chi\to W^+W^-$, 
	compared to $\chi\chi\to W^-e^+\nu$. For the latter process, we show the cross section
	divided by the branching fraction $\Gamma_{W\to e\nu}/\Gamma_W\simeq1/9$ (solid lines). 
	For comparison, we also include the model-independent estimate of Eq.~(\ref{threshold_gen})
	for ${\chi\chi\to W^-(W^+)^*}$ (dotted lines).
	For $m_\chi\lesssim m_W$,
	the 3-body cross section is clearly larger than the lowest-order result; above the threshold,
	on the other hand, the two agree exactly.
	}
	\label{fig:Higgsino_threshold}
\end{figure}

As an illustrative and concrete example, let us consider the process $\chi\chi\to W^-e^+\nu_e$ 
in the limit of pure Higgsino DM. For simplicity, we assume that sleptons are much heavier than 
the neutralino, such that the only contributing
diagrams are of the $V=W^-$ ISR type, with a virtual Higgsino-like chargino and a resonance 
$R={W^+}^*$. In this limit, we find
\be
 \widetilde{\sigma v}_{\chi\chi\to WW^*}=
 \frac{g^4}{512\pi}\frac{\left(16m_\chi^4-8m_\chi^2(m_W^2+m_{23}^2) +(m_W^2-m_{23}^2)^2\right)^\frac32}
 {\left(2m_\chi^2+2m_{\tilde{\chi}^+}^2 -m_W^2-m_{23}^2\right)^2}
\ee
and
\be
\label{Gammatildeexample}
\tilde{\Gamma}_{W^*\to e\nu}=\frac{g^2m_{23}}{48\pi}=\frac{m_{23}}{m_W}{\Gamma}_{W\to e\nu}\,.
\ee
We calculate the full 3-body cross section as  derived in Appendix \ref{app:ExpIntro}, in the pure 
Higgsino limit, and then compare it to the result given in Eq.~(\ref{sigmares_offshell}). 
As shown in Fig.~\ref{fig:Higgsino_threshold}, we obtain excellent agreement even though both the 
directly involved amplitudes and the numerical phase-space integrations are very different in nature 
(the two results for the 3-body cross section, shown as solid lines, lie exactly on top of each other). 
This should of course be expected for a 
process which by construction {\it only} receives contributions from an off-shell final-state particle, but we 
stress that Eq.~(\ref{sigmares_offshell}) is in general much simpler to calculate in praxis for such cases.
For comparison, we also indicate (with dotted lines) the model-independent result given in 
Eq.~(\ref{threshold_gen}); as one can see, even this simplified expression  provides an excellent 
approximation to the full result.

Most importantly, our example illustrates the much more general point
that a 3-body process around or below the kinematic threshold of a large 2-body process can
be significantly enhanced over the {\it total} annihilation rate at lowest order. Above the threshold and rescaled
to the relevant branching ratio for the decay of the resonance $R$, on the other hand, the 3-body cross 
section for a process $\chi\chi\to 1R, R\to23$ equals almost {\it exactly} the 2-body result -- an 
effect which we will discuss in detail in the next Section.

\section{Double counting issues}
\label{sec:doublecount}

We now turn to double counting issues related to unstable final-state 
particles. If the final state of a 2-body annihilation process undergoes a 
subsequent $1\to 2$ decay, in particular, this can also be viewed as a 3-body process 
with the unstable particle (the resonance, in our wording) as an intermediate state. 
While we discussed the situation 
{\it below} the kinematic threshold for the production of the
unstable particle in Section \ref{sec_threshold} as a way of enhancing the total cross section,
we are here interested in the kinematic region {\it above} the threshold.
As before, this is relevant for all massive diboson as well as $\bar tt$ final states considered
here. 

One possibility to avoid over-counting identical kinematic configurations when adding
2-body and 3-body processes would be to altogether disregard the former for 
massive diboson or $\bar t t$ final states.
Interferences between (nearly) on- and off-shell contributions to the amplitude 
would then be correctly accounted for, as
well as the impact of the spin of the resonance. However, this procedure has 
several drawbacks on a practical level, and furthermore turns out to be incorrect for 
2-body processes with identical particles in the final state (such as e.g.~$\chi\chi\to ZZ$), 
as will be discussed in more detail below. We therefore prefer to explicitly subtract 
on-shell contributions to the 3-body processes, 
which allows us to keep most of the advantages of the full 3-body computation
while correctly taking into account all symmetry factors. 
In the following we describe this procedure in more detail for both the 
total cross section and the differential yield of e.g. gamma rays. 

\subsection{Narrow width approximation and total cross section}

For 3-body processes dominated by an on- or off-shell resonance, the {\it total} cross section can 
be written as in Eq.~(\ref{sigmares_offshell}). If the intermediate particle corresponds to 
a nearly on-shell resonance with $\Gamma\ll M$, furthermore, the Breit-Wigner propagator can be 
approximated as
\begin{equation}\label{eq:nwa}
 \frac{1}{(m_{23}^2-M^2)^2+M^2\Gamma^2} \to \frac{\pi}{M\Gamma} \delta(m_{23}^2-M^2)\,.
\end{equation}
This narrow-width approximation (NWA) yields the on-shell contribution of the resonance $R$,
and we denote the corresponding, approximated cross section by $\sigma v^{NWA}$.
Strictly speaking, for the approximation to work well, the kinematic boundaries have to be 
sufficiently far away from the pole, $|m^2_{23}-M^2| \gg M\Gamma$, and
all contributions from the matrix element and phase-space factors apart from the Breit-Wigner 
propagator should be smooth functions of $m_{23}^2$ in the vicinity of the pole,
which we assume in the following.
With this replacement in Eq.~(\ref{sigmares_offshell}), we immediately recover the well-known result
\begin{equation}\label{eq:sigvNWA}
  \sigma v_{2\to 3}^{NWA} =  {\cal S} \times \sigma v_{\chi\chi\to 1,R} \times BR_{R\to 23} \;,
\end{equation}
where $BR_{R\to 23}=\Gamma_{R\to 23}/\Gamma$ is the branching ratio for 
the  resonance $R$ to decay into particles $2$ and $3$, 
$\Gamma_{R\to 23}=S_{1\to2}/(2M) \int d\Phi_2 \overline{|{\cal M}_d|^2}$ is the 
partial decay width, and 
$\sigma v_{\chi\chi\to 1R}=S_{2\to2}/P^2 \int d\Phi_2 \overline{|{\cal M}_p|^2}$ 
is the 2-body cross section. 

In general, more than one resonance can contribute to a given 3-body process, 
and one has to sum over all those contributions (in principle 
there can be interference effects for overlapping resonances with 
$|M_1-M_2|\lesssim \Gamma_1+\Gamma_2$; we will assume this is not the case). 
The narrow-width limit for the processes in Eq.~(\ref{eq:finalstates}) is thus given by
\begin{eqnarray}\label{eq:subtraction_total}
 \sigma v^{NWA}_{W^+f\bar F} &=& \sigma v_{WW}BR(W\to f\bar F) + \sigma v_{W^+H^-}BR(H^-\to f\bar F) \\
 \sigma v^{NWA}_{W^+b\bar t} &=&  \sigma v_{W^+H^-}BR(H^-\to b\bar t) + \sigma v_{t\bar t} BR (t\to W^+b)\\
 \sigma v^{NWA}_{Z f\bar f}  &=& 2\,\sigma v_{ZZ}BR(Z\to f\bar f) +\sigma v_{ZH}BR(H\to f\bar f) \nonumber\\
 && +\sigma v_{Zh}BR(h\to f\bar f) \\
 \sigma v^{NWA}_{H^+f\bar F} &=& \sigma v_{W^-H^+}BR(W\to f\bar F) \\
 \sigma v^{NWA}_{A f\bar f} &=& \sigma v_{Ah}BR(h\to f\bar f) + \sigma v_{AH}BR(H\to f\bar f)\\
 \sigma v^{NWA}_{H f\bar f} &=& \sigma v_{AH}BR(A\to f\bar f) + \sigma v_{ZH}BR(Z\to f\bar f)\\
 \sigma v^{NWA}_{h f\bar f} &=& \sigma v_{Ah}BR(A\to f\bar f) + \sigma v_{Zh}BR(Z\to f\bar f)\;,\label{eq:subtraction_total_last}
\end{eqnarray}
where $f$ denotes any SM fermion, and $F$ its $SU(2)_L$ doublet 
partner. The branching fractions $BR$ are given by the tree-level decay 
widths, divided by the total width appearing in the corresponding Breit-Wigner propagators. 
As stated in the second line, third-generation quarks have to be treated separately 
because of the contribution from top decay. 
Note that these results justify why the interference effects mentioned above can 
indeed be neglected: those 
would  be potentially relevant only in small regions of the MSSM parameter 
space, where the charged Higgs is degenerate in mass with the $W$ boson or the top 
quark (or, instead, one of the heavy neutral Higgses close in mass to the $Z$ boson 
or the SM Higgs $h$).

The total annihilation cross section is then given by
\be
\label{svtot}
  \sigma v = \sigma v_{2\to 2} + \sigma v_{2\to 3} - \sigma v^{NWA}_{2\to 3}\;,
\ee
where each term corresponds to the sum over all possible 2- and 3-body final states, 
respectively.\footnote{%
We neglect loop corrections to $\sigma v_{2\to 2}$ because only 3-body final states
can lift the $m_f^2/m_\chi^2$ suppression.
For very heavy neutralinos, however, enhancements 
$\propto \frac{\alpha}{\pi}\ln^2(m_W^2/m_\chi^2)$ from both soft/collinear IB 
and one-loop corrections to $\sigma v_{2\to 2}$ can become 
sizeable. For EW corrections these logarithmic terms will in general
give a non-zero contribution, 
unless the initial state is a singlet under $SU(2)_L\times U(1)_Y$, such as for a pure 
Bino \cite{Bloch:1937pw, Kinoshita:1962ur}
(the latter also applies to $U(1)$ and $SU(3)$ IB; see Ref.~\cite{Bringmann:2015cpa} for an efficient 
model-independent way of taking the relevant loop contributions into account).
Resummation of logarithmically enhanced contributions has been discussed e.g. in
\cite{Bauer:2014ula,Ovanesyan:2014fwa,Baumgart:2014vma} for pure Winos and Higgsinos.
In addition, for neutralinos with a significant Wino fraction and TeV mass, Sommerfeld enhancement
can play an important role \cite{Hisano:2003ec, Beneke:2014gja, Beneke:2016jpw}.
A joint treatment of all these effects for the general MSSM is beyond the scope of this work,
but would be desirable in view of future indirect detection probes.
}
 In the following, we refer to the difference 
$\sigma v^{\rm sub}_{2\to 3} \equiv \sigma v_{2\to 3} - \sigma v^{NWA}_{2\to 3}$
as the (NWA-subtracted) contribution from 3-body final states, with a similar definition 
for individual 3-body final states. 
We note that $\sigma v^{\rm sub}_{2\to 3}$  can be 
negative (although $\sigma v>0$, of course). To match our conventions for the 
computation of 3-body cross sections, the 2-body cross sections appearing above 
should be evaluated in $s$-wave approximation. Finally, we stress that, even when 
summing over all possible 3-body final states, $\sigma v^{NWA}_{2\to 3}$ is not equal 
to the sum over 2-body cross sections with diboson and $t\bar t$ final states, as one 
may have naively expected. This is partially because the Higgs resonances
can also decay into pairs of bosons, and partially due to a mismatch in the combinatorial 
factors, which can be traced back to ambiguities in the
narrow-width limit. For example, for $\chi\chi\to ZZ^* \to Zf\bar f$, each of the $Z$ bosons 
could act as intermediate resonance, which intuitively explains the factor ${\cal S}=2$ 
encountered in this case. These ambiguities would disappear if one were to treat 
both $Z$ bosons on equal footing, i.e.~consider $4$-body final states, which however  is
impractical for a general MSSM computation. 

\subsection{Spectrum of stable particles}
\label{sec:NWAdiff}

In general, our final state particles $p$ will fragment and decay into potentially
observable particles $P$, such as gamma rays or antiprotons.  For a given 3-body 
annihilation channel, and a conventional normalization to the corresponding yield from 
the 2-body rate, the spectrum can  be written as
\begin{equation}
\label{spectrum}
\frac{dN_P} {dE_P} = \frac{1}{\sigma v^\mathrm{tot}_{2\to2}}\sum_p \int^{E_{p}^{\rm max}}_{E_{p}^{\rm min}}  \frac{dN^{p \rightarrow P+ ...}_P} {dE_P} \frac{d\sigma v_{2\to3}} {dE_p} \, dE_p\,.
\end{equation}
Here, ${dN^{p \rightarrow P+ ...}_P}/{dE_P}$ describes the number of stable 
particles  $P$, per energy bin, that result from the inclusive  process 
$p\rightarrow P + ...$  of a particle $p$ decaying in flight (with energy $E_p$), and
the sum  has in principle to be performed over all helicity states separately 
(because $dN/dE_P$ can differ for different helicities of $p$). Assuming $CP$ 
conservation and that the decaying particles have very narrow widths, a very useful 
approximation in practice consists
in considering instead unpolarized cross sections and replace 
${dN^{p \rightarrow P+ ...}_P}/{dE_P}\to \frac12{dN^{\bar{p} p \rightarrow P+ ...}_P}/{dE_P}$, where the inclusive process $\bar{p} p\rightarrow P + ...$  is evaluated for a 
CMS energy of $2 E_p$ (see e.g.~\cite{Bringmann:2013oja}). 
This quantity can easily be obtained from event generators like {\sf Pythia} 
\cite{Sjostrand:2007gs} and, unlike ${dN^{p \rightarrow P+ ...}_P}/{dE_P}$, has the further 
advantage of being manifestly color neutral.\footnote{%
Note that for quark final states Eq.~(\ref{spectrum}) is not {
correct beyond leading order, where partons fragment independently}, because
it ignores flux tubes.
In general, for a final state 
consisting of a (color-neutral) boson $B$ and a quark pair $\bar qq$, which may have different masses, 
one should instead consider
\begin{equation}
\label{spectrumqq}
{\sigma v^\mathrm{tot}_{2\to2}}\frac{dN_P} {dE_P} = \int^{E_{B}^{\rm max}}_{E_{B}^{\rm min}}  \frac{dN^{B \rightarrow P+ ...}_P} {dE_P} \frac{d\sigma v_{2\to3}} {dE_B} \, dE_B+
\int^{E_{\bar q}^{\rm max}}_{E_{\bar q}^{\rm min}}  \int^{E_{q}^{\rm max}}_{E_{q}^{\rm min}} 
\frac{dN^{\bar qq \rightarrow P+ ...}_P} {dE_P} \frac{d\sigma v_{2\to3}} {dE_{\bar q}dE_q} \, dE_{\bar q}\, dE_q\,.
\end{equation}
The fragmentation function ${dN^{\bar qq \rightarrow P+ ...}_P}/{dE_P}$ that appears here can
be obtained by boosting to the back-to-back system of the quarks, defined as 
$\mathbf{p}_q^\mathrm{bb}=-\mathbf{p}_{\bar q}^\mathrm{bb}$, then evaluate the fragmentation function 
supplied by, e.g., 
{\sf Pythia} for a CMS energy of $E_q^\mathrm{bb}+E_{\bar q}^\mathrm{bb}$, and finally 
boosting back to the DM frame. {
As we only consider leading order effects here,
and the expected difference in $dN_P/dE_P$ is anyway very small,
we restrict our numerical implementation to Eq.~(\ref{spectrum}) also for quark final states.}
}

The above expression depends on the {\it differential} cross section 
$d\sigma/dE_p$, rather than the total cross section discussed in the previous subsection,
implying that we need to re-discuss how to correctly take into account double-counting 
issues. Consider for example the process $\chi\chi\to H^+f\bar F$. We want to remove 
the contribution already contained in $\chi\chi\to H^+W^-$, say in the differential cross 
section for the fermion $f$,
\begin{equation}\label{eq:subSpectrum}
 \left(\frac{d\sigma v_{H^+f\bar F}}{dE_f}\right)^{sub} = \frac{d\sigma v_{H^+f\bar F}}{dE_f} - \left(\frac{d\sigma v_{H^+f\bar F}}{dE_f}\right)^{NWA} \;.
\end{equation}
The question is, what to use for the NWA term.
The simplest assumption would be to replace the branching ratio in Eq.~(\ref{eq:sigvNWA}) 
by the differential spectrum, i.e. 
\begin{equation}
\label{NWAdiff}
\left(\frac{d\sigma v_{H^+f\bar F}}{dE_f}\right)^{NWA} = \sigma v_{H^+W^-} \, \frac{dN_{W^-\to f\bar F}}{dE_f}
\end{equation}
where the last factor is the spectrum of $f$ per decay of the $W$, 
as seen in the CMS frame and normalized  such that
$
\int dE_f \, dN_W/dE_f = BR(W\to f\bar F).
$
Obviously, it is straightforward to generalize Eqs.~(\ref{eq:subSpectrum},\ref{NWAdiff}) to other final 
states, analogous to Eqs.~(\ref{eq:subtraction_total}--\ref{eq:subtraction_total_last}).

Unlike for the total cross section 
\cite{Uhlemann:2008pm}, however,  the replacements (\ref{decorr}) for vector 
and fermion resonances are in general {\it not} correct for the {\it differential} cross section.
Instead, the latter can be affected by the correlation of the helicities/spins of the 
resonance between the production and decay processes, even in the limit 
$\Gamma/M\to 0$. Fortunately, conservation of $CP$ and total angular momentum
uniquely fixes the polarization states of the vector resonances for the case of Majorana
pair annihilation in the $s$-wave limit (see, e.g., \cite{Asano:2012zv}):
vector bosons in $WW$ and $ZZ$ final states are necessarily transversely polarized, 
and those in $H^\pm W^\mp$, $HZ$ and $hZ$ final states longitudinal 
(while $AZ$ final states are not possible, for the same reason).
 Therefore, spin correlations can fully be taken into account by using 
the appropriate energy spectra for {\it polarized} vector bosons in Eq.~(\ref{NWAdiff}),
noting that  the branching fractions are in fact independent of polarization
(see Appendix \ref{app:corr} for a more detailed discussion).
For the example above, this implies that one should use $d N_{W_L\to f\bar F}/dE_f$ 
instead of $d N_{W\to f\bar F}/dE_f$;
for e.g.~$\chi\chi\to WW^*\to Wf\bar F$, on the other hand, the corresponding 
narrow-width contribution contains $d N_{W_T\to f\bar F}/dE_f$.
%
For the decay of a top resonance, $\chi\chi\to t^*\bar t \to W^+b\bar t$, 
conservation of angular momentum and $CP$ 
requires $t^*$ and $\bar t$ to have the same helicity. 
We note that the decay spectrum for polarized tops, $dN_{t_h\to W^+b}/dE_b$, differs for 
the two helicities $h=\pm 1/2$, due to the parity-violating $W$ coupling. 
Nevertheless, since both polarizations are produced with 
equal cross section, as a consequence of $CP$ conservation, the relevant contribution to 
the total narrow-width spectrum is given by
\begin{equation}
\sum_{h=\pm1/2} \sigma v_{t_h\bar t_h} \frac{dN_{t_h\to W^+b}}{dE_b} = \frac{\sigma v_{t\bar t}}{2}\left(\frac{dN_{t_{1/2}\to W^+b}}{dE_b}+\frac{dN_{t_{-1/2}\to W^+b}}{dE_b}\right) = \sigma v_{t\bar t}\frac{dN_{t\to W^+b}}{dE_b}\,,
\end{equation}
where $\sigma v_{t \bar t}/2$ is the usual cross section summed over all final-state 
helicities.
Thus, also for top resonances, no polarization effects occur in the $CP$ conserving MSSM.


In summary, the differential 3-body cross section to be used in 
Eq.~(\ref{spectrum}) is given by
\begin{equation}\label{eq:subSpectrumGeneral}
 \left(\frac{d\sigma v_{2\to3}}{dE_p}\right)^\mathrm{sub} \equiv \frac{d\sigma v_{2\to3}}{dE_p} - \left(\frac{d\sigma v_{2\to3}}{dE_p}\right)^\mathrm{NWA} \;,
\end{equation}
where for {\it fermionic} final state particles ($p=f,\overline{f},F,\overline{F}$) we have
\begin{eqnarray}\label{eq:sub_spectrum_f_total}
  \left(\frac{d\sigma v_{W^+f\bar F}}{dE_p}\right)^\mathrm{NWA} 
  &=& \sigma v_{WW} \frac{dN_{W_T\to f\bar F}}{dE_p} + \sigma v_{W^+H^-}\frac{dN_{H^-\to f\bar F}}{dE_p} \\
  \left(\frac{d\sigma v_{Z f\bar f}}{dE_p}\right)^\mathrm{NWA} 
  &=& 2\,\sigma v_{ZZ} \frac{dN_{Z_T\to f\bar f}}{dE_p} +\sigma v_{ZH}\frac{dN_{H\to f\bar f}}{dE_p}+\sigma v_{Zh}\frac{dN_{h\to f\bar f}}{dE_p} 
\end{eqnarray}
\begin{eqnarray}
  \left(\frac{d\sigma v_{H^+f\bar F}}{dE_p}\right)^\mathrm{NWA} 
  &=& \sigma v_{W^-H^+} \frac{dN_{W_L\to f\bar F}}{dE_p} \\
  \left(\frac{d\sigma v_{A f\bar f}}{dE_p}\right)^\mathrm{NWA} 
  &=& \sigma v_{Ah} \frac{dN_{h\to f\bar f}}{dE_p}+\sigma v_{AH} \frac{dN_{H\to f\bar f}}{dE_p} \\
  \left(\frac{d\sigma v_{H f\bar f}}{dE_p}\right)^\mathrm{NWA} 
  &=&  \sigma v_{AH} \frac{dN_{A\to f\bar f}}{dE_p}+\sigma v_{ZH} \frac{dN_{Z_L\to f\bar f}}{dE_p} \\
  \left(\frac{d\sigma v_{h f\bar f}}{dE_p}\right)^\mathrm{NWA} 
  &=&  \sigma v_{Ah} \frac{dN_{A\to f\bar f}}{dE_p}+\sigma v_{Zh} \frac{dN_{Z_L\to f\bar f}}{dE_p}\,.
\label{eq:sub_spectrum_f_last}  
\end{eqnarray}
Here, the decay spectra are normalized to fermionic branching ratios, 
e.g.~$\int dE_f dN_{W_\lambda\to f\bar F}/dE_f=BR(W\to f\bar F)$. For scalars, or when neglecting 
correlations, these spectra are flat, e.g. 
\begin{equation}
\label{dNdEunpol}
\frac{dN_{h\to b\bar b}}{dE_b} = \frac{BR(h\to b\bar b)}{E_b^{\rm max}-E_b^{\rm min}}\,,
\end{equation}
where $E_b^{\rm max/min}$ are the maximal and minimal allowed energy of the $b$ in the 
decay of the (boosted) Higgs. This is the usual box-shaped spectrum in a cascade decay.
The non-trivial spectra are in principle straight-forward to derive (see Appendix 
\ref{app:corr}), but not needed in our numerical implementation (see below).
For {\it bosonic final states} ($p=Z,W^\pm,h,H,H^\pm,A$), on the other hand,
there is no polarization effect and only the energy allowed by the 2-body kinematics contributes. 
This implies that one has to replace
\be
\sigma v_{pX} \frac{dN_{X_{(h)}\to f\bar F}}{dE_p} \longrightarrow \sigma v_{pX} BR(X\to f\bar F) \,\delta(E_p-E_p^{\chi\chi\to pX})
\ee
in Eqs.~(\ref{eq:sub_spectrum_f_total} -- \ref{eq:sub_spectrum_f_last}), where $E_p^{\chi\chi\to pX}=m_\chi+\left(m_p^2-m_X^2\right)/4m_\chi$.
Annihilation channels involving {\it top quarks}, finally, are slightly special:
\begin{eqnarray}\label{eq:sub_spectrum_t_total}
  \left(\frac{d\sigma v_{W^+b\bar t}}{dE_b}\right)^\mathrm{NWA} 
  &=& \sigma v_{W^+H^-} \frac{dN_{H^-\to t\bar b}}{dE_b} + \sigma v_{\bar t t}\frac{dN_{t\to W^+b}}{dE_b} \\
  \left(\frac{d\sigma v_{W^+b\bar t}}{dE_{\bar t}}\right)^\mathrm{NWA} 
  &=& \sigma v_{W^+H^-} \frac{dN_{H^-\to t\bar b}}{dE_{\bar t}} +\sigma v_{\bar t t}BR(t\to bW)\,\delta(E_{\bar t}-m_\chi)\\
  \left(\frac{d\sigma v_{W^+b\bar t}}{dE_W}\right)^\mathrm{NWA} 
  &=& \sigma v_{\bar tt} \frac{dN_{t\to W^+b}}{dE_W} + \sigma v_{W^+\!H^-}BR(H^-\!\!\to\! \bar t b)\,\delta(E_W\!-\!E_W^{\chi\chi\to W^+H^-})~~~
  \label{eq:sub_spectrum_t_last}
\end{eqnarray}

Now let us consider these NWA corrections to the energy distribution of final state 
particles $p$ in the context of Eq.~(\ref{spectrum}), i.e.~the spectrum of {\it stable}
particles $P$. From each of the terms on the r.h.s.~of 
Eqs.~(\ref{eq:sub_spectrum_f_total} -- \ref{eq:sub_spectrum_f_last}), and a given 
channel $\chi\chi \to Y f\bar F$, we pick up a contribution of the form
\begin{align}
&\frac{\sigma v_{YX}}{\sigma v^\mathrm{tot}_{2\to2}} \left\{ \sum_{p=f,\bar F}\int \frac{dN^{p \rightarrow P+ ...}_P} {dE_P}
\frac{dN_{X_{(h)\to f\bar F}}}{dE_p}  \, dE_p
+ BR(X\to f\bar F) \left.\frac{dN^{Y \rightarrow P+ ...}_P} {dE_P}\right|_{E_Y=E_Y^{\chi\chi\to YX}}
\right\}\nn\\
&\quad={BR(\chi\chi\to YX)}\left\{  \frac{dN^{X_{(h)} \to f\bar F  \to P+ ...}_P} {dE_P}
+ BR(X\to f\bar F) \frac{dN^{Y \rightarrow P+ ...}_P} {dE_P}
\right\}_{E_{X,Y}=E_{X,Y}^{\chi\chi\to YX}}
\end{align}
\begin{align}
&\quad\to{BR(\chi\chi\to YX)}BR(X\to f\bar F)\left\{  \frac{d\hat N^{X_{(h)}  \to P+ ...}_P} {dE_P}
+ \frac{dN^{Y \rightarrow P+ ...}_P} {dE_P}\right\}_{E_{X,Y}=E_{X,Y}^{\chi\chi\to YX}}
\\ 
&\quad={BR(\chi\chi\to YX)}BR(X\to f\bar F) \frac{d\hat N^{XY \rightarrow P+ ...}_P} {dE_P}\,.
\label{specNWAfinal}
\end{align}
Here, we can replace $X_{(h)}\to X$ in the last step because, as stressed before, the 
helicity of $X$ in $\chi\chi\to XY$ is uniquely fixed by conservation of angular momentum 
and $CP$ (for $Y$, on the other hand, the helicity has been fixed that way right from the 
start). Furthermore, we introduced the notation $d\hat N_P^{X_{(h)}  \to P+ ...}/dE_P$ for the yield of species $P$ considering only fermionic
decays of $X$. Similarly, for computing $d\hat N^{XY \rightarrow P+ ...}_P/dE_P$ we take into account only
fermionic decays of $X$ (while for $Y$ all decay modes are included).
Note that the second step ($\to$) is then only valid under the assumption 
that $dN/dE_P$ has the same shape for a single decay channel $X_{(h)} \to f\bar F$ 
as for the sum over all fermionic final states, which can be a rather poor approximation for a 
given channel (involving, e.g., neutrinos). The {\it total} spectrum of a 
given stable state particle $P$, however, is correctly recovered when {\it summing}
Eq.~(\ref{specNWAfinal}) over all possible 3-body channels involving the boson $Y$ and a pair of fermions. 
Numerically, we implement this sum for all fermionic final states for decays
of $X=Z,W,h,H,H^\pm,A$.
For cases where a nearly on-shell $t^*$ quark gives a large contribution, even 
the single-channel yield is well approximated because the decay $t\to Wb$ dominates.

Let us conclude this Section with two comments regarding the correct use of final 
state helicities for the determination of yields of stable particles. {\it i)} 
Eq.~(\ref{spectrum}) has indeed to be summed over all helicities of $p$ that 
contribute to the cross section $\sigma_{2\to3}$; it is thus in general {\it not} 
sufficient to determine only the yields ${dN^{\bar{p} p \rightarrow P+ ...}_P}/{dE_P}$
that are required for 2-body processes (for which the helicities of $p$ and $\bar p$ are fixed by the
requirement $J^P=0^-$). {\it ii)} The yields from the NWA subtraction, on the other hand, 
do result from final states with helicities fixed by the {\it same} symmetry argument 
as in the 2-body case. This implies that double counting is fully 
avoided in our prescription even if the yields ${dN^{XY \rightarrow P+ ...}_P}/ {dE_P}$ 
are throughout approximated by using unpolarized final state particles $X$ and $Y$:
In that case, the procedure described above consistently removes the double counting
related to the yields produced from decay and fragmentation of $Y$. For $X$, on the
other hand, the full 3-body matrix element automatically takes into account
polarization effects in the decay $X\to f\bar F$, while the NWA term subtracts the
yield for unpolarized decays. This means that, in this case, the NWA-subtracted 3-body
contribution accounts precisely for the difference, and correctly replaces the unpolarized
by the polarized yield for the $X$ decay after adding two- and 3-body contributions.

\section{Results for the MSSM}
\label{sec:mssm}
In order to demonstrate the impact of our results on realistic models, we work 
in the framework of simplified phenomenological MSSM versions, introduced in 
Section \ref{sec:models},
that are however generic enough to capture the relevant phenomenology discussed in this
work. 
We assess in turn the consequences for the overall annihilation cross section
(Section \ref{sec:svtot}) and yields (Section \ref{sec:yield+}), before discussing in detail
selected example spectra in Section \ref{sec:spectra}.
 As we will see, some of the most relevant part of the parameter space
involves SUSY spectra with degenerate particle states, that are typically more difficult to test in
proton collider experiments.

\subsection{Theoretical benchmark models}
\label{sec:models}

We introduce four phenomenological ``pMSSM-9'' realisations, each defined by 9 parameters 
at the electroweak scale  as specified below (see also Table~\ref{tab:param1}): 

\begin{description}
\item[MSSM-91] The Higgsino mass parameter $\mu$,
the Bino and Wino masses $M_1$ and $M_2$, the $CP$-odd Higgs boson mass $M_A$,  
the ratio of Higgs vacuum expectation values $\tan \beta$, the squark and slepton mass terms 
$M_{\rm \tilde{q}}$, and $M_{\rm \tilde{\ell}}$, and the
third generation trilinear couplings, $A_t$ and $A_b$ (note that $M_1$ and $M_2$ are not
constrained by the GUT unification relation).
\item[MSSM-92] Here, instead of distinguishing squark and slepton masses, 
we decouple the $3^{\rm rd}$ sfermion generation from the $1^{\rm st}$ and $2^{\rm nd}$ 
unified generations.
\item[MSSM-93] Squark and slepton mass terms are decoupled as in MSSM-91. Here, we allow for 
a separate $3^{\rm rd}$ generation slepton mass, and a common slepton mass
for the  $1^{\rm st}$ and $2^{\rm nd}$ generation, respectively. In this case we assume $A_t = A_b$.
\item[MSSM-94] Adding more freedom to the squark sector, we allow here for an independent 
right-handed stop (UR) mass and left-handed $3^{\rm rd}$ generation squark mass (L), while 
adopting a universal sfermion mass for all other cases. 
\end{description}

\begin{table}
\begin{center}
\begin{tabular}{|c||c|c|c|c|c|c|c|c|c|}
\hline
MSSM-91 & $\mu$ & $M_2$ & $M_1$ & $M_A$ & $\tan\beta$ & $A_t$ & $A_b$ & $M_{\tilde q}$ & $M_{\tilde\ell}$ \\
MSSM-92 & $\mu$ & $M_2$ & $M_1$ & $M_A$ & $\tan\beta$ & $A_t$ & $A_b$ & $M_{{\rm sf}, 1+2{\rm nd\ gen.}}$ & $M_{{\rm sf}, 3{\rm rd\ gen.}}$ \\
MSSM-93 & $\mu$ & $M_2$ & $M_1$ & $M_A$ & $\tan\beta$ & $A_{tb}$ & $M_{\tilde q}$ & $M_{\tilde\ell, 1+2{\rm nd\ gen.}}$ & $M_{\tilde\ell, 3{\rm rd\ gen.}}$ \\
MSSM-94 & $\mu$ & $M_2$ & $M_1$ & $M_A$ & $\tan\beta$ & $A_{tb}$ & $M_{\tilde t_R}$ & $M_{\tilde t_L/\tilde b_L}$ & $M_{{\rm sf,\ rest}}$ \\
\hline
\end{tabular}
\caption{Free parameters for the four types of MSSM models considered in this work. Note that for the last two 
models we assume $A_t=A_b$.}
\label{tab:param1}
\end{center}
\end{table}

In addition, we used in all cases a fixed value of $M_3=5\,$TeV for the gluino mass parameter.
We performed Bayesian scans over the parameter space of these models by 
using \mn\ 
\cite{Feroz:2008xx}, which we interface to \ds\ to compute all 
relevant quantities that enter in the likelihood evaluations.
The joint likelihood that we adopt takes the form
\begin{equation}
\ln \likeJ =  \ln \like_{\Ohsq} + \ln \like_{m_{\rm Higgs}}  + \ln \like_{\rm SUSY} +  \ln \like_{\rm LUX} \, ,     
\label{eq:like}
\end{equation}
where $\like_{\Ohsq}$ refers to the constraint on the cold DM relic abundance from 
cosmic microwave background (CMB) observations; $\like_{m_{\rm Higgs}}$ imposes the 
mass of the lightest SUSY Higgs to agree with the Higgs boson mass;
$\like_\text{SUSY}$ includes constraints from sparticles searches at colliders; 
and $\like_{\rm LUX}$ accounts for the constraints on 
DM-nucleon interactions from the LUX direct detection experiment.

The relic abundance is computed including co-annihilations 
\cite{Edsjo:1997bg,Edsjo:2003us}, using a central value of $\Ohsq = 0.1198$ \cite{Ade:2015xua} 
and a generous error of $20$\% to account for both experimental and, more importantly, 
theoretical uncertainties in the $\Ohsq$ prediction.
We impose the predicted mass of the lightest Higgs to match the measured Higgs mass, 
$m_h=125.09 \pm 0.24$\,GeV \cite{Aad:2015zhl}.
Since we are mainly interested in suppression lifting mechanisms, rather than a detailed 
phenomenological analysis of the MSSM parameter space, we adopt a conservative set of further 
constraints to roughly indicate some of the more relevant experimental constraints.
Apart from  LEP and TeVatron constraints as implemented in \ds {} \cite{ds4},
we impose LHC constraints on stop, sbottom, light squark and slepton masses assuming direct
production \cite{Khachatryan:2015vra,CMS:2014wsa,CMS:2014yma, Chatrchyan:2013xna,CMS:2013nia, Khachatryan:2014doa, CMS:2013ida,Aad:2014nra,Aad:2013ija,Aad:2014vma,Aad:2014yka},
as well as null results from direct DM detection
\cite{Akerib:2013tjd}. 

\begin{table}
\begin{center}
\begin{tabular}{lclc|cl }
\hline
\hline
Model parameters & Low-mass & High-mass \\
\hline
$\mu$ [GeV] & $(70, 2000)$ &  $(500, 4000)$ \\
$M_1 $ [GeV] & $(70, 2000)$ &  $(500, 4000)$ \\
$M_2 $ [GeV] & $(70, 2000)$ &   $(500, 4000)$ \\
$M_{\tilde q}, M_{\tilde\ell}\ (M_{{\rm sf},i})$  [GeV] & $(70, 4000)$ & $(500,4000)$ \\
$A_t/M_{\tilde t_R}, A_b/M_{\tilde t_R}$  & $(-3, 3)$ & $(-3, 3)$  \\
$M_A $ [GeV] & $(70, 2000)$  & $(500, 4000)$ \\
$\tan\beta$ & $(5, 40)$ & $(5, 40)$ \\
\hline
\end{tabular}
\end{center}
\caption{Parameter ranges of the scans performed. The range for $M_{{\rm sf},i}$ applies to all combinations of independent
sfermion mass parameters for the four models.}
\label{tab:param}
\end{table}

For each model, we performed scans with two different parameter ranges: one for low-mass 
neutralinos,  roughly $\sim$ 50 -- 2000 GeV, and one for high masses, $\sim$ 500 -- 3000 GeV, 
adopting logarithmic priors on the parameters, except for the trilinear 
couplings $A$. Table~\ref{tab:param} contains the parameters and the ranges
of the ``low-mass'' and ``high-mass'' scans.
We emphasise that we do not perform global scans of the MSSM in light of the most recent results 
from various experiments (for this, see instead Ref.~\cite{Athron:2017yua}), nor is it our purpose to do 
so here. 
In particular, we note that important additional constraints
can arise from electroweakino and MSSM Higgs searches, that are however highly dependent on the 
specific configuration of masses
and decay channels, and therefore beyond the scope of this work.
Rather than identifying the most probable parameter regions of our pMSSM-9 models, taking into
account all  experimental constraints, our focus is simply to provide a 
phenomenological proof
of concept concerning the impact of 3-body final states on the annihilation of DM particles
and on the resulting cosmic-ray fluxes.

\begin{figure}[t!]
	\centering
		\includegraphics[width=0.4\columnwidth]{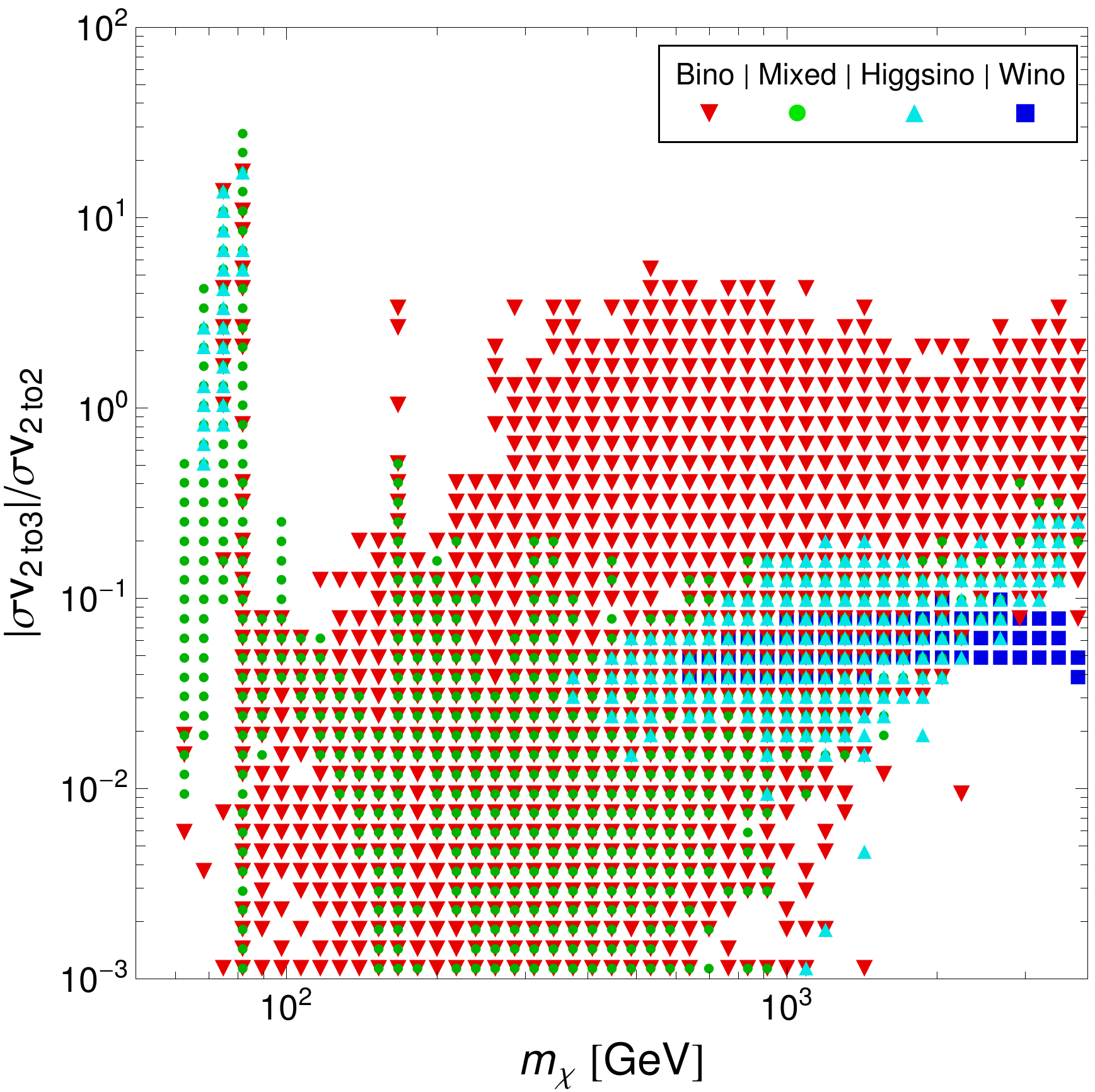}
		\hspace*{1cm}
		\includegraphics[width=0.4\columnwidth]{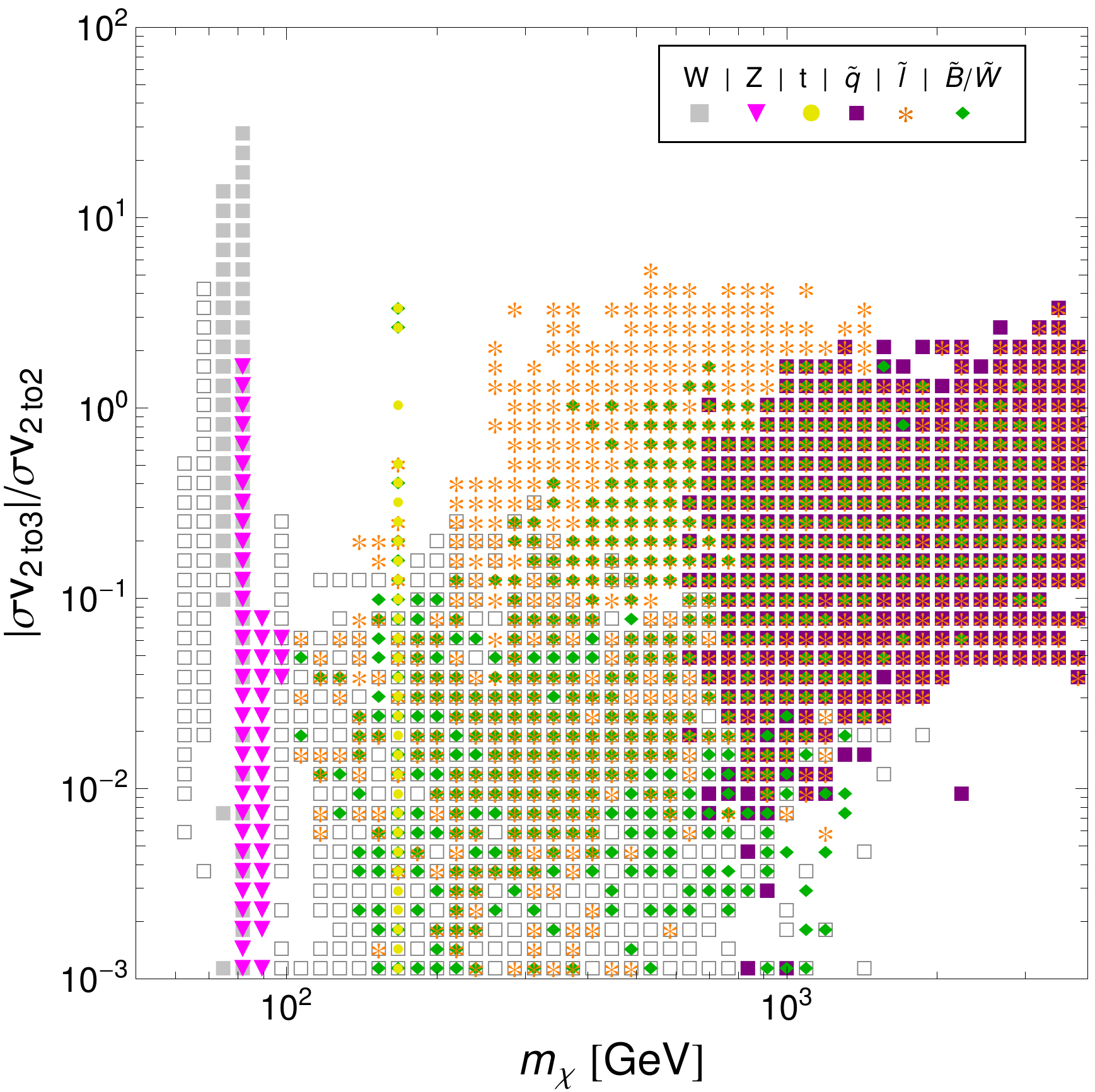}
	\caption{\textit{Left panel}: Ratio between the (NWA subtracted) 3-body and total 
	         2-body cross section for the MSSM 
	         models defined in Section \ref{sec:models}, as a function of the lightest neutralino mass $m_\chi$.
		Different neutralino compositions are indicated as ``Bino'' ($|N_{11}|^2 > 0.8$) in red, ``Wino'' 
		(\mbox{$|N_{12}|^2 > 0.8$}) 
		in blue, ``Higgsino'' ($|N_{13}|^2 + |N_{14}|^2 > 0.8$) in cyan, and ``mixed'' (otherwise) in green.
		\textit{Right panel}: Same, but now broken down to models that are either {\it (a)} close to a threshold $X=W,Z,t$
		(with $-4 \, \Gamma_X<m_\chi -m_X<  \Gamma_X$, where $\Gamma_X$ is the total width of $X$) or  {\it (b)}
		with a degenerate sfermion $\tilde f$ in the mass spectrum ($m_\chi/m_{\tilde f}>0.9$) or  {\it (c)} Bino-like neutralinos
		with Wino coannihilation ($m_\chi/M_2 > 0.9$ and $|N_{11}|^2 > 0.98$, without degenerate sfermions).
                 }
	\label{fig:svratio}
\end{figure}

\subsection{Total annihilation rate}
\label{sec:svtot}

In our extensive scans, we identified many MSSM models where the total annihilation 
cross section is significantly enhanced by including the 3-body final states we consider here.
As an illustration, we show in Fig.~\ref{fig:svratio} the ratio between the 
 (NWA subtracted) 3-body and total 2-body cross section, in function of the neutralino mass.
In the left panel, we furthermore indicate the neutralino composition, while in the right panel we 
indicate the models where the neutralino is either close to a threshold or there exists an almost 
mass-degenerate 
SUSY particle.  As anticipated, models with large enhancements indeed broadly fall into those two 
classes. In particular, cross section enhancements by ${\cal O}(10)$ factors are 
 possible for models with neutralino 
 masses close to the kinematical threshold of producing $W^+W^-$, $ZZ$ and $\bar tt$ final states;  on 
 the other hand, we did not find any models where the total cross section is significantly enhanced below 
 the $Zh$ threshold. For models with neutralino masses close to the threshold for annihilation into a 
 heavy Higgs boson and a SM particle
($ZH^0, A^0h, W^\pm H^\mp$) we identified up to $\sim 10\%$ enhancements.

The second class of enhancement mechanisms are models  with small mass splittings
between the neutralino and other SUSY particles. For neutralino masses $m_\chi\gtrsim200$\,GeV, 
degenerate sleptons can significantly increase $\sigma v$, while for even heavier neutralinos, 
$m_\chi\gtrsim600$\,GeV, this also becomes possible for 
degenerate squarks. A final, somewhat less expected class of models where $\sigma v_{2\to3}$
can be of at least the same size as $\sigma v_{2\to2}$ are heavy Bino-like neutralinos with almost 
mass-degenerate Winos. Here, the relic density is set by Wino coannihilation, while the
DM annihilation rate  is highly suppressed due to either the small neutralino mixing (for $WW$ or 
$ZZ$ final states) or helicity factors $m_f^2/m_\chi^2$ (for fermionic final states). The lifting of the 
latter by 3-body annihilation processes is thus very relevant in this case, even for models that do not 
feature degenerate sfermions.

\begin{figure}[t!]
	\centering
		\includegraphics[width=0.4\columnwidth]{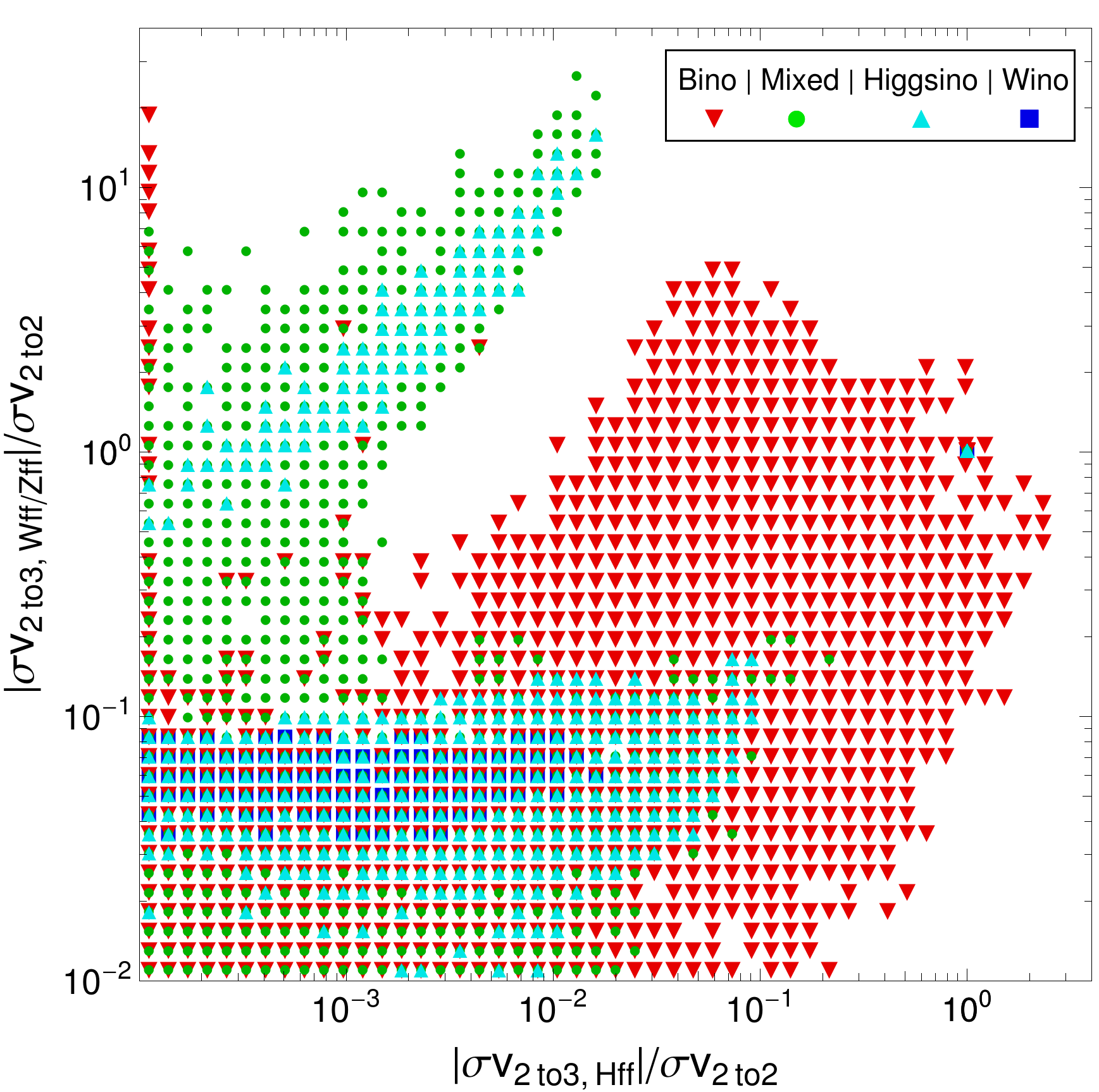}
	\caption{Cross section for annihilation into 3-body final states containing a gauge boson 
		versus that containing {\it any} of the Higgs bosons in the final state, normalized to the 
		total 2-body cross section. Different neutralino compositions are indicated as in Fig.~\ref{fig:svratio}.}
	\label{fig:svratio_comp}
\end{figure}

One of the main new additions of this work is the full inclusion of final states containing a Higgs boson. 
To demonstrate their impact, we show in Fig.~\ref{fig:svratio_comp} the annihilation cross section to 
$H\bar ff/h\bar ff/A\bar ff/H^\pm \bar fF$ as compared to the one for $Z \bar ff$ and $W^\pm \bar f F$ 
final states. We find that Higgs boson IB alone can increase the lowest-order cross section by up to a 
factor 
of $\sim3$, but in general the relative importance of the Higgs and gauge boson channels is very model 
dependent. Models where the former dominates (bottom right corner of the plot) feature neutralino 
masses in the TeV range, 
and a stop not much heavier than the neutralino (within a few hundred GeV). The dominant annihilation
channels are then $Htt, Att, H^\pm tb$, in particular via $\tilde t$ exchange in the $t/u$ channel, where 
the large Yukawa coupling to top quarks explains why the corresponding gauge boson final states are 
not equally pronounced. We note that these cases are examples where lifting of isospin suppression is 
more important than lifting of Yukawa suppression (see Section \ref{sec:lifting}). 

A further class of models with large contributions of Higgs final states is characterized by large values 
of $\mu$ and $\tan\beta$, which lead to an enhancement of the Higgs coupling to sfermions. However, 
this enhances also the mixing between left- and right-handed sfermions and hence implies only a mild 
Yukawa suppression; nevertheless the three-body processes can be particularly important for the 
antiproton yield as discussed in more detail for benchmark model D3  further down.
 
For masses below the SM 
thresholds,  annihilation 
into Higgs plus top final states is kinematically forbidden.
For Binos, as clearly seen in the top left part of 
Fig.~\ref{fig:svratio_comp}, this can result in large IB enhancements without Higgs contribution. 
For neutralinos with a significant Higgsino fraction, on the other hand, there is a correlation between gauge 
boson (mostly $W \bar f F$) and Higgs (mostly $h\bar f f$) final states (top middle part of 
Fig.~\ref{fig:svratio_comp}). These models correspond to neutralino masses below the $WW$ threshold, 
and the 3-body final states arise dominantly from virtual $W$ or $Z$ boson decays, $WW^*\to W\bar f F$ 
and $hZ^*\to h\bar f f$, respectively. The latter are suppressed compared to the former by the 
gaugino fraction, but otherwise of comparable size, which leads to the correlation\footnote{
Note that this also explains the observed separation: the upper left part in the figure corresponds to 
threshold models, for which the Higgs contribution can be arbitrarily small in the pure Bino limit. The middle 
part corresponds 
to models where $t$-channel sfermion exchange is important. In this case, there is a $h\bar f f$ contribution 
even in the pure Bino limit, that arises from VIB emission of the Higgs boson, as discussed above. On the 
other hand, the gap between the pure Bino models on the very left part and the mixed and Higgsino-like 
models in the top middle part is due to (large, but still limited) statistics of our sample. The reason is that 
the ratio of $h\bar f f$ and $W/Z \bar F f$ cross section is extremely sensitive to the Higgsino fraction 
(roughly $\propto (Z_{13}^2+Z_{14}^2)^3$), and therefore varies very rapidly with the input parameters.
}
seen in Fig.~\ref{fig:svratio_comp}.

\subsection{Yield enhancement}
\label{sec:yield+}

The zero-velocity annihilation cross section is already a good indicator for the reach of indirect 
detection experiments, but  observationally more relevant is the resulting number of
stable particles.
Astrophysical background spectra are generally rather soft, i.e.~they fall quickly with energy, such
that  from the point of view of indirect DM searches mostly 
annihilation products with relatively large energies are relevant 
(with the notable exception of CMB constraints that 
are mostly sensitive to the total energy deposition, see e.g.~\cite{Liu:2016cnk}).

\begin{figure}[t]
	\centering
		\includegraphics[width=0.32\columnwidth]{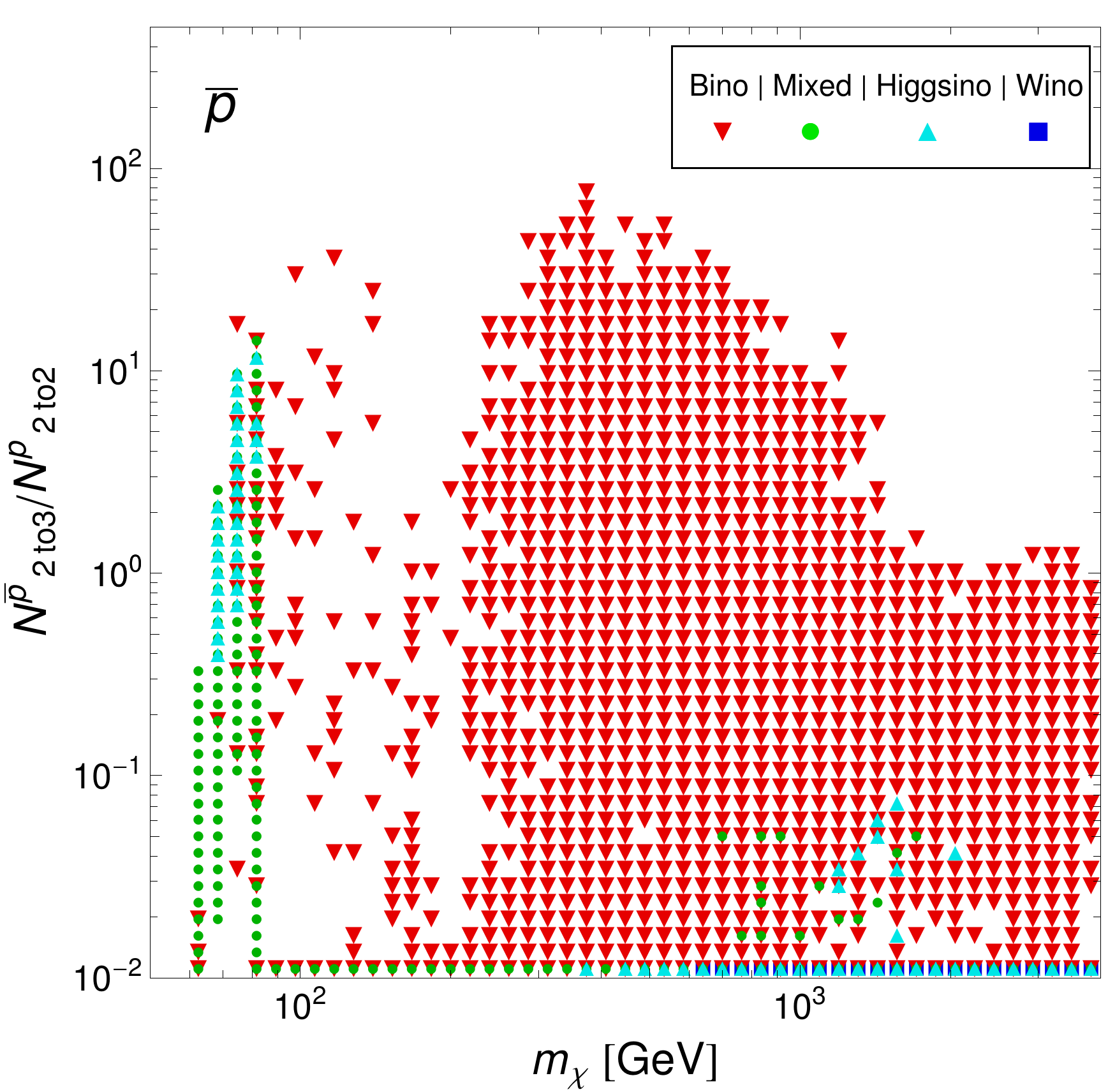}
		\includegraphics[width=0.32\columnwidth]{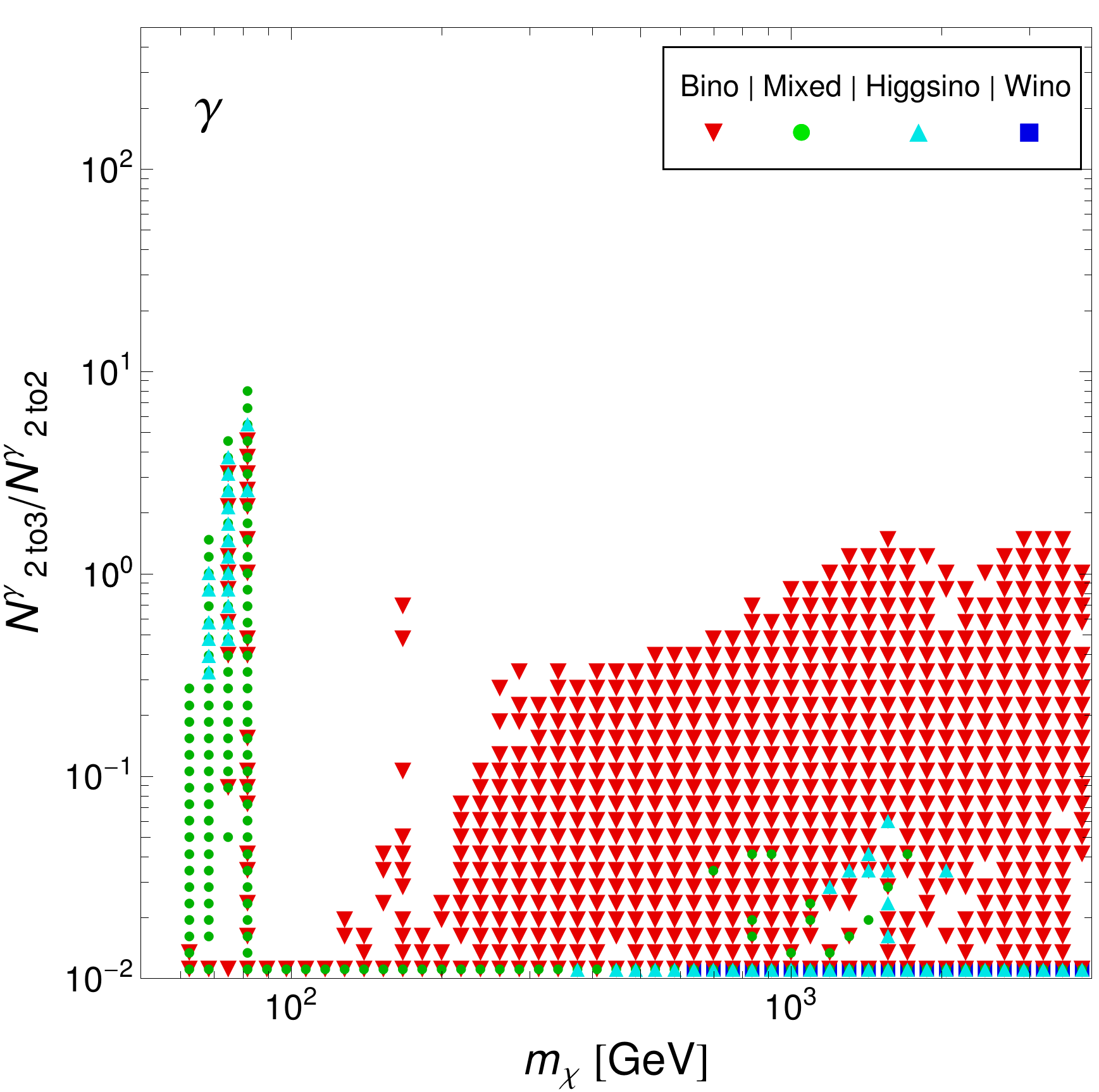}
		\includegraphics[width=0.32\columnwidth]{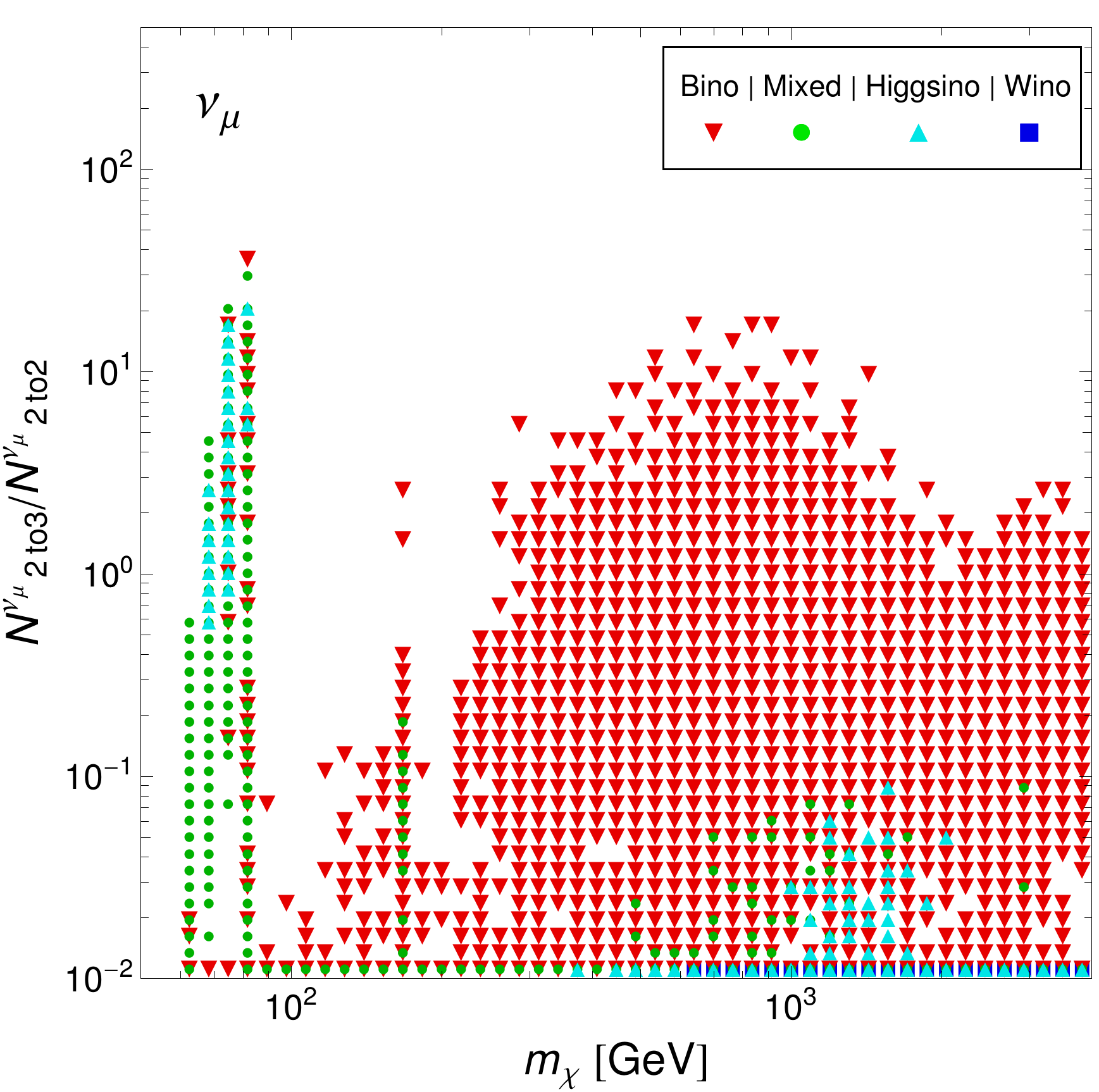}
	\caption{Ratio of integrated yields above $E_X > m_\chi/10$ from (NWA-subtracted) 3-body 
	final states and the 2-body result, for species X = $\bar{p}, \gamma, \nu_\mu$, plotted against the 
	neutralino mass $m_\chi$. 
	The other light lepton yields ($\nu_\tau$ and $e^+$)  are qualitatively very similar to the $\nu_\mu$
	case. Different neutralino compositions are indicated as in Fig.~\ref{fig:svratio}.}
	\label{fig:yieldratio}
\end{figure}

\begin{figure}[t]
	\centering
		\includegraphics[width=0.32\columnwidth]{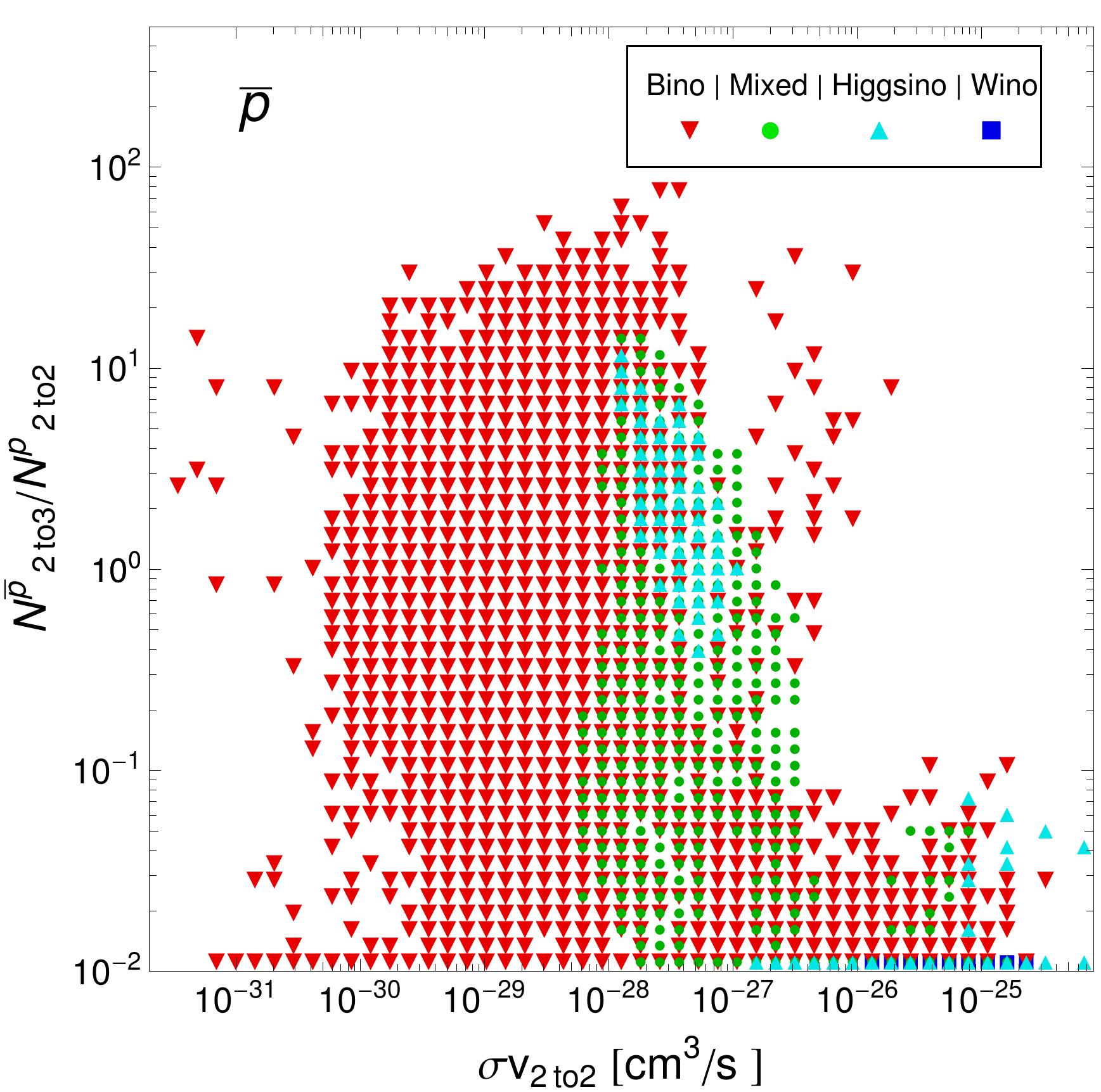}
		\includegraphics[width=0.32\columnwidth]{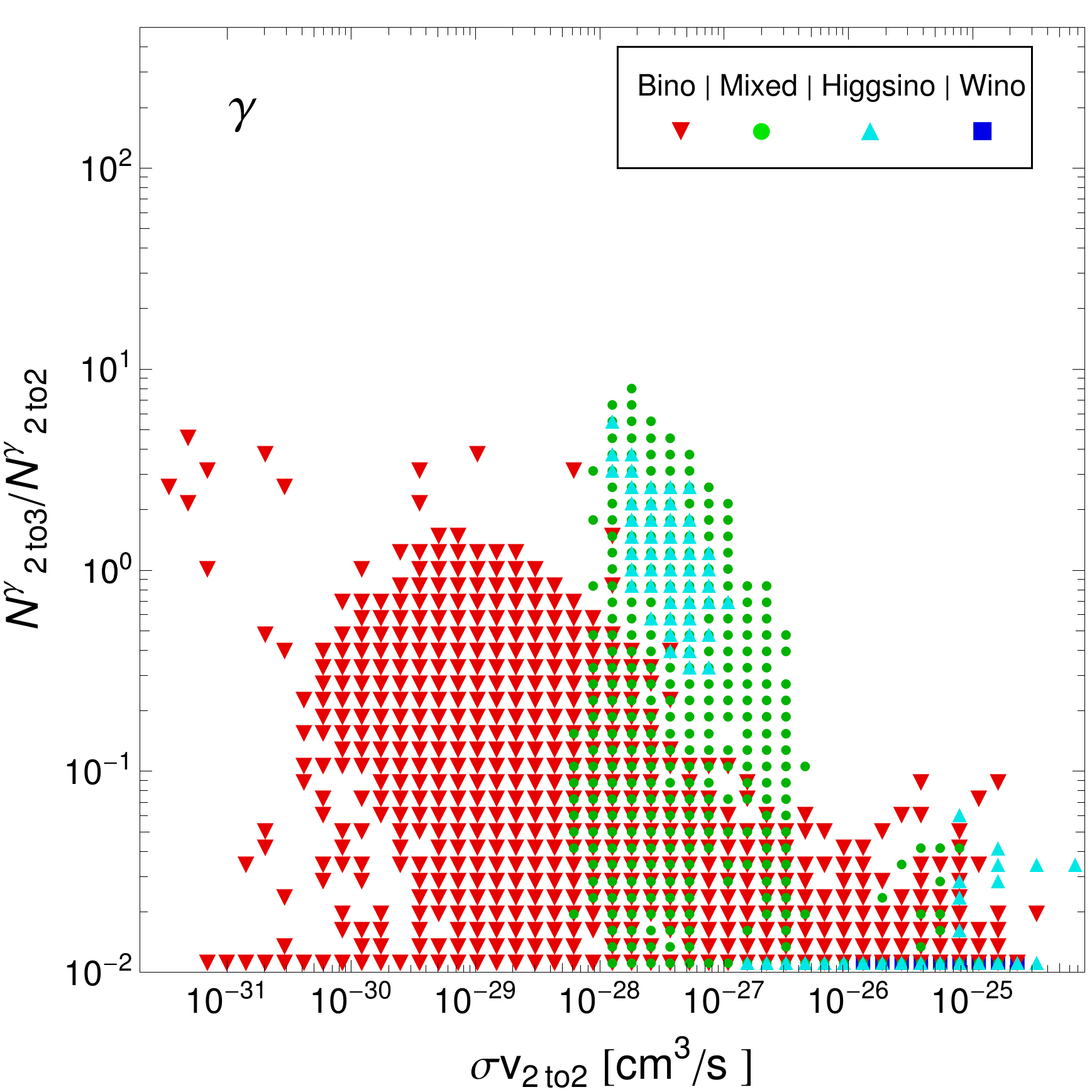}
		\includegraphics[width=0.32\columnwidth]{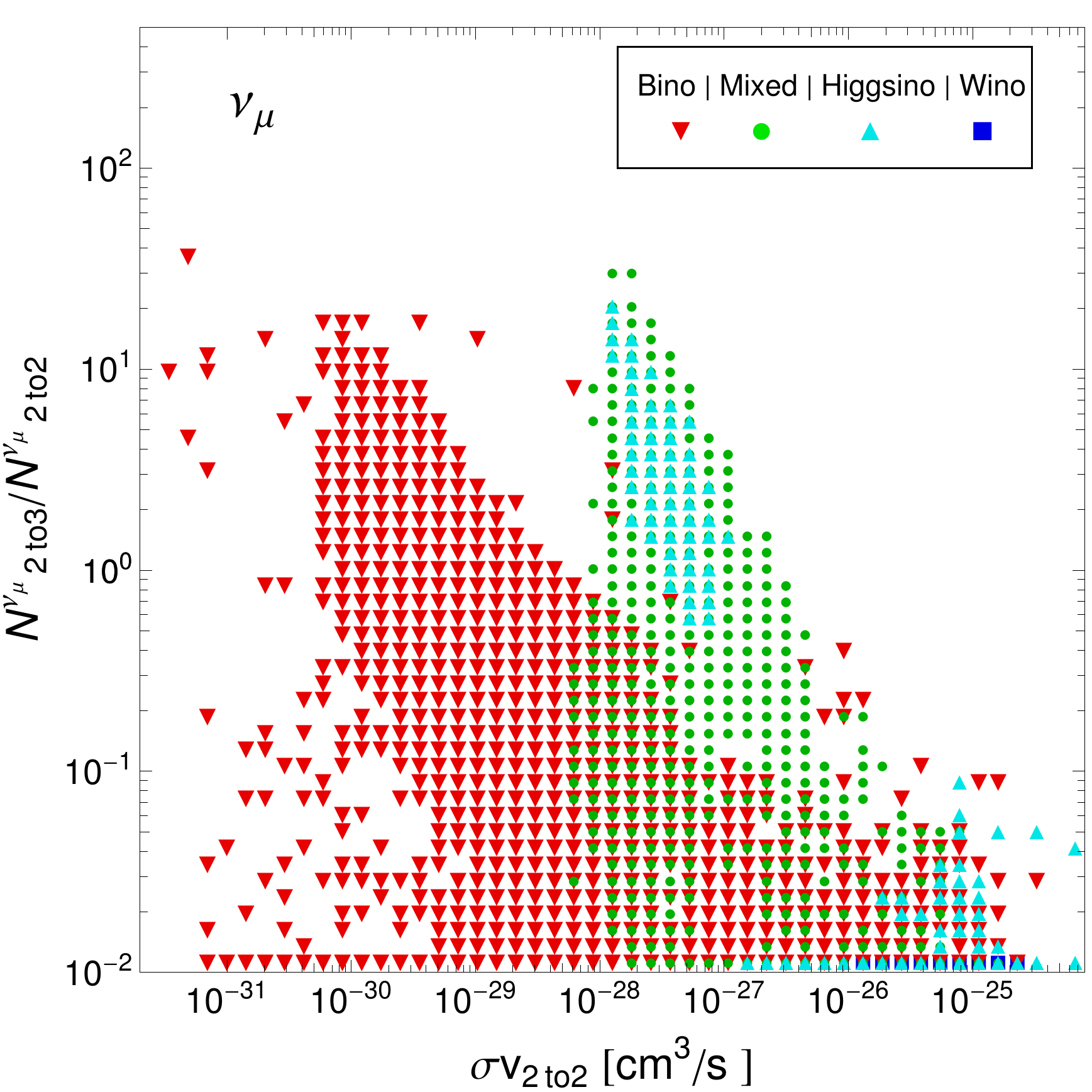}
	\caption{Same as Fig.~\ref{fig:yieldratio}, but now showing the integrated yields, for $E_X > m_\chi/10$,  against the 
	total 2-body cross section instead of the neutralino mass.
	\label{fig:yieldratio_sv2to2}
	}
\end{figure}

\begin{figure}[t!]
	\centering
		\includegraphics[width=0.4\columnwidth]{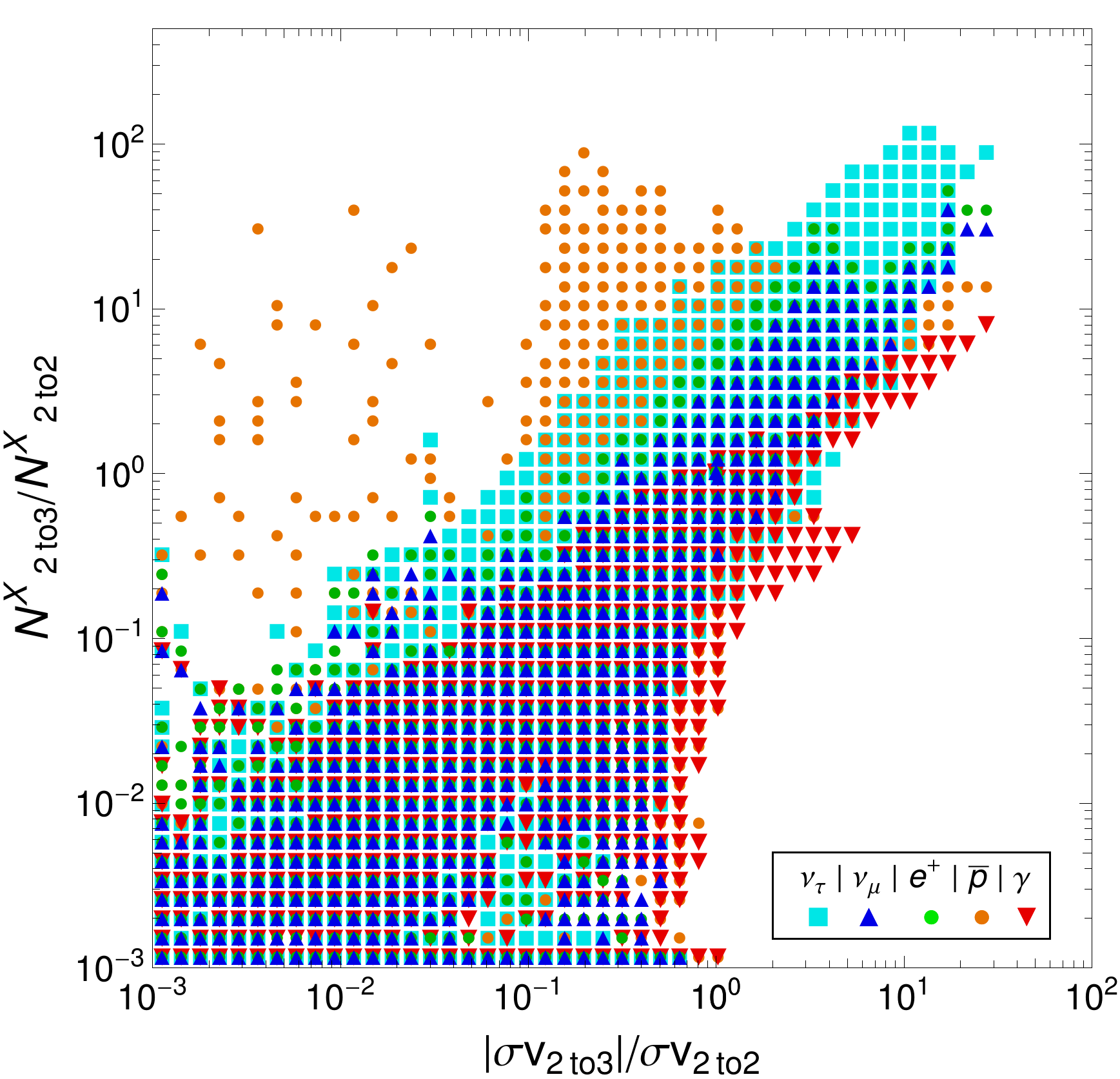}
		\hspace*{1cm}
		\includegraphics[width=0.4\columnwidth]{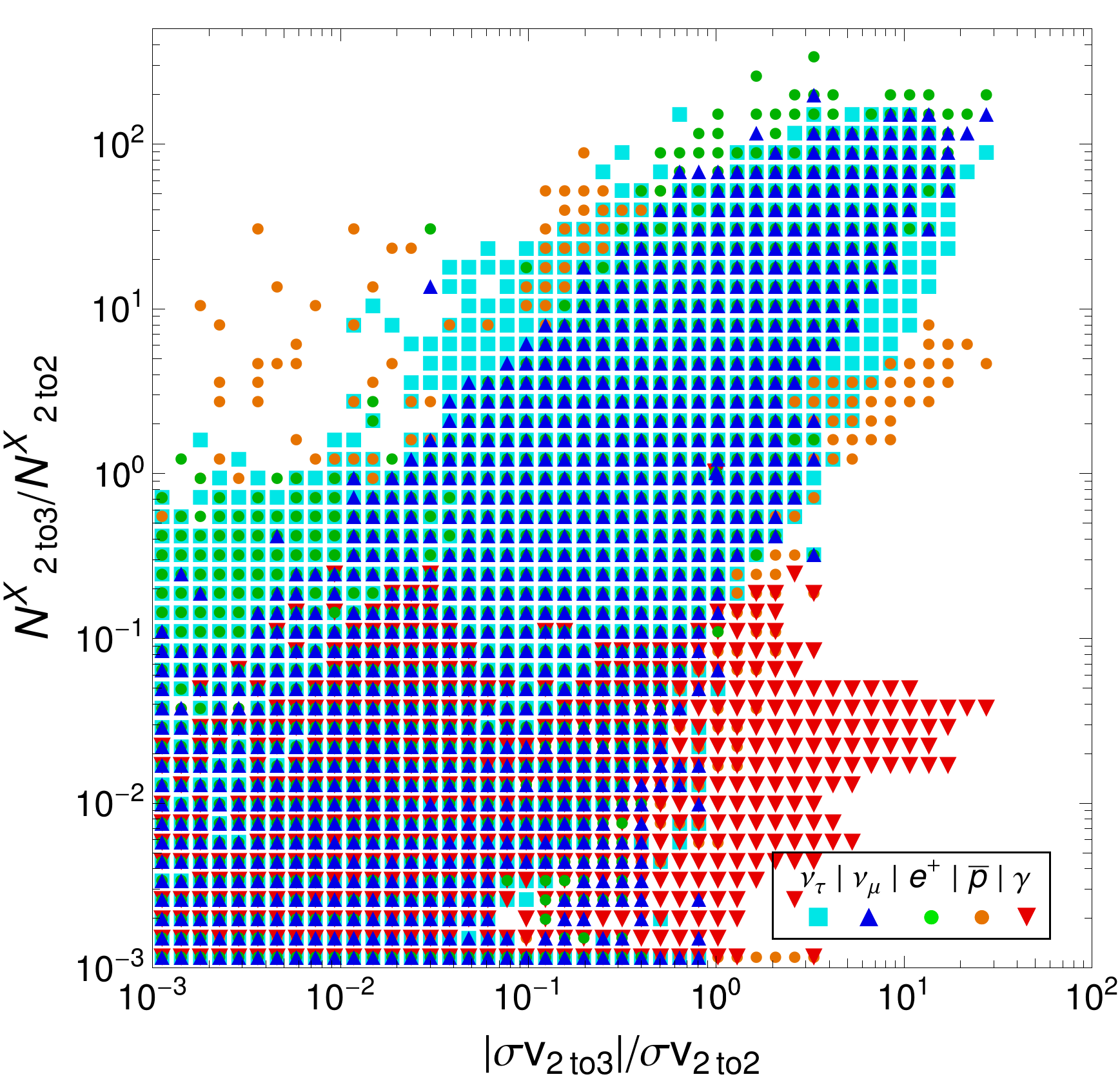}
	\caption{Same yield ratios as in Fig.~\ref{fig:yieldratio}, but now for all species 
	$X = \bar{p}, \gamma, \nu_\mu, \nu_\tau, e^+$ and plotted against the ratio of cross sections.
	\textit{Left panel}: Integrated yields above an energy threshold of $E_X > m_\chi/10$.
	\textit{Right panel}: Integrated yields above an energy threshold of $E_X > m_\chi/2$.
	}
	\label{fig:yieldratio_svratio}
\end{figure}

In Figs.~\ref{fig:yieldratio}--\ref{fig:yieldratio_svratio}, we therefore consider the integrated yield of all 
relevant stable particles X = $\bar{p}, \gamma, \nu_\mu, \nu_\tau, e^+$ above some threshold energy.
Fig.~\ref{fig:yieldratio}, in particular, shows the ratio of the 3-body yield to the typically considered
yield expected from  2-body final states, for $E_X > m_\chi/10$, as a function of the neutralino 
mass and for different neutralino compositions. In  Fig.~\ref{fig:yieldratio_sv2to2}, we show the same 
quantity, but now as a function of the total 2-body annihilation rate. This immediately allows us to 
make the interesting observation that the 
yields are most strongly enhanced for models with cross sections somewhat below the `thermal' cross 
section of $\sim 10^{-26}{\rm cm}^3{\rm s}^{-1}$; including the effect of electroweak corrections, as 
already pointed out in Ref.~\cite{Bringmann:2013oja}, may thus turn out to be the decisive ingredient 
to make these models accessible by current and near-future indirect detection experiments. We 
complement these two sets of figures by Fig.~\ref{fig:yieldratio_svratio}, which demonstrates how the yield 
enhancement correlates with the cross section enhancement discussed in the previous subsection. 
This is done both for the yield enhancement at low energies (left) and at high energies close to 
the kinematic threshold (right), in order to get a first qualitative indication of how the spectral 
shape is affected. In the following, we discuss the various relevant stable particles in turn.

\begin{description}
\item[Antiprotons.]
For the bulk of the models considered here, the enhancement in the $\bar p$ yield scales as expected 
linearly with the enhancement in $\sigma v$. For $\tau^+\tau^-$ final states, however, 
antiproton production is not kinematically allowed. Including the $\tau^+\tau^-Z$ and $\tau^\pm\nu_\tau W^\mp$ 
channels can therefore drastically enhance the $\bar p$ yield even if the corresponding enhancement 
of the cross section is at most moderate.
This effect, clearly seen in Fig.~\ref{fig:yieldratio_svratio}, is responsible for the largest 
enhancements (for Bino-like models) in the left panel of  Fig.~\ref{fig:yieldratio}. It is also reflected in 
the left panel of Fig.~\ref{fig:yieldratio_sv2to2}, which shows that the enhancement is most pronounced 
for models with small 2-body annihilation rates.
\\
Antiprotons are only produced in the fragmentation and decay of the annihilation products, and therefore 
cannot obtain energies $E_{\bar p}\sim m_\chi$. The additionally emitted gauge or Higgs boson must 
first decay to quarks, inducing an additional step in the $\bar p$ production 
chain compared to 2-body quark final states.
The largest enhancement of the antiproton yield from 3-body final states will thus on 
average occur mostly at small energies, an effect clearly seen when comparing the two panels of 
Fig.~\ref{fig:yieldratio_svratio}.

\item[Photons.] Compared to antiprotons, gamma-ray yields lack the ``atypical'' 
enhancement from $\tau^+\tau^-$ final states. Consequently, the enhancement in the yield
is generally weaker, and more strongly correlated with the one in $\sigma v$. 
As for antiprotons, 
the spectrum is only enhanced at somewhat lower energies because the additional photons only result 
from steps further down in the decay chain. Unlike for the $\bar p$ case, on the other hand, our improved 
treatment of the NWA prescription for differential rates detailed in Section \ref{sec:NWAdiff} can actually 
{\it decrease} the photon yield at large energies (see Section \ref{sec:spectra} for example spectra).
The main reason for the large difference between the two panels of Fig.~\ref{fig:yieldratio_svratio}, 
however, are rather the monochromatic photon final states $\gamma Z$ and 
$\gamma\gamma$. While the annihilation rate into these 
states is loop-suppressed, they can still dominate the {\it differential} photon yield at the highest 
energies.\footnote{
Note that, throughout the manuscript, we include final states that vanish
at tree-level when referring to two-body final states. This is not only the usual convention adopted 
in \ds, but serves to stress our emphasis on the differences between 2-body and 3-body final states
(rather than between tree-level and next-to-leading order results).
}
The models where the photon-yield enhancement due to the inclusion of 3-body final states 
is largest -- which, from Fig.~\ref{fig:yieldratio} are those close to the $W$ threshold -- thus at the same
time feature particularly large annihilation rates into $\gamma Z$ and $\gamma\gamma$, too.
 
\item[Leptons.] Unlike photons and antiprotons,  leptons can appear directly
in the final states considered here. This leads to a yield enhancement in particular at high 
energies, $E_\ell\gtrsim m_\chi/2$,  and with a strong correlation with the $\sigma v$ enhancement. 
Observationally,  the resulting characteristic spectral features are especially relevant for 
positrons \cite{Bergstrom:2013jra} in view of
the excellent energy resolution of the AMS experiment \cite{Accardo:2014lma},  
but also the  neutrino spectra can be striking signatures to look for \cite{Fukushima:2012sp, Ibarra:2013eba, Ibarra:2014vya} if neutralino annihilation 
in the sun is sizeable (for more details, see the benchmark spectra discussed below).
\\
As expected, $e^+$ and $\nu_e$ yields are in general very similar, and the same holds for $\nu_\mu$. 
Large yield enhancements are in particular found {\it (i)} for models below the $W$ threshold, dominated
by $\chi\chi\to W^*W \to \ell\nu W$, and {\it (ii)} for TeV models with almost degenerate sleptons
and dominant 3-body rates $Z\ell\ell$ and $W\ell\nu$.
Compared to the other 
leptons,  tau neutrinos can receive a somewhat larger yield enhancement, and generally feature a 
spectrum that  is less pronounced at small energies. The reason for both effects is that the decay of the 
copiously produced pions,
both from 2-body final states and due to the additional final state boson, results in many low-energy 
1$^{\rm st}$ and 2$^{\rm nd}$ generation leptons, but hardly any 3$^{\rm rd}$ generation leptons.
\end{description}

\subsection{Indirect detection spectra}
\label{sec:spectra}

\begin{table}
\begin{center}
{\scriptsize
\begin{tabular}{|c|c|c|c|c|c|c|c|c|c|c|c|c|}
\hline
\!\!\!\!\!Bench\!\!\!\! & 
\multirow{2}{*}{\!\!\!\!\!Type\!\!\!}&
\multirow{2}{*}{\parbox{0.8cm}{\centering $\mu$  \\ (GeV)}} &
\multirow{2}{*}{\parbox{0.8cm}{\centering$M_2$ \\ (GeV)}} &
\multirow{2}{*}{\parbox{0.8cm}{\centering$M_1$ \\ (TeV)}} &
\multirow{2}{*}{\parbox{0.8cm}{\centering $M_A$  \\ (GeV)}} &
\multirow{2}{*}{\!\!$\tan \beta$\!\!}
 & \multicolumn{2}{|c|}{$A$ ($M_{\tilde{q}}$)} 
 & \multicolumn{2}{|c|}{$M_{\tilde{q}}$ (GeV)} 
 & \multicolumn{2}{|c|}{$M_{\tilde{\ell}}$ (GeV)}  \\\cline{8-9}\cline{10-11}\cline{12-13}
   mark&  & & &  &  & & $A_t$ & $A_b$ & {\tiny 1$^\mathrm{st}$/2$^\mathrm{nd}$}& {\tiny 3$^{\rm rd}$} & 
   {\tiny 1$^\mathrm{st}$/2$^\mathrm{nd}$} & {\tiny 3$^{\rm rd}$} \\
 \hline
\hline
{\bf D1} & 93 &  -2022 & 1257 & 1125 & 2841 & 12.4 & -2.00 & 2.00 & 1715 & 1715 &  1180 & 1147 \\[2pt]
\multirow{2}{*}{\bf \!\!D2}&  \multirow{2}{*}{\!\!94} & \multirow{2}{*}{3531} & \multirow{2}{*}{3458} & \multirow{2}{*}{3396} & \multirow{2}{*}{2562} & 
\multirow{2}{*}{22.5} & \multirow{2}{*}{1.94} & \multirow{2}{*}{1.94} & \multirow{2}{*}{3599}  & 3665$^{\rm R}$ & 
\multirow{2}{*}{3599}  & \multirow{2}{*}{3599}   \\
 &&  &  &  &  &  &  &  &   & 5419$^{\rm L}$ &   &    \\
{\bf D3}&  93 & 3586 & 411.9 & 380.2 & 1766 & 11.7 & -1.43 &  -1.43 & 3714 & 3714 & 485.3 &  470.1 \\[2pt]
{\bf T1} & 91 & -102.0 & 601.7 & 119.8 &  521.1 & 12.5 &  -2.00 & -1.81 &  1350 & 1350 & 1072 & 1072 \\[2pt]
{\bf T2} & 91 & -586.4 & 189.0 &  172.6 & 981.4 & 8.66 &  -1.56 & 2.34 & 1733 & 1733 &  698.7 & 698.7 \\[2pt]
{\bf W} & 93 & -3717 & 1346 & 1319 & 3160 & 9.01 & -1.91  & -1.91  & 2872  & 2872  & 1748 &  1885 \\[2pt]
{\bf H} &  93 & 3492 & 3976 & 3371& 1418 & 9.47 & 1.57 & 1.57 & 3629 & 3629 & 3405 & 3783 \\
\hline
\hline
\end{tabular}
}
\caption{Benchmark models for which we show the resulting spectra of stable particles in 
	Figs.~\ref{fig:modC}--\ref{fig:special}, with model parameters for the various pMSSM-9 types 
	introduced in Section \ref{sec:models}. $A_{t/b}$ are given in units of $M_{\tilde t_R}$. 
	Model D2 takes as input the right-handed stop mass (R), 
	the left-handed third generation squarks mass (L), 
	and otherwise assumes a common mass scale for all other sfermion mass terms.
	See Table \ref{tab:pheno} for some phenomenological properties, and main text for more details. 
	}
\label{tab:benchmark}
\end{center}
\end{table}

The additional final state boson may not only enhance the yields significantly, 
as described above, but also change the shape of the resulting cosmic-ray spectra in a characteristic 
way. The maybe most striking examples that we identified are sharp spectral features near the kinematic 
endpoint of neutrino and positron spectra, resembling in fact the often discussed cases of 
positron \cite{Bergstrom:2008gr} and gamma-ray \cite{Bringmann:2007nk} spectra for photon VIB. In this 
Section, we discuss in more detail a few example models where the spectrum of at least one type of 
stable particles changes significantly once 3-body processes are taken into account.

\begin{table}
\begin{center}
{\footnotesize
\begin{tabular}{|c|c|c|c|c|c|c|c|c c|}
\hline
Bench
 & $m_\chi$ & \multirow{2}{*}{$\tilde{\chi}^0$} & $m_{\tilde{q}}$ ($\tilde{q}$)  & $m_{\tilde{l}}$ ($\tilde{l}$)  & $m_{\tilde{\chi}^i}$
 & \multirow{2}{*}{$\displaystyle \frac{\sigma v_{3b}}{\sigma v_{2b}}$} & \multirow{2}{*}{$\displaystyle \frac{\sigma v_{3b,h}}{\sigma v_{3b,W/Z}}$} 
 & \multirow{2}{*}{$\displaystyle \frac{N_{3b,X}}{N_{2b,X}}$} & \multirow{2}{*}{$X$} \\
 mark & [GeV] & & [GeV] &  [GeV] &   [GeV]& & && \\
 \hline
\hline
{\bf D1} & 1125 & $\tilde{B}$ &  1551 ($\tilde{t}_1$) & 1129 ($\tilde{\tau}_1$)& 1254 & 0.58 & 0.12 & 61 & $\nu_\mu$\\
{\bf D2}& 3396 &  $\tilde{B}$ & 3478 ($\tilde{b}_1$) & 3397 ($\tilde{\tau}_1$) & 3458 & 2.7  & 0.64 & 8.9 & $\nu_\tau$\\
{\bf D3}& 380.1 & $\tilde{B}$ &  3585.5 ($\tilde{t}_1$) & 385.3 ($\tilde{\tau}_1$) & 411.4 & $0.17$  & $0.52$ & 80 & $\bar p$\\ 
{\bf T1} & 80.34 & $\tilde{B}/\tilde{W}$  & 1176 ($\tilde{t}_1$) & 1070  ($\tilde{\nu}_e$) & 113.1 & 25 & $< 10^{-3}$ & 43 & $e^+$\\
{\bf T2} & 172.4 & $\tilde{B}$ & 1604 ($\tilde{t}_1$) & 693.7 ($\tilde{\tau}_1$ ) & 188.8 & 3.0  & $< 10^{-3}$ & 2.8 & $\nu_\tau$\\ 
{\bf W} &1319 & $\tilde{B}$ & 2720 ($\tilde{t}_1$) & 1746 ($\tilde{\nu}_e$) & 1345 & 0.39 & 0.12 & 26 & $\nu_\mu$ \\ 
{\bf H} & 3371 & $\tilde{B}$ &  3404 ($\tilde{b}_1$) & 3400 ($\tilde{\mu}_1$) & 3976 & 3.1 & 4.9 & 2.8 & $\nu_\tau$\\ 
\hline
\end{tabular}\\[2ex]
\begin{tabular}{|c|cc|cccc|}
\hline
Bench
 &  \multicolumn{2}{|c|}{\multirow{2}{*}{$2\to 2$}} & \multicolumn{4}{|c|}{\multirow{2}{*}{$2\to 3$}}  \\
 mark & & & &  &&  \\
 \hline
\hline
{\bf D1} &  $gg$ (48\%) & $b b$ (42\%) &
   $W\nu\ell$ (37\%) &$Z\ell\ell$ (33\%) & $Z\nu\nu$ (9\%) &   $Wtb$ (5\%) \\ 
{\bf D2}&  $H W/Z,hA$ (54\%)  & $gg$ (25\%) &
   $W\nu\ell$ (30\%) & $h\ell\ell$ (17\%) & $Z\ell\ell$ (16\%) &   $Htb$ (10\%) \\ 
{\bf D3}& $\tau \tau$ (99\%)  & $\gamma\gamma$ (0.3\%)&
    $W\nu\ell$ (42\%) & $h\ell\ell$ (37\%) & $Z\ell\ell$ (20\%) & \\
{\bf T1} &  $c c$ (37\%) & $gg$ (31\%) & 
    $WQq$ (63\%) & $W\nu\ell$ (33\%) & $Zqq$ (2.6\%) & \\
{\bf T2} &  $W W$ (59\%) & $gg$ (21\%) &
    $W t b$ (99\%) &&&\\ 
{\bf W} & $gg$ (56\%) & $bb$ (28\%) & 
     $W\nu\ell$ (31\%) & $Z\ell\ell$ (28\%) & $Wtb$ (14\%) & $Z\nu\nu$ (7.5\%) \\ 
{\bf H} &  $H W/Z, hA$ (62\%) & $gg$ (25\%) &
     $Htb$ (42\%) & $Htt$ (19\%) & $Att$ (17\%) & $Wtb$ (6.8\%) \\ 
\hline
\end{tabular}
}
\caption{Characteristic properties of the benchmark models defined in Table~\ref{tab:benchmark}. 
The upper table shows neutralino mass and composition ($\tilde B=$ Bino-like, 
$\tilde B/\tilde W=$ mixed Bino/Wino), identity and mass of the lightest squark and slepton, 
next-to lightest neutralino, ratio of 3-body to 2-body cross section, ratio of 3-body cross sections
involving Higgs bosons to that involving weak gauge bosons, and the ratio of 3-body to 
2-body yields for various species $X$ (integrated above $E_X>m_\chi/10$ for D3, T1, T2 and 
above $E_X>m_\chi/2$ for D1, D2, W, H). The lower table shows the dominant two- and 3-body 
annihilation channels (for leptonic channels we sum over all three generations, while for quarks 
we quote separately the final states involving top quarks).
\vspace*{-0.8cm}
	}
\label{tab:pheno}
\end{center}
\end{table}

For this purpose, we define seven pMSSM-9 benchmark models in Table~\ref{tab:benchmark},
and collect in Table~\ref{tab:pheno} the phenomenological properties that are most relevant for our 
discussion. In particular, we include two threshold models, T1 and T2, with neutralino masses just below 
the $W$ and $t$ mass, respectively. Three of the benchmark models, D1 to D3, show a mass spectrum 
where at least one of the sfermions is degenerate in mass with the neutralino, while model W describes a 
TeV-scale Bino DM candidate where the correct relic density is obtained due to 
coannihilations with an almost degenerate Wino. Model H, finally, is a model example with a 
particularly large rate to 3-body final states containing a Higgs boson.

\begin{description}
\item[Degenerate mass spectra.]
Models with \textit{all sleptons} degenerate in mass with the neutralino show a significant overall 
enhancement of the yield in leptonic channels, $\ell=e^\pm, \nu_\mu, \nu_\tau$, caused by sharp 
spectral features at the kinematic end-point of those spectra. In full analogy to the
positron spectrum from VIB $e^+e^-\gamma$ final states \cite{Bergstrom:2008gr},
the annihilation in these models is dominated by $t$-channel diagrams with $\ell$ appearing 
directly in the final state (and additionally in the decay of the $W$ or $Z$ boson in the three-body final state);
 these diagrams lift the helicity suppression, and are maximized when 
the corresponding sleptons are degenerate with the neutralino. 
As an example of this type of models, we show in Fig.~\ref{fig:modC} the spectrum of benchmark D1. 
For leptonic final states (left panel) we can clearly see these sharp spectral features, leading to a yield
enhancement as large as $\mathcal{O}(100)$ at high energies for {\it all} leptons. $SU(2)$ corrections 
thus further enhance the $e^+$ feature associated to photon IB, indicated separately with dotted lines,
which the AMS experiment is highly sensitive to~\cite{Bergstrom:2013jra}. In addition, similar features 
appear also in neutrino final states,  giving rise to a potential smoking-gun signature for annihilating 
Majorana DM at neutrino 
telescopes~\cite{Ibarra:2013eba,Bell:2012dk,Fukushima:2012sp}.\footnote{
If only first and second generation sleptons were degenerate in mass with the 
neutralino, a corresponding feature for $\nu_\tau$  (but not for $e^\pm$ and $\nu_\mu$) 
would be absent. The (non-)detection of such features can thus in addition be a 
powerful tool to robustly discriminate between such scenarios. 
}
In the right panel of Fig.~\ref{fig:modC}, we show instead the impact of radiative corrections on the 
gamma-ray and antiproton spectra. The impact of photon IB on the former is as expected 
large \cite{Bringmann:2007nk}, while $SU(2)$ corrections lead to much less significant, though still 
noticeable, spectral distortions. The antiproton spectrum only receives an overall enhancement 
directly related to the total $\sigma v$ enhancement.
%
\begin{figure}[t!]
	\centering
		\includegraphics[width=0.45\columnwidth]{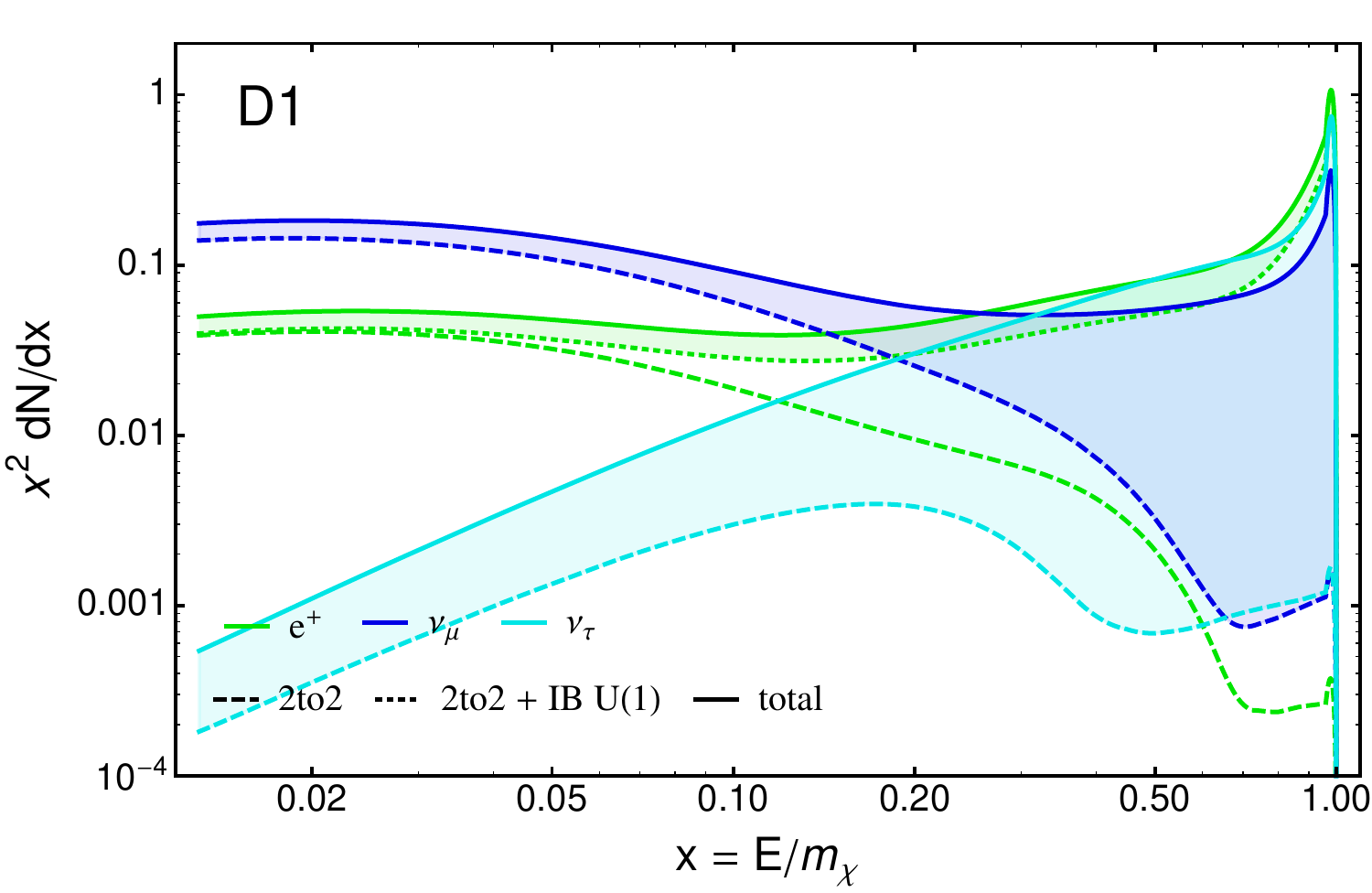}
		\hspace*{0.4cm}
		\includegraphics[width=0.45\columnwidth]{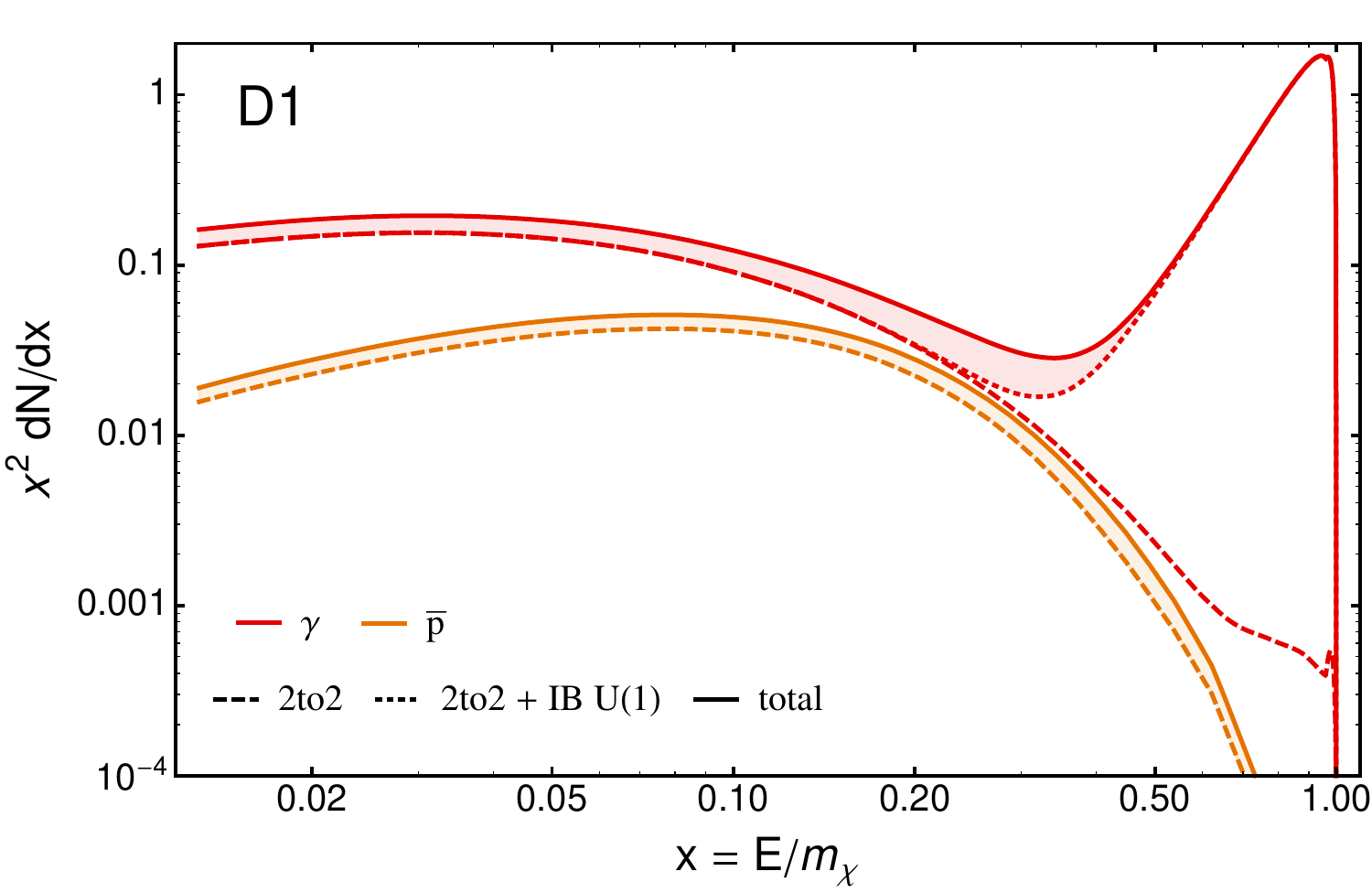} 
	\caption{Spectral energy distribution of a model where \textit{all} sleptons are degenerate (D1 
		in Table \ref{tab:benchmark}). In the \textit{left panel}, 
		we display leptonic final states $e^\pm$, $\nu_\mu$, and $\nu_\tau$ (green, blue and cyan line respectively) and, in 
		the \textit{right panel} photon (red) and antiproton (orange) spectra. 
		Solid lines indicate the total (NWA-corrected 2-body plus 3-body) contribution, the dashed lines the 2-body result.
		The dotted lines represent the contribution from 2-body final states plus that from {\it photon} 
		bremsstrahlung alone.
		Shaded areas thus highlight the effect of including $SU(2)$ corrections.}
	\label{fig:modC}
\end{figure}
%
\begin{figure}[t!]
	\centering
		\includegraphics[width=0.45\columnwidth]{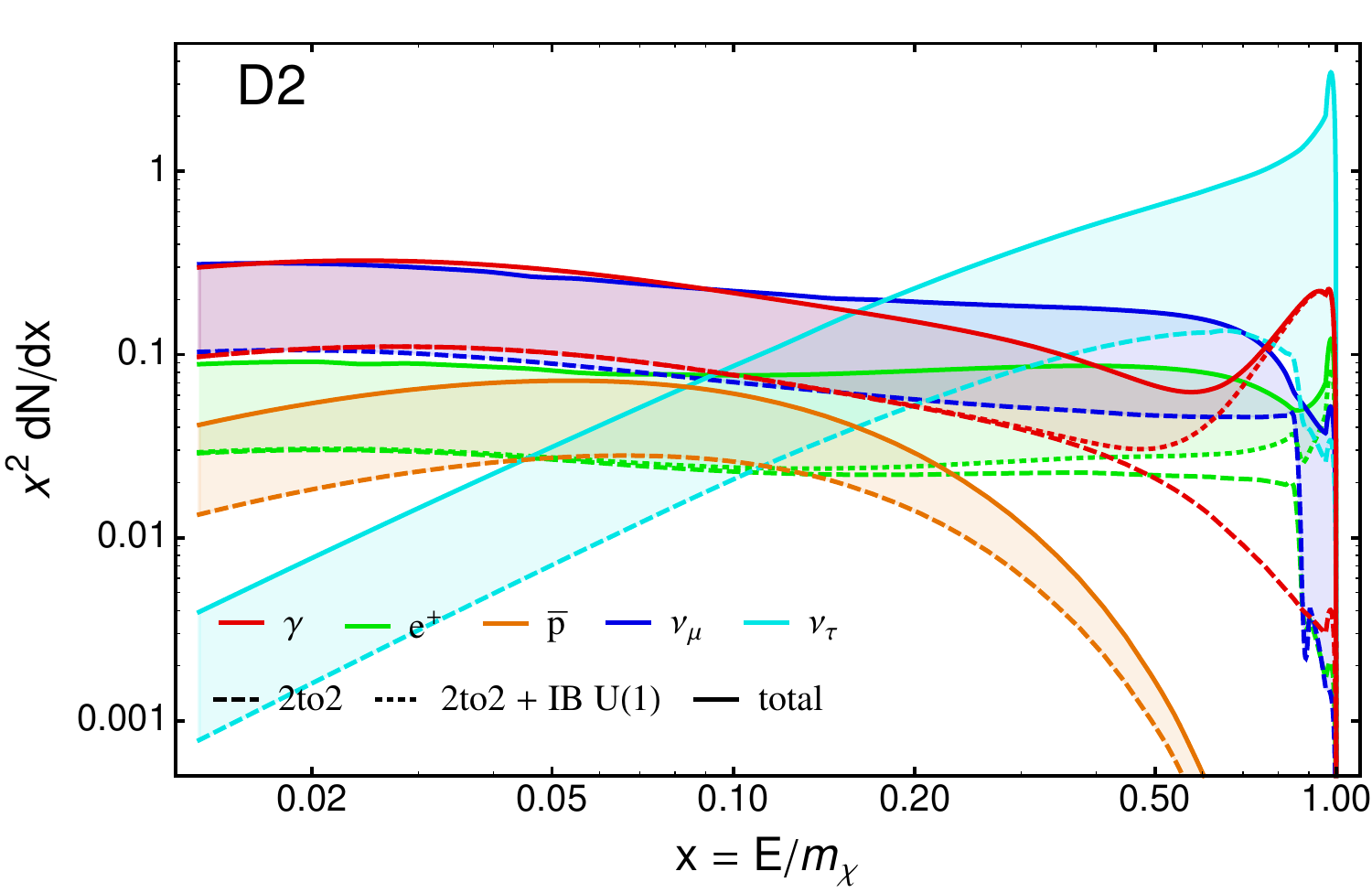} 
		\hspace*{0.4cm}
		\includegraphics[width=0.45\columnwidth]{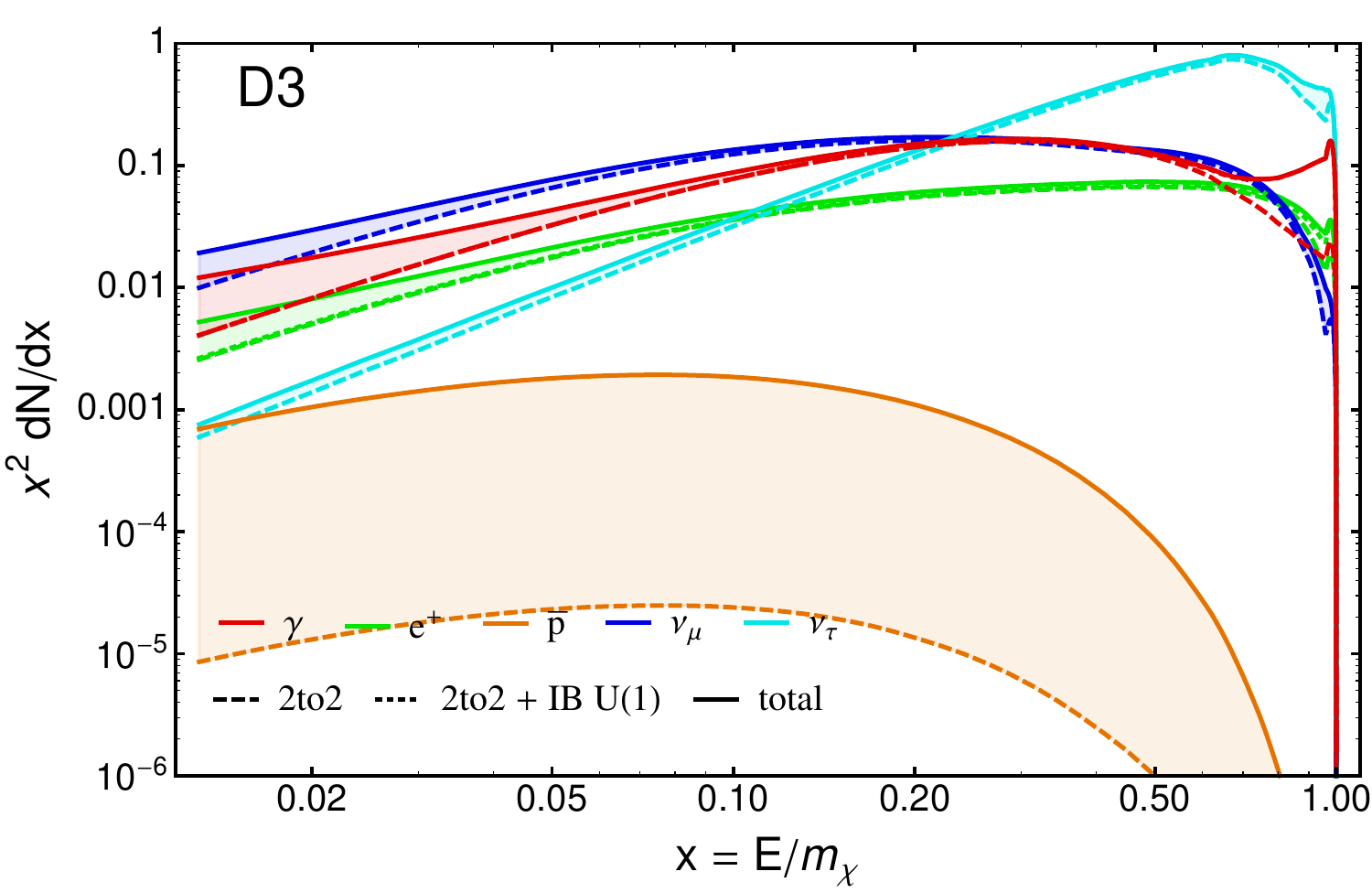} 
	\caption{Same as Fig.~\ref{fig:modC}, but now for models where both squarks and sleptons are degenerate 
	        in mass with the neutralino (model D2, left panel) or with a degenerate stau that is a mixture of $\tilde\tau_L$ and $\tilde\tau_R$ (model D3, right panel).}
	\label{fig:D23}
\end{figure}
%
\\[0.5ex]
If, on the other hand, only {\it squarks} are degenerate, then the lepton spectrum does not show any 
significant distortion but just an overall enhancement proportional to the one in $\sigma v$,
while the photon spectrum again hardens slightly. In the left panel of Fig.~\ref{fig:D23} we show
for illustration the case of benchmark model D2, which features {\it both} degenerate squarks
and sleptons and hence, as expected qualitatively very similar spectra compared to 
those of D1.\footnote{
One noticeable feature is the step-like behaviour of lepton yields from 2-body annihilation 
(dashed lines). This drop by more than an order of magnitude, at $x\gtrsim 0.85$, is due 
the channels $H^\pm W^\mp$, $hA$ and $HZ$ which contribute about $50\%$ to the 
2-body cross section: the $W$/$Z$ decays constrain the resulting lepton energy to 
$E_\ell < E_{W/Z} = m_\chi (1-(m_H^2-m_{W/Z}^2)/(4m_\chi^2)) \sim 0.85m_\chi$
 (for $m_H=2.6\,$TeV as in D2). The 3-body yields (solid lines) smear 
out the abrupt step (for $\nu_\mu$) or lead to a pronounced bump at $x\sim 1$ (for $\nu_\tau$).
}
\\[0.5ex]
The spectra of benchmark model D3 (right panel of Fig.~\ref{fig:D23}) are again qualitatively
very different and feature a significant
enhancement only in the antiproton channel. The reason, as already discussed in Section \ref{sec:yield+}, 
is that the 2-body annihilation in D3 is largely dominated by $\tau\tau$ final states. 
Therefore, the 3-body final states, specifically $h\tau\tau$ and $W\tau\nu$, lead
to a very large enhancement of the $\bar p$ flux -- even though this only leads to a 
relatively small ($\sim 20\%$) correction in $\sigma v$.\footnote{
This model features a very large $\mu$-term ($\sim 3.5$\,TeV) and $\tan(\beta)\sim 12$,
leading to a large stau mixing and hence only a mild helicity suppression for $\chi\chi\to\tau^+\tau^-$.
Corrections to leptonic channels are thus small in this specific case, despite an almost degenerate 
stau. We also note that $h\tau\tau$ final states, via Higgs VIB, are enhanced with respect to gauge 
boson IB in this type of model. The reason is that the mixing contribution 
$\propto  g\mu v_{EW}\tilde\tau_L\tilde\tau_R$ to the stau mass is directly linked to a large 
$\tilde\tau_L \tilde\tau_R h$ coupling to the Higgs boson (since the VEV and the Higgs field appear in the 
combination $v_{EW}+h$ in the Lagrangian).
}

\begin{figure}[t!]
	\centering
		\includegraphics[width=0.45\columnwidth]{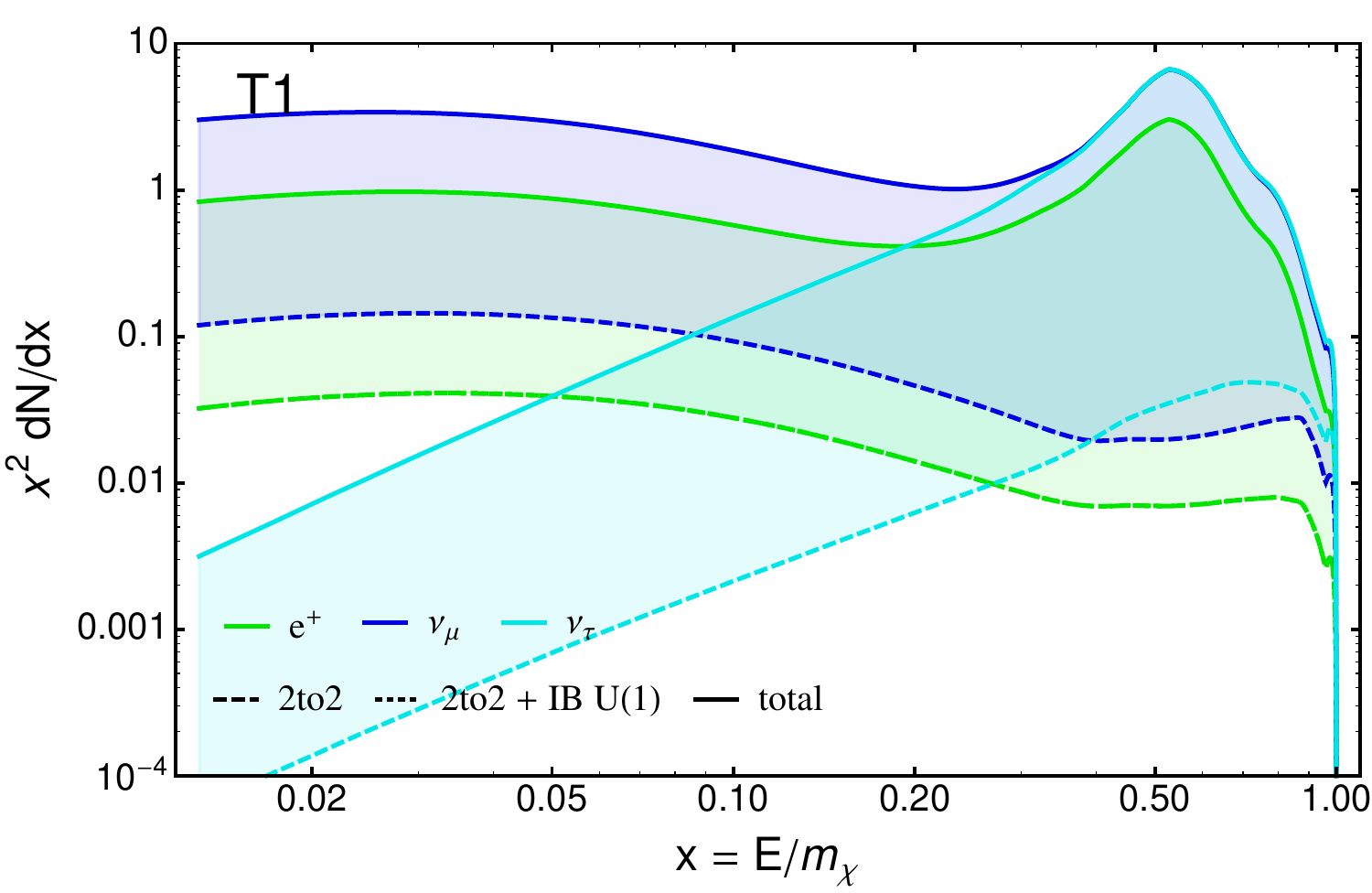} 
		\hspace*{0.4cm}
		\includegraphics[width=0.45\columnwidth]{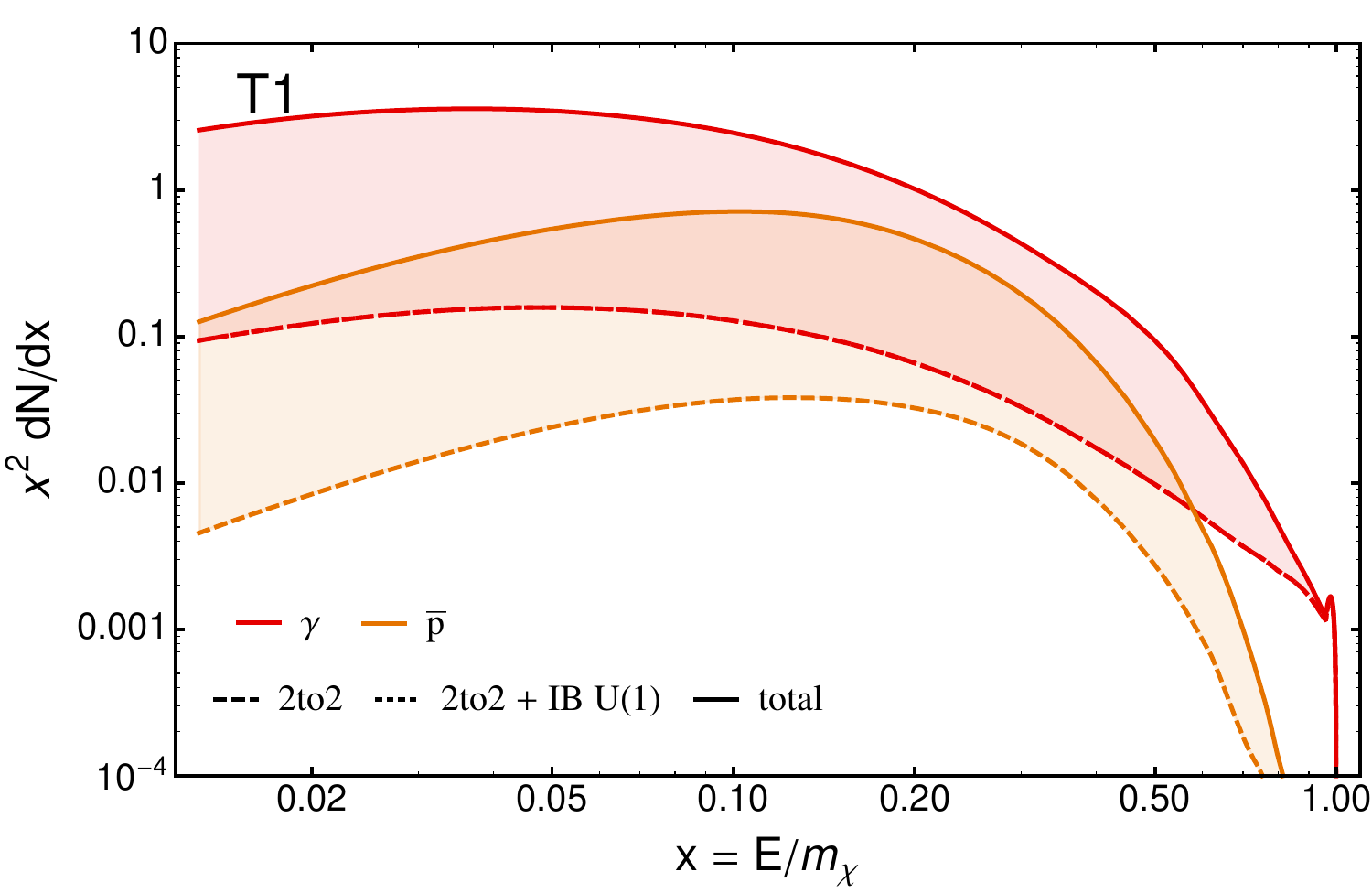} \\[5pt]
		\includegraphics[width=0.45\columnwidth]{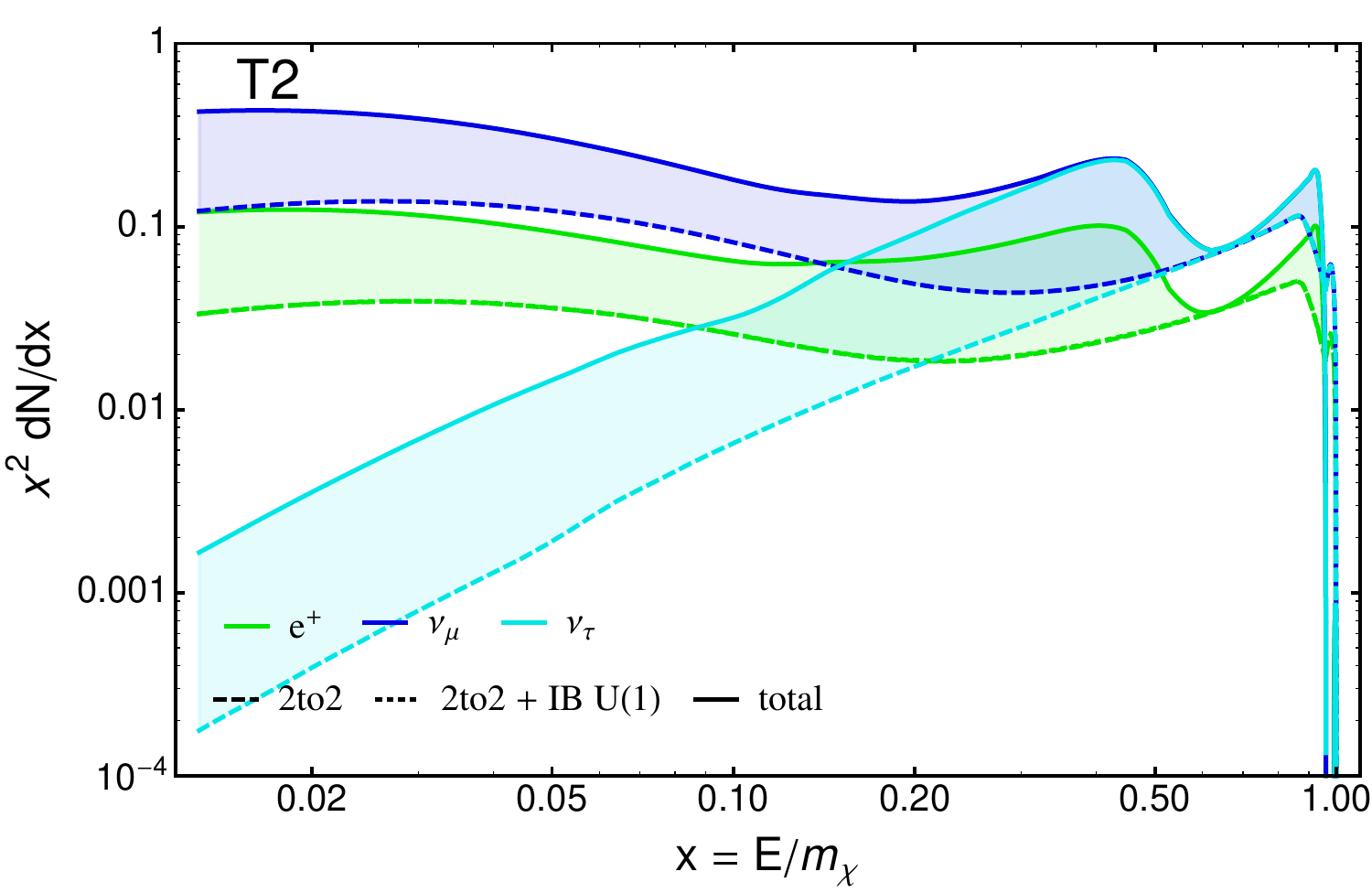} 
		\hspace*{0.4cm}
		\includegraphics[width=0.45\columnwidth]{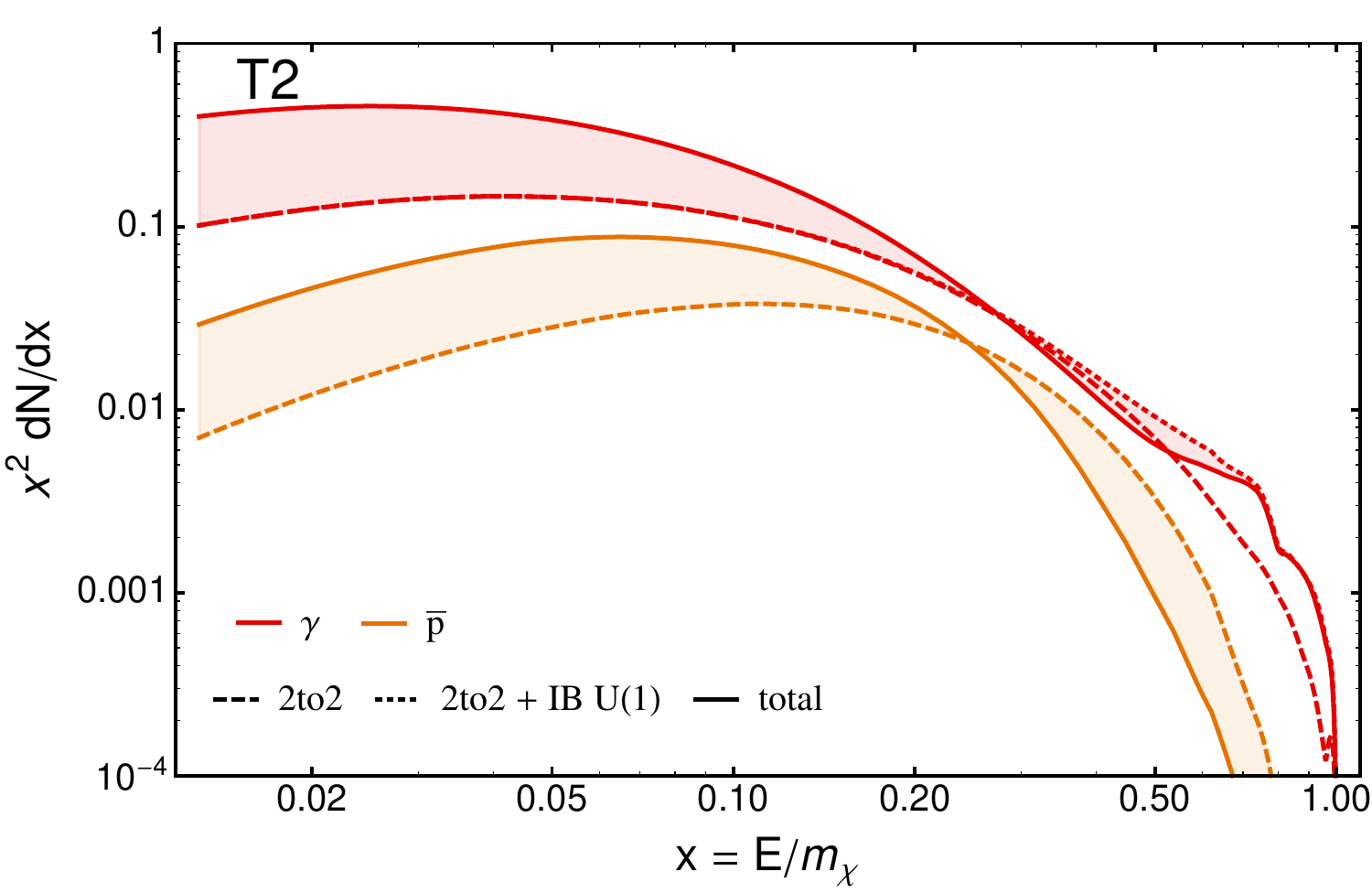} 
	\caption{Same as figure~\ref{fig:modC}, but for a $W$ threshold model (T1, top)
	and a top-quark threshold model (T2, bottom). The features in the
        shape of the lepton and gamma
        ray spectra are due to an interplay of various effects as discussed in detail in the text.
	}
	\label{fig:thresholds}
\end{figure}

\item[Threshold models.]
In Fig.~\ref{fig:thresholds}, we display the spectra for two models just below the $W$ and 
top quark threshold, corresponding to models T1 and T2 from Tab.~\ref{tab:benchmark}, respectively. 
Models with $m_\chi\sim m_W$ show a strong enhancement in all channels and 
at all energies. In particular, the 3-body contributions induce pronounced bump-like features for 
$\nu_\mu$, $\nu_\tau$, $e^\pm$ spectra at $x \sim 0.7$, which result from the decay 
$W^{(*)}\to \ell\nu_\ell$. As expected from the discussion in Section \ref{sec_threshold}, these 
enhancements are most significant for $m_\chi$ slightly {\it below} $m_W$.
Photon and antiproton spectra, on the other hand, are somewhat softer than the 
2-body spectra, but can be greatly enhanced at all energies.
\\[0.5ex]
We also find enhancements in all channels when $m_\chi\sim m_t$, and observe a peculiar 
double structure in the spectrum of leptonic channels for T2: a bump-like feature 
at $x \sim 0.4$ and a sharp spectral feature at higher energies. 
The first is directly related to the dominant off-shell top decay close to the 
threshold, $\chi\chi\to t \bar t^{(*)}\to W t b$.
The line-like feature close to $x=1$, on the other hand, arises from leptonic decays of 
the transversely polarized $W$-bosons in the process $\chi\chi\to W \bar W \to W \ell \nu_\ell$.\footnote{ 
Note that the NWA-subtracted cross section for $W\ell\nu_\ell$ is much smaller than for the 
kinematically accessible $\chi\chi\to W^+W^-$. The enhancement at $x\sim1$ originates thus exclusively
from our 3-body computation taking the (transverse) $W$ polarization 
into account (see Appendix \ref{app:corr}, specifically Eq.~(\ref{eq:Wpol})). 
This type of correction 
can occur whenever $m_\chi\gg m_W(m_Z)$ and $\chi\chi\to W^+W^-,ZZ$ 
proceeds with a significant rate. 
} 
\\[0.5ex]
For the antiproton and gamma spectra, finally, the inclusion of the 3-body result causes relatively
large deviations as the 2-body yields can both increase and decrease, 
depending on the energy.\footnote{
The peculiar shape of the gamma-ray spectrum, in particular, can be explained as follows: 
for low energies $x\lesssim 0.2$ there is a strong enhancement from $Wtb$ final states, 
while for intermediated energies $x\sim 0.2-0.6$, the spectrum is 
slightly suppressed compared to the 2-body case. At high energies $x\gtrsim 0.6$, 
$WW\gamma$ and $\bar\ell\ell\gamma$, i.e.~photon IB final states dominate; the sharp drop 
around $x\sim 0.8$ is due to the kinematic endpoint of the $WW\gamma$ contribution, 
while photons from $\bar\ell\ell\gamma$ dominate for $x>0.8$.
}

\begin{figure}[t!]
	\centering
		\includegraphics[width=0.45\columnwidth]{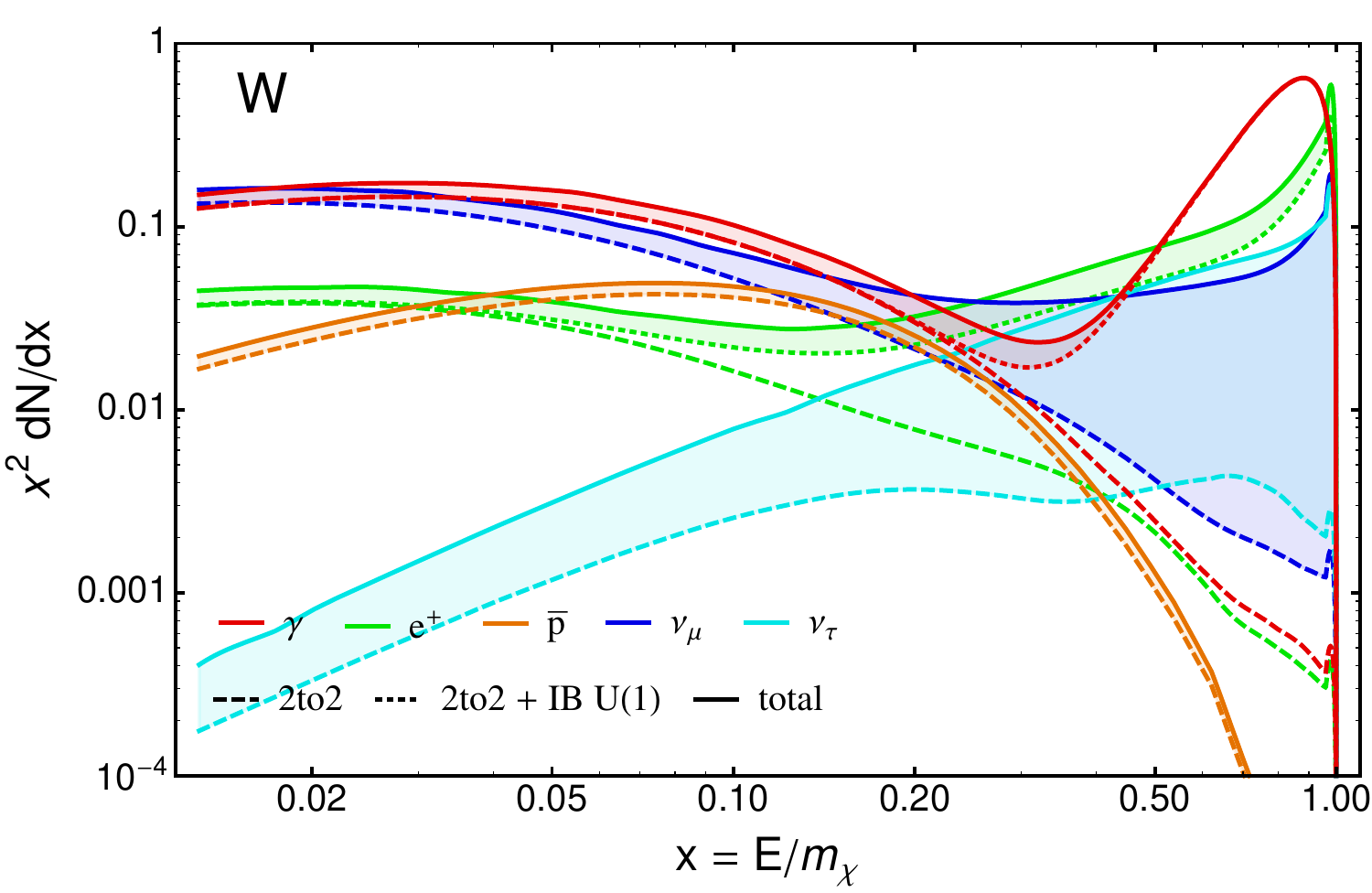} 
		\hspace*{0.4cm}
		\includegraphics[width=0.45\columnwidth]{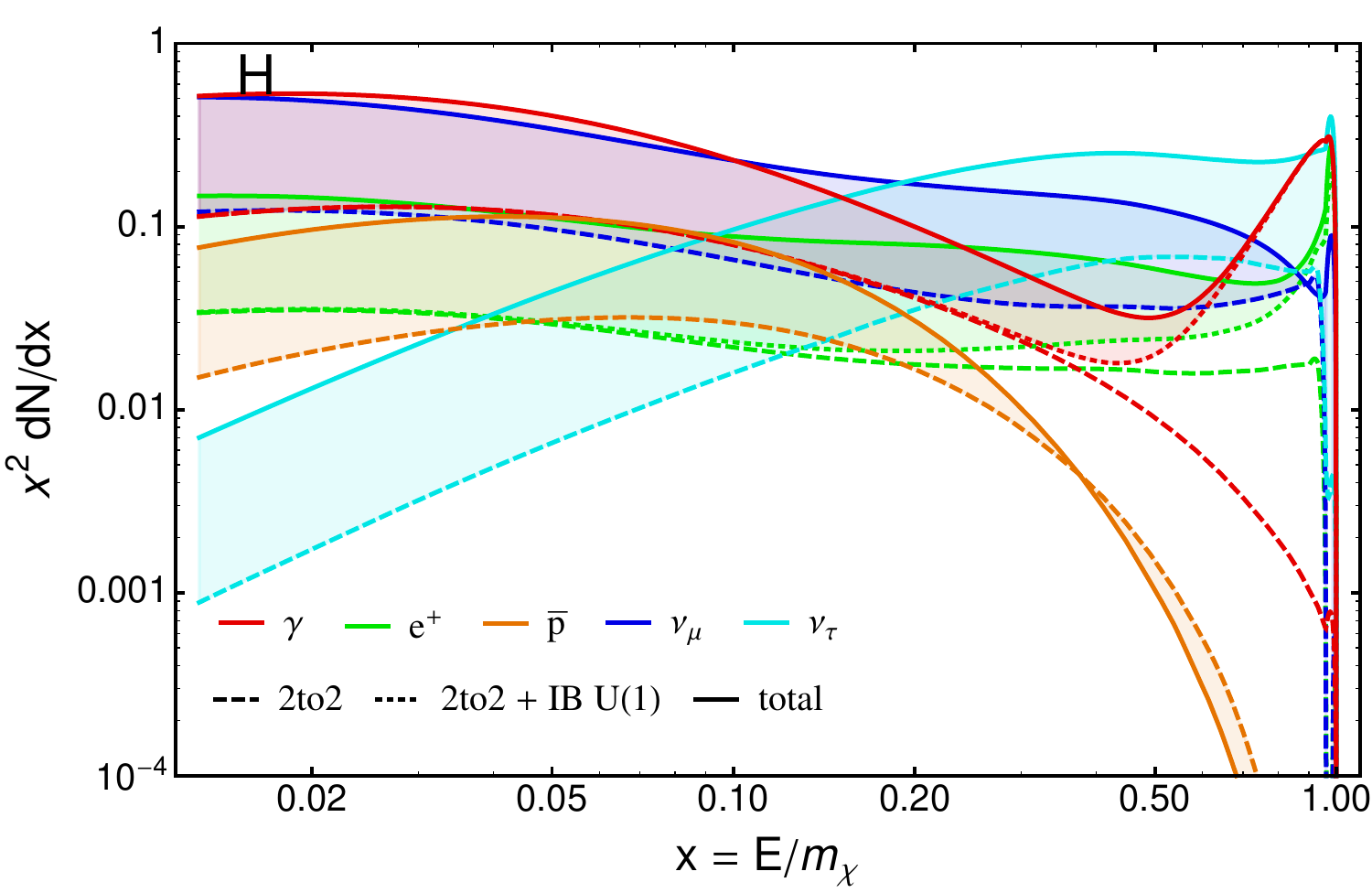} 
	\caption{Same as figure~\ref{fig:modC}, but for a Bino-like neutralino with almost 
	degenerate Wino (W, left) 	and a model with large $H f \bar{f}$ contribution (H, right). 
	The features in the
        shape of the lepton and gamma
        ray spectra are due to an interplay of various effects as discussed in detail in the text. 
		\textit{Left panel}: Bino-like neutralino with almost degenerate Wino (benchmark model W).
		Final state channels: photons (red), antiprotons (orange), positrons (green), $\nu_\mu$ (blue), and $\nu_\tau$ (cyan).
		Solid lines indicate the total (2-body and 3-body) contribution, the dashed lines the 2-body process.
		\textit{Right panel}: Large $H f \bar{f}$ contribution (benchmark model H).}
	\label{fig:special}
\end{figure}

\item[Special cases.]
In Fig.~\ref{fig:special}, we show two interesting cases that do not fall in either of the categories 
above. In the left panel, we present benchmark model W, a Bino-like neutralino degenerate with 
the Wino.  The (small) 2-body annihilation rate is dominated by $gg$ final states, followed by $\bar ff$.
The 3-body process thus lifts the helicity suppression of the latter and can be important even
if the sfermions are not highly degenerate in mass with the neutralino. Because the contribution to the 
neutrino and positron spectra still come dominantly from $W \nu \ell$ final states, they show sharp 
spectral features like in models with even more degenerate sleptons. 
The right panel
of Fig.~\ref{fig:special}, instead, corresponds to a model with a large ($\sim85\%$) contribution to the cross
section from channels that involve the MSSM Higgs bosons and top quarks (benchmark model H). The neutralino 
mass is rather heavy ($\sim 3.3$ TeV) such that even $t\bar t$ final states suffer from a certain amount 
of helicity suppression. Due to the large top Yukawa coupling, the suppression is lifted preferably via 
Higgsstrahlung. For this model, leptons are dominantly produced indirectly, and correspondingly lepton 
spectra are enhanced broadly at all energies. The small additional spike at very 
high energies results from the $W/Z$ decay from $W\bar F f$ ($10\%$) and $Z\bar f f$ ($5\%$) 
final states.
\end{description}
 
 \begin{figure}[t!]
	\centering
		\includegraphics[width=0.45\columnwidth]{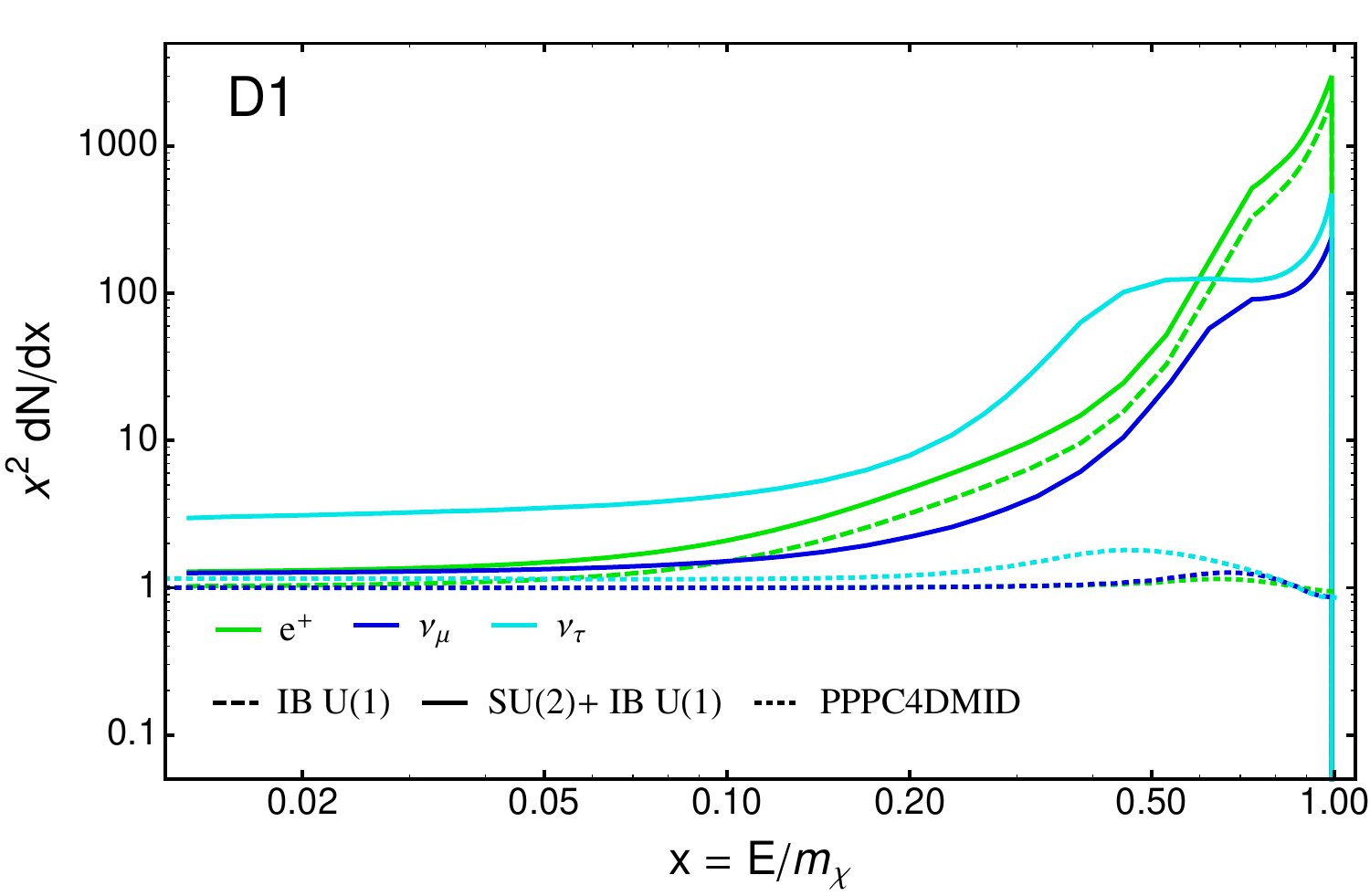}
		\hspace*{0.4cm}
		\includegraphics[width=0.45\columnwidth]{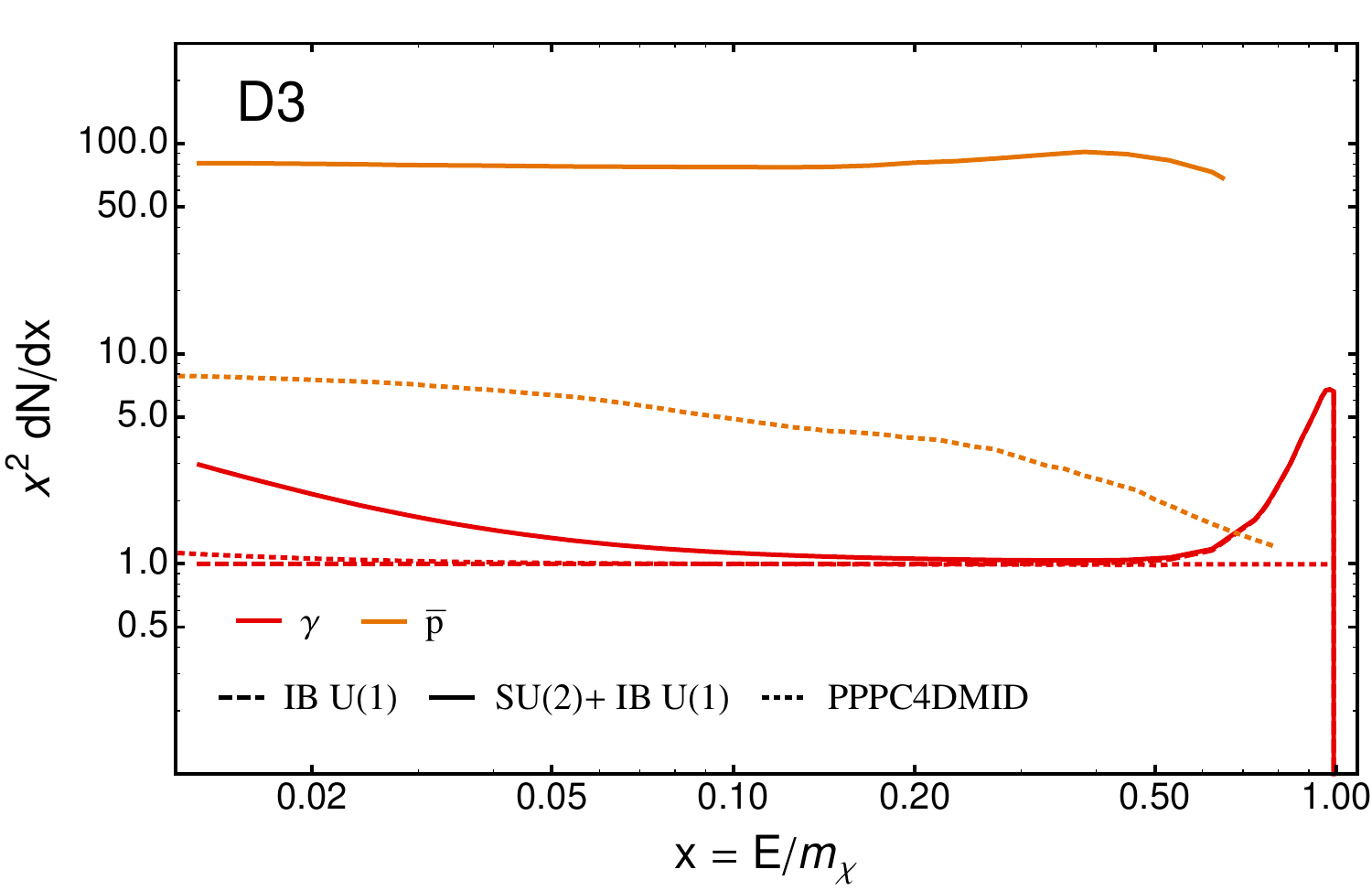}\\[5pt]
		\includegraphics[width=0.45\columnwidth]{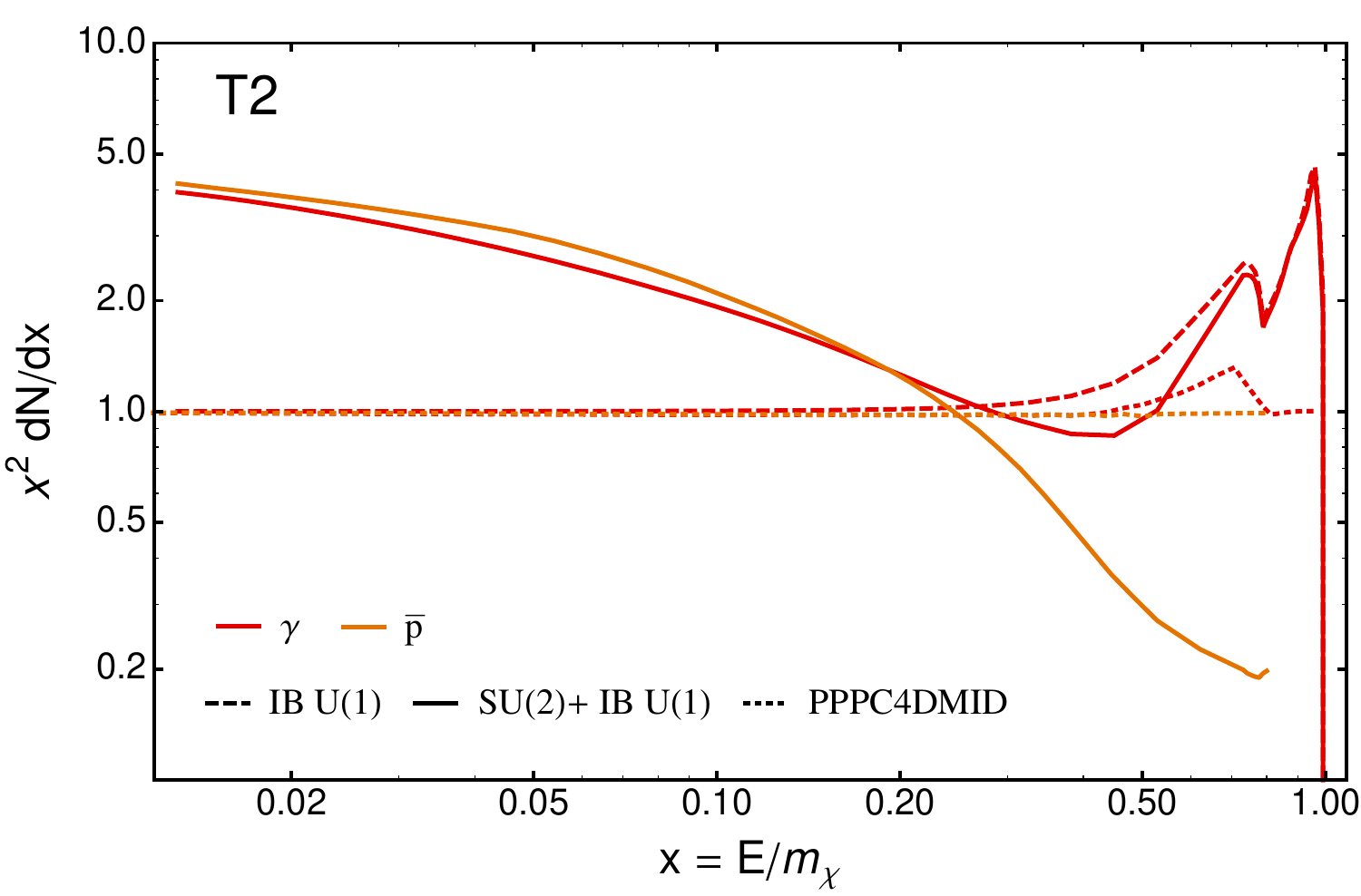} 
		\hspace*{0.4cm}
		\includegraphics[width=0.45\columnwidth]{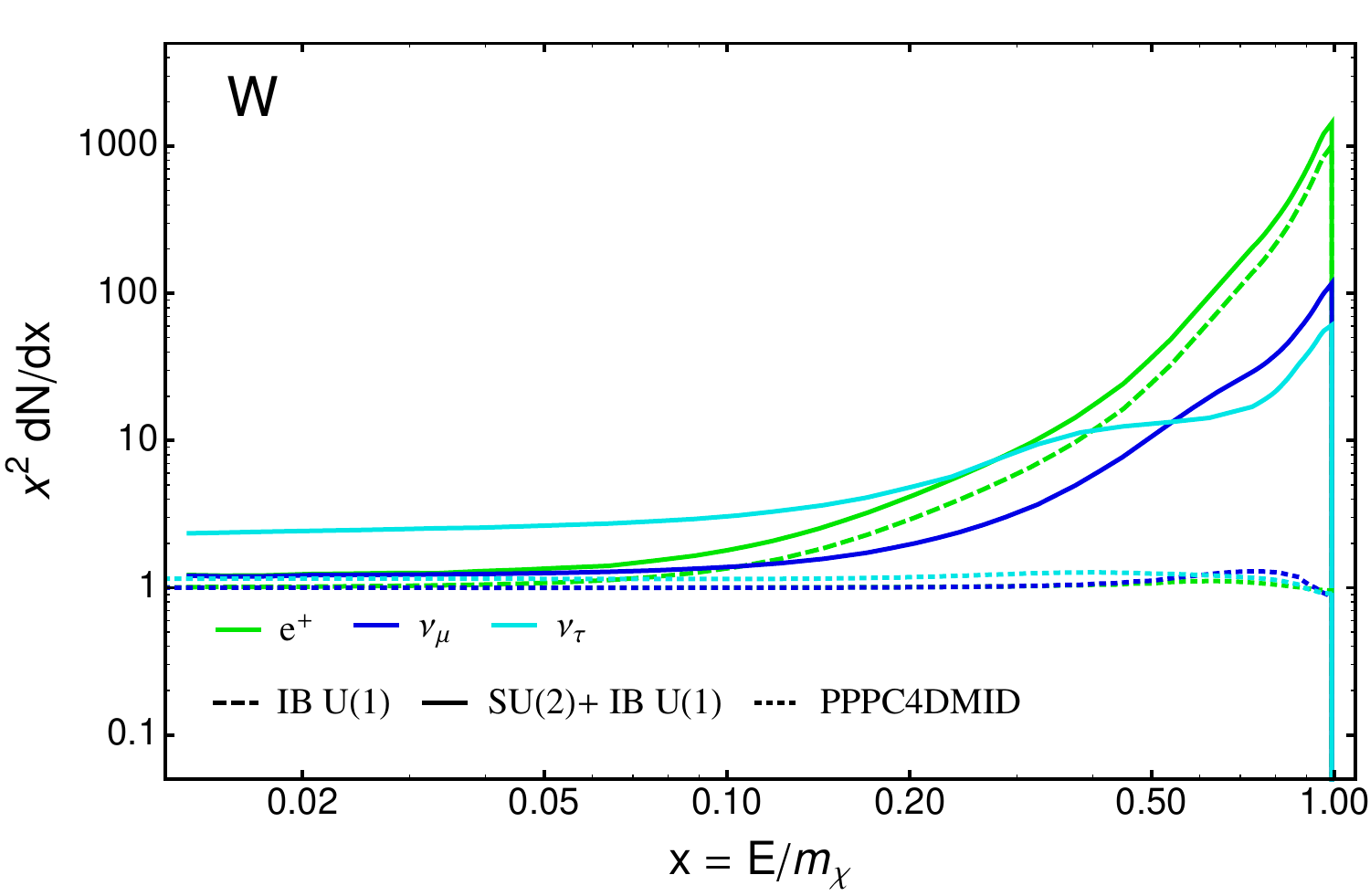} \\[5pt]
		\includegraphics[width=0.45\columnwidth]{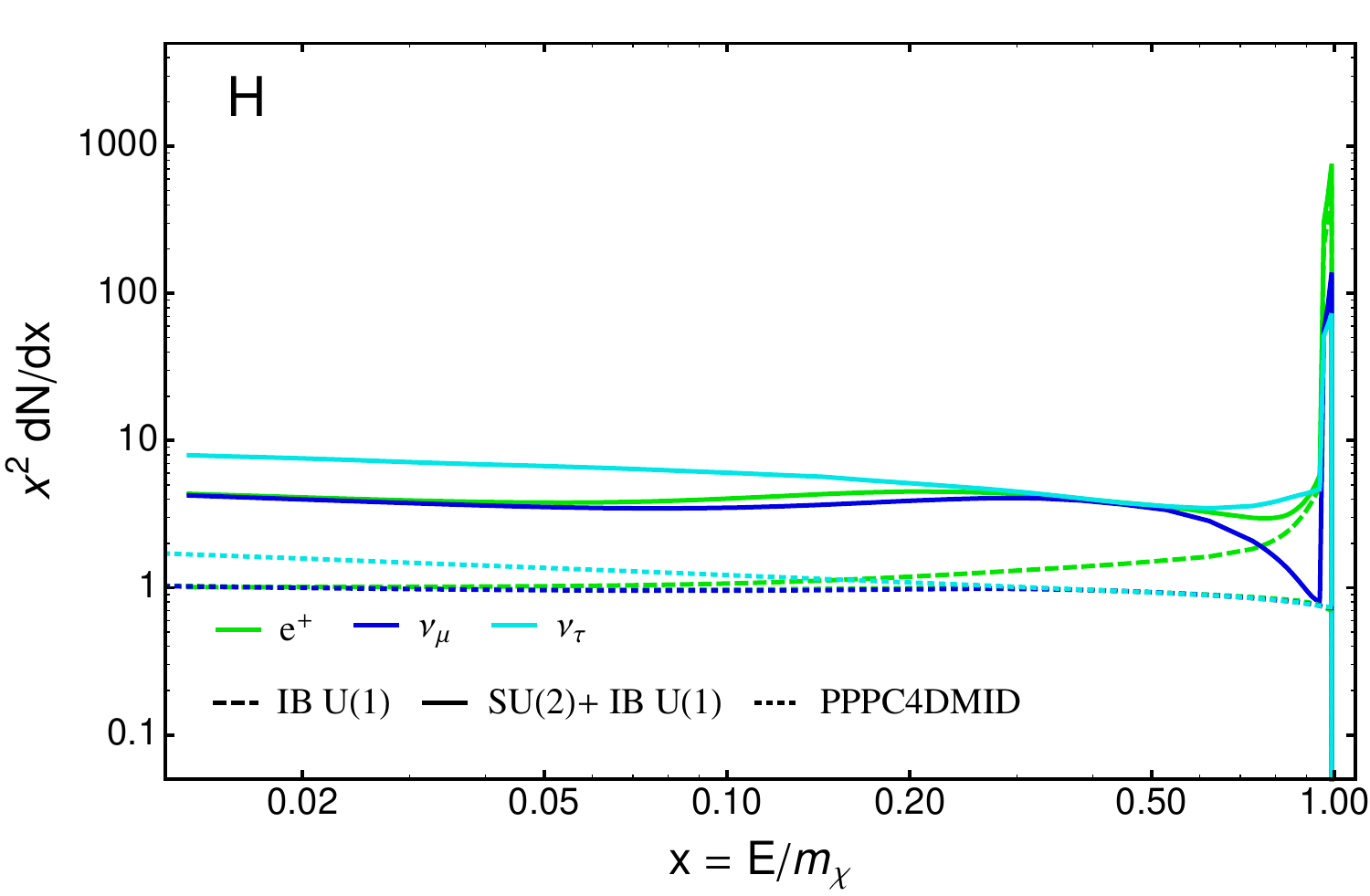} 
		\hspace*{0.4cm}
		\includegraphics[width=0.45\columnwidth]{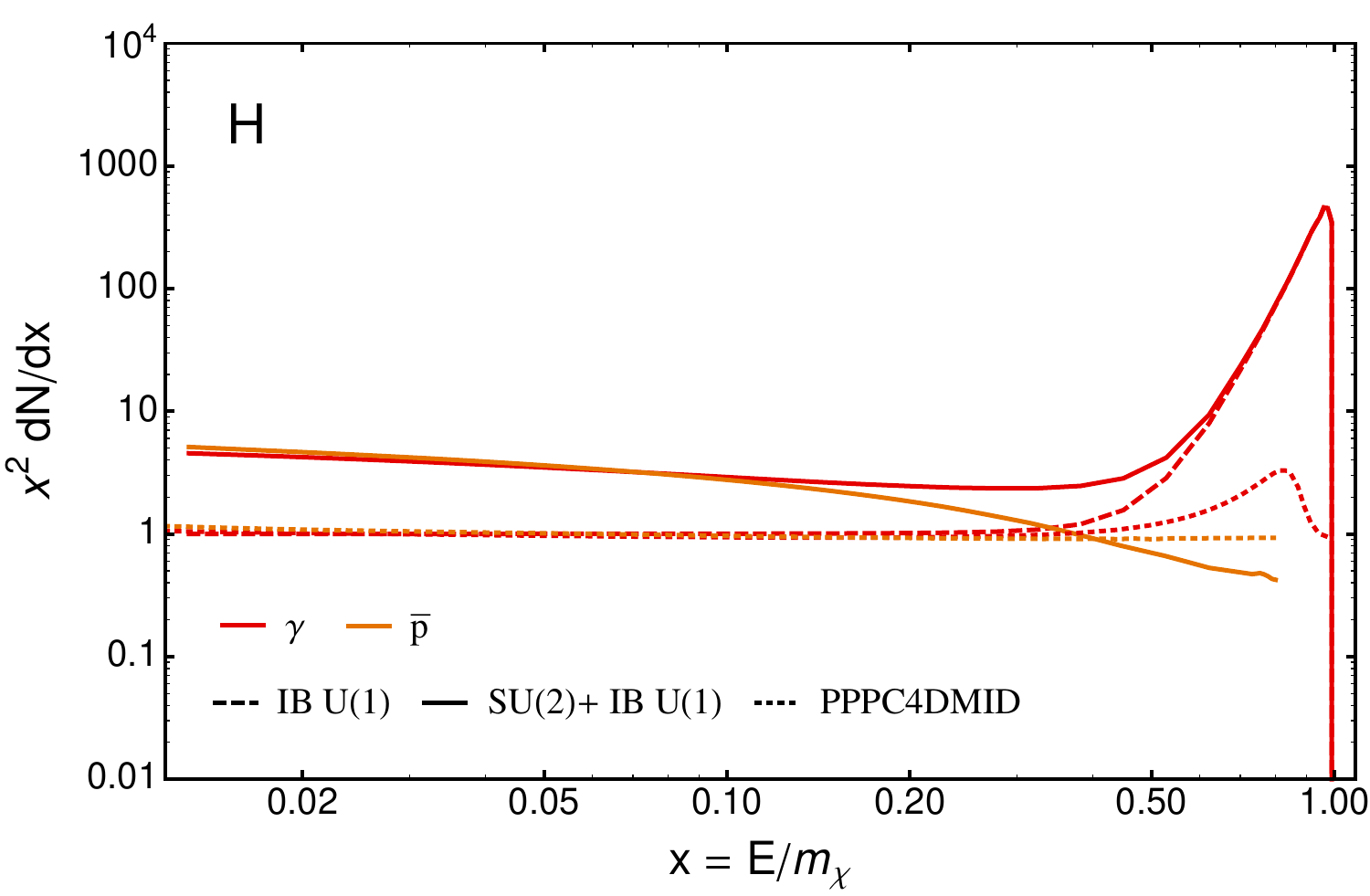} \\
	\caption{Ratio of 3-body to 2-body yields, for some of the benchmark models from 
	Tab.~\ref{tab:benchmark}. Solid lines show the full ratios, while dashed lines indicate the 
	contribution from photon IB
	alone. Dotted lines show the result from the method implemented in the `Poor Particle Physicist 
	Cookbook for Dark Matter Indirect Detection' (PPPC4DMID) \cite{Cirelli:2010xx}; note that the 
	relatively small resulting corrections are not restricted to these benchmark models but generic for  
	neutralino annihilation. We cut the $\bar{p}$ spectrum at $x \sim0.8$ in order to avoid numerical
	artefacts from very small, and hence poorly sampled, 2-body yields provided by \ds.
	See text for more details.}
	\label{fig:PPPC}
\end{figure}

In Fig.~\ref{fig:PPPC}, finally, we show for a subset of our benchmark models the {\it ratios} of  
3-body to 2-body yields, illustrating some of the features 
discussed above on a model-by-model basis from a slightly different angle. We note in particular the strong 
enhancement of high-energy lepton spectra for model H,  which is explained -- similar to
the situation for model D2 -- by a sharp drop in the 2-body yield  from $W^\pm H^\mp$ and
$ZH$ due to the maximal lepton energy from $W/Z$ decays that is kinematically possible.

A widely used phenomenological approach to take into account electroweak corrections to DM 
annihilation spectra, often referred to as `model-independent' in the literature,  is based on splitting 
functions inspired by a parton picture~\cite{Ciafaloni:2010ti,Cirelli:2010xx}. These effectively
result from assuming point-like interactions being 
responsible for the 2-body annihilation channels, such that 3-body final states are dominated 
by gauge (or Higgs) bosons that are soft or collinear with the  final state particle they are 
radiated from. In Fig.~\ref{fig:PPPC}, we therefore also indicate the ratios that result from 
this approach.\footnote{
Since only the SM Higgs boson is implemented in \cite{Cirelli:2010xx}, we have approximated the 
missing yields by replacing the heavy MSSM Higgs bosons with Goldstone bosons, 
i.e.~longitudinal gauge boson components. 
{\it For the sake of Fig.~\ref{fig:PPPC}} we thus implemented, e.g., $hA$ final states 
as 50\% $hh$ and 50\% $Z_LZ_L$ final states.
}
As one can see, the resulting changes in the yield differ at times drastically. 
In fact, for the pMSSM models studied here, we find much more generally that 
{\it the full annihilation spectra from final states containing fermions deviate substantially 
from the model-independent 
approximation whenever electroweak corrections induce 
even $\mathcal{O}(10\%)$ changes to the 2-body rates.} 
Also in the case of TeV DM models, the difference between 2-body and  
corrected yields is  much larger than what one would expect when adopting the 
PPPC \cite{Cirelli:2010xx} implementation.

For neutralino annihilation, this is quite straight-forward to understand, and for this reason
we also expect these conclusions to hold in even more generic MSSM models. In 
the `model-independent'
prescription, in particular, none of the enhancement mechanisms described in detail in 
Section~\ref{sec:enhancement} is captured. Final states containing leptons are thus necessarily 
still suppressed by $m_\ell^2/m_\chi^2$, while quark and gluon final states are hardly 
affected by electroweak corrections due to soft or collinear radiation~\cite{Ciafaloni:2010ti} 
given the strong dynamics leading to 
essentially immediate fragmentation. Furthermore, MSSM specific final states (heavy 
Higgs bosons) are not included, and even for the final states that are included the energy 
distribution of the final state particles can differ substantially when taking into account the full 
matrix element or just assuming a point-like interaction, respectively. The latter is for example
visible in the antiproton spectrum of model D3: as the 2-body annihilation is dominated by $\tau$ lepton final 
states, the main contribution must come from the final state boson; in the PPPC case the softer
antiproton spectrum can then be traced back to the fact that the emitted boson has a much softer spectrum.

\section{Conclusions}
\label{sec:conc}

In this article, we have studied in detail the annihilation of Majorana DM particles into a
pair of fermions and an electroweak gauge or Higgs boson in the final state. We have
revisited the arguments why the annihilation to fermionic 2-body final states is helicity suppressed,
and pointed out that this can be traced back to a combination of two fundamentally distinct 
effects, dubbed Yukawa- and isospin suppression, which can independently be lifted if the final state boson carries isospin. Furthermore, 
we have consistently generalized a standard way of avoiding `double-counting' of contributions 
from two- and 3-body final states, which consists in subtracting the latter in the 
narrow-width approximation, to differential cross sections and the yield of stable 
particles relevant for indirect DM searches. The latter constitutes one of our main results,
which we believe will prove useful also in other contexts.

As a concrete application, we have performed the first full analytical 
calculation of all differential cross sections for the internal bremsstrahlung of electroweak 
gauge bosons, as well as any MSSM Higgs boson, for fermion final states from 
neutralino annihilation. We have performed a detailed analysis of these results in light of
our general discussion, recovering specific examples pointed out previously in the literature, 
and extending them, but also pointing out qualitatively new processes. In order to 
estimate the size of the corrections reported here, we have performed dedicated scans over 
the parameter space of various MSSM realizations.

We find that both  lifting of the Yukawa and lifting of the isospin suppression can significantly
increase the total neutralino annihilation rate. Even more importantly, however, 
the resulting spectra of positrons, neutrinos, antiprotons and gamma rays can differ
substantially from those obtained for 2-body final states -- even when including the 
`model-independent' electroweak corrections implemented in PPPC \cite{Cirelli:2010xx}.
We stress that this is a generic result which is not restricted to specific cases but holds
whenever electroweak corrections to fermionic channels are (at least) comparable to the 
2-body results, e.g.~for TeV DM models. Given that
the supersymmetric neutralino still is {\it the} prototype WIMP DM candidate, our results 
thus underline the importance of performing full computations consistent with the model that 
is being studied. In other words, such radiative corrections are intrinsically highly 
model-dependent, and even the often adopted `model-independent' approach of 
Ref.~\cite{Cirelli:2010xx}
can be argued to simply rest on one rather specific model realization (in the sense that the underlying 
assumptions essentially describe a point like interaction, which is a good approximation 
under roughly the same conditions under which an effective operator analysis is valid
 to leading order in perturbation theory.\footnote{Note that while some of the effects of helicity 
 suppression lifting could in principle be described via additional higher-dimensional effective operators 
 in specific regions of parameter space \cite{DeSimone:2013gj},
it turns out that within the MSSM the part of parameter space in which this is possible typically does 
not coincide with the one where the corrections are
most relevant.})

To conclude, we have shown that the way electroweak corrections to DM annihilation are 
commonly estimated can lead to rather misleading results for a given DM model. The consistently
computed spectra of stable particles from DM annihilation can be much larger or offer 
striking spectral features, either of which may significantly help to indirectly detect DM
in forthcoming experiments. We stress that this holds for {\it all} yields relevant for indirect
DM detection, i.e.~both gamma rays, charged cosmic rays ($\bar p$ and $e^+$) and neutrinos.
The routines needed to compute all relevant rates and particle yields for neutralino annihilation 
in the MSSM will be included in the shortly upcoming public release 6.0 of the \ds\ package 
\cite{ds6}.

\vfill
\acknowledgments
We thank Maria Eugenia Cabrera Catalan, Marco Cirelli, Feng Luo, Are Raklev, 
Roberto Ruiz de Austri Bazan and Filippo Sala for useful discussions.

\newpage
\appendix
\section{Neutralino annihilation amplitudes}\label{app:HelicityAmplitudes}

In this Appendix, we review our analytical approach of calculating the matrix elements 
by means of an expansion in helicity amplitudes (\ref{app:ExpIntro}). 
For illustration of our full results, we then consider a number of  phenomenologically interesting 
limiting cases concerning the composition of the lightest neutralino (\ref{app:AmplitudeExpansion}).

\subsection{Expansion of Amplitudes in the Helicity Basis}
\label{app:ExpIntro}
For the analytical calculation of amplitudes we closely follow the procedure of 
Ref.~\cite{Bringmann:2013oja}, presented in detail in 
chapter 4 and corresponding appendices of Ref.~\cite{Calore:2013sqa}.
We thus modify the  generic MSSM 
model file shipped with the {\sf FeynArts} mathematica package \cite{Hahn:2000kx} such as to agree with the 
conventions adopted in \ds\ \cite{Edsjo:1997hp,ds6}. We then use {\sf FeynArts} to generate all possible 
Feynman diagrams for neutralino annihilation into 3-body final states containing a fermion,
an anti-fermion and a Boson. In the next step, given that we want to restrict ourselves to the $v\to0$ limit, we 
project out the singlet state ($J^P=0^-$) of the annihilating neutralino pair with total momentum $p$ by 
replacing the two external Majorana spinors in the amplitude with 
$P_{^1S_0}\equiv\frac{\gamma^5}{\sqrt{2}}(m_\chi-\slashed{p}/2)$. Finally, we  expand the amplitude 
for each diagram in terms of helicity amplitudes, applying a method used originally 
for neutralino annihilation to 2-body final states~\cite{1997PhRvD..56.1879E} and extended to 
3-body final states in~\cite{Bringmann:2013oja, Calore:2013sqa}.

Let us review those final steps in a bit more detail. By applying the $P_{^1S_0}$ projector, 
in particular, we can reduce any of the matrix elements considered here to the generic form
\begin{equation}
\label{eq:matrixpro}
\mathcal{M}  \propto  \bar{u}_{r}(k_{1}) \Gamma_{rs} v_{s}(k_{2}) \epsilon_{\mu}^{*}(k_{3}) \,,
\end{equation}
where only the final state spinors appear and  $\Gamma$ is a  
4$\times$4 matrix. We use {\sf Feyncalc}~\cite{Hahn:2009bf} to decompose $\Gamma$ into the 
standard basis of matrices where the corresponding Dirac bilinears are real and have 
definite transformation properties under the Lorentz group (i.e.~scalar, vector, tensor, 
pseudo-vector and pseudo-scalar, respectively). In order to assign helicities, we work in the 
{\it back-to-back} frame of the outgoing fermion-antifermion pair,  which we define by 
$\mathbf{k_1}=-\mathbf{k_2}$. For states of definite helicity $h=\pm1/2$, this implies that we can 
use \cite{Jacob:1959at}
\be
  \chi_h(\hat{\mathbf{k}}_{1,2}) = \chi_{-h}(-\hat{\mathbf{k}}_{1,2}) 
\ee
for the two-component spinors that appear in the explicit Dirac representations of both $u$ and $v$.
Choosing the fermion momentum $\mathbf{k}_1$ to be aligned with the $z$-axis, we thus obtain
\be
u_+=
\left(
\begin{array}{c}
\eta^+_{k_1}\\ 0\\\eta^-_{k_1}\\ 0
\end{array}
\right),
\qquad
u_-=
\left(
\begin{array}{c}
0\\ \eta^+_{k_1}\\ 0\\ -\eta^-_{k_1}
\end{array}
\right),
\qquad
v_+=
\left(
\begin{array}{c}
\eta^-_{k_2}\\ 0\\ -\eta^+_{k_2}\\ 0
\end{array}
\right),
\qquad
v_-=
\left(
\begin{array}{c}
0\\ \eta^-_{k_2}\\ 0\\ \eta^+_{k_2}
\end{array}
\right),
\ee
where we have introduced $\eta_{k_{1,2}}^\pm\equiv\left({k^0_{1,2}\pm m_{1,2}}\right)^{1/2}$.
In this frame, we can furthermore choose the momentum of the final state boson,
$\mathbf{k}_3$, to lie in the $y-z$ plane, spanning an angle $\theta$ with the $z$-axis. For the 
case of a massive vector boson, the 3 possible polarization states $\epsilon_\lambda$ of definite 
helicity are thus given by
\be
\epsilon_\pm=\frac1{\sqrt{2}}
\left(0, \mp1, -i\cos\theta, i\sin\theta \right),\qquad
\epsilon_0=
\frac{1}{m_3}\left( \left|\mathbf{k}_3\right|, 0, k^0_3 \sin\theta, k^0_3 \cos\theta \right) \, .
\ee

\begin{table}[t!]
\begin{center}
\footnotesize
{
\begin{tabular}{|c|c|c|c|c|} 
    \hline  
$(\bar u\Gamma v)_h$ & $h=(0, 0)$ & $h=(1, -1)$ &$h=(1, 0)$ &$h=(1, 1)$\\
\hline 
$\bar{u}v $ & 0 & 0 & $p_{+}$& 0
\\
$\bar{u} \gamma^{\mu}  v  $  & 0 &  $E_{+} \, e_{+}^{\mu}$ & $ p_{-} \, e_{u}^{\mu} - \, E_{-} \, e_{0}^{\mu}$ & $ E_{+} \, e_{-}^{\mu}$
\\
\multirow{2}{*}{$\bar{u}  \sigma^{\mu \nu}  v $}  
& \multirow{2}{*}{$ i p_{+}( e_{-}^{\mu} e_{+}^{\nu} - e_{+}^{\mu} e_{-}^{\nu})$}
& $i p_{-} (e_{+}^{\mu} e_{0}^{\nu}  -  e_{0}^{\mu} e_{+}^{\nu} )$
& \multirow{2}{*}{$ i E_{+}( e_{0}^{\mu} e_{u}^{\nu} - e_{u}^{\mu} e_{0}^{\nu})$}
& $ i p_{-} (e_{-}^{\mu} e_{0}^{\nu}-e_{0}^{\mu} e_{-}^{\nu})$
\\
&&$+ iE_{-}( e_{u}^{\mu} e_{+}^{\nu}  -  e_{+}^{\mu} e_{u}^{\nu}) $
&&$+ iE_{-} (e_{u}^{\mu} e_{-}^{\nu}- e_{-}^{\mu} e_{u}^{\nu}) $
\\
$\bar{u}  \gamma^{\mu} \gamma^{5} v $  & $p_{-} \, e_{0}^{\mu} - E_{-} \, e_{u}^{\mu}$ &  $-p_{+} \, e_{+}^{\mu}$ & 0 & $p_{+} \, e_{-}^{\mu}$
\\
$\bar{u} \gamma^{5}  v  $  &  $ -E_{+}$ & 0 & 0 & 0 
\\ 
\hline 
\end{tabular}
}
\caption{Decomposition of basis Dirac bilinears into helicity eigenstates of the
two final state fermions. For a definition of the quantities appearing here, see Eqs.~(\ref{eq:Epdef}--\ref{eq:helbasis}).}
\label{tab:helicitydec}
\end{center}
\end{table}

Singlet and triplet spin states of the two-fermion system can now be constructed from the 
individual helicity states as in Eqs.~(\ref{eq:helstates1}, \ref{eq:helstates3}), i.e.~we can
decompose each bilinear as
\bea
(\bar{u}\Gamma v)_{(0,0)}  &=&\left(\bar{u}_+\Gamma v_+ - \bar{u}_-\Gamma v_- \right)/\sqrt{2}\,,\\
(\bar{u}\Gamma v)_{(1,-1)}  &=&\bar{u}_-\Gamma v_+\,, \\
(\bar{u}\Gamma v)_{(1,0)}  &=&\left(\bar{u}_+\Gamma v_+ + \bar{u}_-\Gamma v_- \right)/\sqrt{2}\,,\\
(\bar{u}\Gamma v)_{(1,+1)}  &=&\bar{u}_+\Gamma v_-\,.
\eea
In Table~\ref{tab:helicitydec}, we show the result of this decomposition for each of the 16 basis Dirac bilinears, where
for ease of notation we have introduced the following kinematic quantities (note the different normalization convention 
with respect to \cite{Calore:2013sqa}):
\bea
\label{eq:Epdef}
E_\pm&\equiv&  \sqrt{2}\left(\eta^+_{k_1}\eta^+_{k_2}  \pm \eta^-_{k_1}\eta^-_{k_2}  \right) \,,\\
p_\pm&\equiv&   \sqrt{2}\left(\eta^+_{k_1}\eta^-_{k_2}  \pm \eta^-_{k_1}\eta^+_{k_2}  \right) \,.
\eea
Four-vectors and tensors, furthermore, are more conveniently expressed in the {\it helicity basis},
\be
\label{eq:helbasis}
  \left\{ \tilde e_{(\mu)}\right\}
  =\left\{\tilde e_u, \tilde e_+,\tilde e_-,\tilde e_0\right\}
  \equiv 
  \left\{
  \left(1,0,0,0\right),
  \left(0,\frac{-1}{\sqrt{2}},\frac{-i}{\sqrt{2}},0\right), 
  \left(0,\frac{1}{\sqrt{2}},\frac{-i}{\sqrt{2}},0\right), 
  \left(0,0,0,1\right)
  \right\}\,,
\ee
which is an orthonormal basis choice just like the canonical coordinate basis $\left\{e_{(\mu)}\right\}$
with $\left(e_{(\mu)}\right)^\nu=\delta^\nu_\mu$. This implies, e.g.,~that the components of a four-vector $V$ 
for these basis choices are related by $V^\mu= A^\mu_{~\nu} \tilde V^\nu$, where 
$A_{\mu\nu}\equiv e_{(\mu)}\cdot \tilde e_{(\nu)}$. 

With the above decompositions of fermion and vector boson polarizations even the full analytic expressions 
for the amplitudes turn out to be relatively easily manageable. We evaluate the amplitude for every 
helicity configuration, for each diagram separately, and simplify it further by explicitly 
contracting the remaining polarisation vectors (when applicable), basis vectors and four-momenta.
We then sum over all diagrams to obtain the total helicity amplitudes $\mathcal{M}^{(h, \lambda)}$,
where $h$ is the helicity of the fermion-antifermion 
pair in the back-to-back system and  $\lambda$ the polarisation state of the emitted vector boson. 
Finally, we obtain the total amplitude squared  by averaging over initial ($r, s$) spins and summing over 
final ($r', s'$)  degrees of freedom:
\bea
\overline{\left|\mathcal{M}\right|}^2 \equiv \frac{1}{4} \sum_{r, s, r', s', \lambda}\Big| \, \mathcal{M}_{\chi \chi \rightarrow  \bar{F} f X}\Big|^2  \equiv
\frac{1}{4} \sum_{h, \lambda} \,\Big|\mathcal{M}^{(h, \lambda)}_{\chi \chi \rightarrow  \bar{F} f X}\Big|^2\,,
\label{eq:helsum}
\eea
where $X$ is either a vector boson ($W/Z$) or a scalar Higgs (in which case no polarisation is present).
For convenience, we then transform 
back to the CMS frame. This allows us to compute the total 3-body cross section by integrating 
over the phase space,
\begin{equation}
\label{svddiff}
\frac{d (\sigma v_{2\to3})}{dE_1 dE_2} = \frac{1}{16 \, m_{\chi}^{2}} \, \frac{1}{(2 \pi)^{3} } {\left|\mathcal{M}\right|}^2\,,
\end{equation}
where $E_1$ and $E_2$ are the CMS energies of any two final state particles.
For details concerning the numerical implementation, we refer to Appendix \ref{app:num}.

\subsection{Results for expanded amplitudes}
\label{app:AmplitudeExpansion}

Let us consider our analytical results in the limit of heavy neutralino masses, which amounts to 
taking the ratio of the electroweak VEV and the neutralino mass, $\delta_v\equiv v_{EW}/m_\chi$,
and expanding the full results for the amplitude around $\delta_v=0$. Besides allowing for compact
analytic expressions, this limit is particularly useful for deriving the scaling behaviour of the 
amplitudes not only with $\delta_v\equiv v_{EW}/m_\chi$, but also with the Yukawa couplings $y_f$ and 
the gauge coupling $g$.
We express the results of this procedure in terms of the ratio of the 3-body to the corresponding 
2-body amplitudes, the latter of which are suppressed by a factor $m_f= y_f v_{EW}$.  
If the 3-body process lifts Yukawa suppression, the amplitude ratio will thus scale as $\propto 1/y_f$, 
and if it lifts isospin suppression it scales as $\propto 1/\delta_v$. 

We therefore introduce the dimensionless ratio of the helicity amplitudes to the 
spin-summed/-averaged matrix element for the corresponding 2-body process,
\be
  R^{\lambda,h} \equiv m_\chi \frac{{\cal M}^{\lambda, h}}{\sqrt{|\overline{{\cal M}|^2}_{2\to 2}}}\,.
\ee
For the total amplitude squared, the individual helicity contributions have to be summed over,
c.f.~Eq.~(\ref{eq:helsum}). For the sake of our discussion here, we organize this sum in a slightly 
different way and split it into contributions from final state fermions with equal or opposite chirality 
(rather than singlet and triplet states), as well as longitudinal and transverse polarizations (like before).
We note that, since the external fermions are massless in the limit 
that we are considering here, helicity coincides with chirality and hence becomes Lorentz invariant. 
In the back-to-back frame introduced in Appendix \ref{app:ExpIntro}, the helicity 
components $h=+-$ and $h=-+$ then
correspond to chiralities $f_R\bar f_R$ or $f_L\bar f_L$, respectively,
and coincide with the spin-triplet components $(1,+1)$ and $(1,-1)$ discussed there.
The helicity combinations $h=++$ and $h=--$ of the fermion pair, on the other hand, 
correspond to $f_R \bar f_L$ and $f_L\bar f_R$ states, respectively. The sum over these latter
two contributions is then equivalent to the sum over the singlet and the remaining triplet states, 
$(0,0)$ and $(1,0)$ in the notation from above.
 Altogether, this yields the decomposition
\be
 \sum_{\lambda, h} \left|R^{\lambda,h}\right|^2= \left(R^L_{LR/RL}\right)^2+ \left(R^L_{LL/RR}\right)^2 
 +\left(R^T_{LR/RL}\right)^2+ \left(R^T_{LL/RR}\right)^2\,,
\ee
where
\bea
  \left(R^L_{LR/RL}\right)^2 &\equiv & \sum_{\lambda=0\atop h=++,--} \left|R^{\lambda,h}\right|^2 \,,\qquad
  \left(R^L_{LL/RR}\right)^2 \ \equiv \ \sum_{\lambda=0\atop h=+-,-+} \left|R^{\lambda,h}\right|^2 \,,\nn\\
  \left(R^T_{LR/RL}\right)^2 &\equiv & \sum_{\lambda=\pm1\atop h=++,--} \left|R^{\lambda,h}\right|^2 \,,\qquad
  \left(R^T_{LL/RR}\right)^2 \ \equiv \ \sum_{\lambda=\pm1\atop h=+-,-+} \left|R^{\lambda,h}\right|^2 \,.
\label{helsumratios}
\eea
For Higgs final states, the summation over polarizations is absent, and we define
the corresponding ratios $R_{LR/RL}$ and $R_{LL/RR}$ analogously, 
corresponding to $h\bar f_R f_L+h\bar f_L f_R$ and $h\bar f_L f_L+h \bar f_R f_R$ final states,
respectively.

We then start from our full result for the helicity amplitudes, using the 
explicit representations of the generic couplings and mass matrices  that appear there,
and expand them up to $\mathcal{O}(\delta_v^2)$. Note that the limit $\delta_v\to 0$ implies in 
particular that we expand in the fermion mass $m_f\propto y_f v_{EW}$ and in gauge 
boson masses $m_{W/Z} \propto g v_{EW}$. 
In order to simplify the resulting analytic expressions, we set all sfermion masses equal to the neutralino mass, 
noting that larger sfermion masses would suppress $t/u$-channel rates relatively strongly because 
$\sigma v_{2\to 3}^{t-{\rm channel}} \propto m_{\tilde f}^{-8}(m_{\tilde f}^{-4})$ for Bino- (Higgsino/Wino-)like 
neutralinos, as discussed 
previously~\cite{Garny:2011cj, Bell:2011if, Ciafaloni:2011gv, Garny:2011ii,Ciafaloni:2012gs}. 
Furthermore, we use the notation $B \bar F f$ where for neutral bosons ($B=Z,h$) the final state 
fermion types are identical, $\bar F=\bar f$, while for 
charged bosons we adopt in the following the convention that $f$ denotes the up-type fermion 
(e.g.~the top quark in $\chi\chi\to W\bar b t$).
In these cases, we keep for simplicity only the dependence on the Yukawa coupling of the up-type fermion, and
set the other one to zero.

We furthermore consider six distinct scenarios describing the dominant neutralino composition, which 
result from different assumptions about the involved mass hierarchies and which are of particular 
phenomenological interest:
\begin{itemize}
\item Higgsino DM, with small Bino admixture ($\mu\ll M_2,\,M_1\rightarrow\infty$)
\item Higgsino DM, with small Wino admixture ($\mu\ll M_1,\,M_2\rightarrow\infty$)
\item Bino DM, with small Wino admixture ($M_1\ll M_2,\,\mu\rightarrow\infty$)
\item Bino DM, with small Higgsino  admixture ($M_1\ll \mu,\,M_2\rightarrow\infty$)
\item Wino DM, with small Bino admixture ($M_2\ll M_1,\,\mu\rightarrow\infty$)
\item Wino DM, with small Higgsino admixture ($M_2\ll \mu,\,M_1\rightarrow\infty$)
\end{itemize}
From the Bino-, Wino- and Higgsino mass parameters $M_1$, $M_2$ and $\mu$,
we define the dimensionless mass suppression factors $\delta_{M_1}\equiv m_\chi/M_1$, $\delta_{M_2}\equiv m_\chi/M_2$ 
and $\delta_\mu\equiv m_\chi/\mu$.
For all six scenarios listed above, we expand the amplitude ratios to leading order in these mass 
suppression factors. 
Effectively, the neutralino mixing between either Bino or Wino and Higgsino 
then becomes a perturbative `mass insertion' $\propto gv_{EW}$ represented by the 
respective off-diagonal entries in the mass matrix of Eq.~(\ref{eq:massmatrix}).
Furthermore, for definiteness, we also expand to linear order in $\delta_A\equiv m_\chi/M_A$, 
i.e.~we work in the decoupling limit where the heavy Higgs states are much heavier 
than the neutralino or SM-like Higgs boson. We note that it is straightforward to 
generalize these results, and our numerical results anyway include all MSSM Higgs 
bosons and are valid for arbitrary mass hierarchies.

To lowest order in the expansion parameters defined above, isospin and fermion chirality 
have to be conserved in all interaction vertices (assuming that the 
mass splitting between $M_1$, $M_2$ and $|\mu|$ is large compared to $gv_{EW}$). 
One of the implications, as it turns out, is that the gauge-invariant subset of $t$-channel diagrams 
discussed in Section \ref{sec:neutralino-ann} can be further split into two separate gauge-invariant 
sets. The first, which we will denote by $(I)$, does 
not contain any neutralino mixing insertion $\propto gv_{EW}$, and would hence contribute even in the 
limit of a pure neutralino state. The set of diagrams that contain at least one such insertion 
(denoted by $(II)$), on the other hand, require a mixing in the neutralino sector (just like is the 
case for all $s$-channel diagrams).

In Tables \ref{tab:EnhancementsW} -- \ref{tab:EnhancementsHiggs}, we show the results of this expansion for
the helicity-summed ratios $R$ that we have introduced in Eq.~(\ref{helsumratios}), where the different tables 
correspond to the three types of final states ($W\bar Ff$, $Z\bar ff$, and $h\bar f f$, respectively). Each 
table contains the results for all six mass hierarchy scenarios specified above, broken down to contributions 
from each set of gauge-invariant diagrams.\footnote{
We checked explicitly (up to $\mathcal{O}(\delta_v^2)$) that the Ward identities, Eqs.~(\ref{eq:WardIdentityZ}, 
\ref{eq:WardIdentityW}) are satisfied for each set separately. Note that we use Breit-Wigner widths in the 
amplitudes, and while they break gauge invariance at 
$\mathcal{O}(\delta_v)$, they do not contribute to the amplitudes as $\delta_v\rightarrow0$
} 
For the sake of the presentation, we keep only contributions that 
lift the isospin- or Yukawa suppression of the corresponding 2-body process (or both).
In particular, as apparent from Table \ref{tab:liftingMSSM}, the ratio $R^T_{LR/RL}$ cannot lift any of 
these suppressions, and is therefore not included in Tables \ref{tab:EnhancementsW} -- 
\ref{tab:EnhancementsHiggs}. Furthermore, for each of the gauge-invariant sets of diagrams, 
we include only those amplitude ratios that actually 
do lift at least one of the suppression factors. For the remaining entries, a `$0$' indicates 3-body 
amplitudes that vanish to the order we consider, while for entries containing a `$-$' both 2- and 
3-body amplitudes vanish\footnote{
While in this case the ratio would be formally ill-defined, we only identified one example 
where the 2-body amplitude vanishes while the 3-body amplitude does not, 
marked with a $(*)$.
We note that the relevant process,  $\chi\chi\to WW^* \to W\bar F f$ for a Wino-like neutralino,
is phenomenologically not important because 
for $m_\chi>M_W$ it is largely captured by annihilation into $WW$, while Wino-like neutralinos
with $m_\chi<M_W$ are practically excluded.
}.
The 2-body amplitudes, finally, are for convenience summarized in Table \ref{tab:2to2}.

\begin{sidewaystable}
\begin{center}
\[ \chi\chi\to W^{-}\bar F f \]
\begin{tabular}{|c|c|cc|ccc|c|}
\hline
Mass Hierarchy & \multicolumn{3}{c|}{$s$-channel}                                & \multicolumn{4}{c|}{$t+u$-channel} \\ \hline
               & $H^\pm$ med. & \multicolumn{2}{c|}{$Z/h/W$ mediator}  & \multicolumn{3}{c|}{$(I)$}            & $(II)$\\
               & $W_L\bar F_L f_R$    & $W_L\bar F_L f_R$ & $W_T\bar F_L f_L$  & $W_L\bar F_L f_R$ & $W_L\bar F_L f_L$ & $W_T\bar F_L f_L$ & $W_L\bar F_L f_R$   \\ 
               & $W_L\bar F_R f_L$    & $W_L\bar F_R f_L$ &                    & $W_L\bar F_R f_L$ &                   &                   & $W_L\bar F_R f_L$   \\ \hline
& & & &&&& \\[-1ex]
$\mu\ll M_2,\,M_1\rightarrow\infty$ 
 &  $\frac{1}{\sqrt{2} \delta_v (1+s_{2\beta})}$  
 & $0$ 
 & $\frac{g\, C(x_1,x_2)}{ \sqrt{2}y_f \delta_v \delta_{M_2} c_{2\beta} (x_1+x_2-1)}$ 
 & $\frac{1}{2\sqrt{2}\delta_vx_1}$ 
 & $0$
 & $\frac{g \, C(x_1,x_2)}{2\sqrt{2}y_f\delta_v  x_1}$ 
 & $\frac{2}{\sqrt{2} \delta_v(1+t_\beta) x_1}$ \\
 & & &&&&& \\[-0.5ex] \hline
 & & &&&&& \\[-1ex]
$\mu\ll M_1,\,M_2\rightarrow\infty$ 
 & $0$  
 & $0$ 
 & $\frac{g\, C(x_1,x_2)}{ \sqrt{2}y_f \delta_v \delta_{M_1}t_w^2  c_{2\beta}(x_1+x_2-1)}$ 
 & $\frac{1}{2\sqrt{2}\delta_vx_1}$ 
 & $0$
 & $\frac{g \, C(x_1,x_2)}{2\sqrt{2}y_f\delta_v  x_1}$ 
 & $0$ \\
 & & &&&&& \\[-0.5ex] \hline
 & & &&&&& \\[-1ex]
$M_1\ll \mu,\,M_2\rightarrow\infty$ 
 & $\frac{1}{2\sqrt{2} \delta_v }$  & $\frac{1}{\sqrt{2}\delta_v}$ 
 & $0$
 & $\frac{A(x_1,x_2)}{2\sqrt{2}\delta_v}$ 
 & $\frac{g^2 c_{2\beta} K}{2\sqrt{2}y_f  x_1 x_2}$
 & $\frac{g D(x_1,x_2)K}{2\sqrt{2}y_f \delta_v  }$  
 & $\frac{B(x_1,x_2)}{2\sqrt{2} \delta_v}$\\
 & & &&&&& \\[-0.5ex] \hline
 & & &&&&& \\[-1ex]
$M_1\ll M_2,\,\mu\rightarrow\infty$ 
 & - 
 & \multicolumn{2}{c|}{-}
 & $\frac{A(x_1,x_2)}{2\sqrt{2}\delta_v}$ 
 & $\frac{g^2 c_{2\beta} K}{2\sqrt{2}y_f  x_1 x_2}$
 & $\frac{g D(x_1,x_2) K}{2\sqrt{2}y_f \delta_v  }$ 
 & - \\
 & & &&&&& \\[-0.5ex] \hline
 & & &&&&& \\[-1ex]
$M_2\ll \mu,\,M_1\rightarrow\infty$ 
 & $\frac{1}{2\sqrt{2} \delta_v }$ 
 & $\frac{1}{\sqrt{2}\delta_v}$
 & $\frac{g\sqrt{2} C(x_1,x_2)}{y_f \delta_v \delta_{\mu}^2 c_{2\beta}(x_1+x_2-1)}$ 
 & $\frac{1}{2\sqrt{2}\delta_v x_2}$ 
 & $\frac{g^2 c_{2\beta} }{4\sqrt{2}y_f  x_1 x_2}$
 & $\frac{g D(x_1,x_2)}{2\sqrt{2}y_f \delta_v   }$ 
 & $\frac{1}{2\sqrt{2}\delta_v x_2}$ \\
 & & &&&&& \\[-0.5ex] \hline
 & & &&&&& \\[-1ex]
$M_2\ll M_1,\,\mu\rightarrow\infty$
 & -  
 & - 
 & $\frac{g 2 \sqrt{2} C(x_1,x_2)}{y_f \delta_v (x_1+x_2-1)}$ ($^*$) 
 & $\frac{1}{2\sqrt{2}\delta_v x_2}$ 
 & $\frac{g^2 c_{2\beta} }{4\sqrt{2}y_f  x_1 x_2}$
 & $\frac{g D(x_1,x_2)}{2\sqrt{2}y_f \delta_v    }$
 & - \\
 & & &&&&& \\[-0.5ex]\hline
Lifting of & isospin & isospin & Yukawa and isospin & isospin & Yukawa & Yuk and iso & isospin \\\hline
\end{tabular}
\caption{\label{tab:EnhancementsW}
Amplitude ratios for 3-body processes $\chi\chi\rightarrow W_{T(L)}^{-}\bar F_X f_Y$ (with $XY=LL/RR; LR/RL$) 
to 2-body processes $\chi\chi\rightarrow\bar f f$, in the limit $\delta_v=v_{EW}/m_\chi \rightarrow 0$ and for different 
SUSY mass hierarchies. Compared to the definitions introduced  in Eq.~(\ref{helsumratios}), we show here 
$R^{T}_{LL/RR}/\sqrt{x_1+x_2-1}$,  $R^{L}_{LR/RL}/\sqrt{x_1+x_2-1}$ and  $R^{L}_{LL/RR}/\sqrt{(1-x_1)(1-x_2)}$,
respectively. We express everything in dimensionless quantities, where $x_{1(2)}=E_{1(2)}/m_\chi$ are the CMS
energies of the final state fermions and $x_\textrm{X}\equiv m_\chi/M_X$ mass suppression factors. We also define 
$y_f\equiv m_f/v_{EW}$, using thus a convention without factors of $\tan\beta$, and neglect 
$y_{\bar F}\equiv m_F/v_{EW}$ for $F=d,s,b,\ell$. Furthermore, 
$t_w\equiv\tan(\theta_W)$, $s_w\equiv\sin(\theta_W)$, $c_w\equiv\cos(\theta_w)$, $s_\beta\equiv\sin(\beta)$, 
$c_\beta\equiv\cos(\beta)$, $s_{2\beta}\equiv\sin(2\beta)$ and
$c_{2\beta}\equiv\cos(2\beta)$. $q_f$ is the fermion charge, and $Y_{f_L}\equiv q_f-t_{3f}$, $Y_{f_R}\equiv q_f$ 
are the hypercharges (with $t_{3f}=+1/2$ for $f=u,c,t,\nu$). For more compact expressions, we also introduced 
$K \equiv Y_{f_L}^2/(Y_{f_L}^2+Y_{f_R}^2)$, $A\equiv K/x_2+(1-K)/x_1$, $B\equiv 2Y_{f_R}/x_1-2Y_{f_L}/x_2$, 
$C \equiv \sqrt{(x_1-1)^2+(x_2-1)^2}/(x_1+x_2-2)$, $D \equiv \sqrt{(x_1-1)^2+(x_2-1)^2}/(x_1x_2)$.
A `$0$' indicates 3-body amplitudes that vanish to the order we consider, while for entries containing a `$-$' 
both 2- and 3-body amplitudes vanish.
For the field marked by ($^*$), finally, we normalize by the amplitude for the $t$-channel process as the 
corresponding 2-body amplitude vanishes in this case. 
}
\end{center}
\end{sidewaystable}

\begin{sidewaystable}[ph!]
\begin{center}
\[ \chi\chi\to Z\bar f f \]
\begin{tabular}{|c|c|cc|cc|c|}
\hline
Mass Hierarchy & \multicolumn{3}{c|}{$s$-channel}                       & \multicolumn{3}{c|}{$t+u$-channel} \\ \hline
               & $H$ mediator & \multicolumn{2}{c|}{$Z/h$ mediator} & \multicolumn{2}{c|}{$(I)$}    & $(II)$\\
               & $Z_L\bar f_L f_R$  & $Z_L\bar f_L f_R$ & $Z_T\bar f_L f_L$       & $Z_L\bar f_L f_R$ & $Z_T\bar f_L f_L$ & $Z_L\bar f_L f_R$   \\ 
               & $Z_L\bar f_R f_L$  & $Z_L\bar f_R f_L$ & $Z_T\bar f_R f_R$       & $Z_L\bar f_R f_L$ & $Z_T\bar f_R f_R$ & $Z_L\bar f_R f_L$   \\ \hline
 &  & & &&& \\
$\mu\ll M_2,\,M_1\rightarrow\infty$ 
 & $\frac{1-s_{2\beta}}{2\sqrt{2} \delta_v (1+s_{2\beta})}$ 
 & $\frac{1}{\sqrt{2} \delta_v (x_1+x_2-1)}$ & $\frac{g\, C(x_1,x_2) J}{ 2 y_f \delta_v \delta_{M_2} c_w  c_{2\beta}(x_1+x_2-1) }$
 & $\frac{(x_1+x_2)}{2\sqrt{2}\delta_v x_1x_2}$ & $\frac{g F(x_1,x_2)}{4 \delta_v y_f c_w  }$
 & $\frac{(t_\beta-1)E(x_1,x_2)}{2\delta_v (t_\beta+1)}$ \\
 & & &&&& \\\hline
  &   & &&&& \\
$\mu\ll M_1,\,M_2\rightarrow\infty$
 & $\frac{1-s_{2\beta}}{2\sqrt{2} \delta_v (1+s_{2\beta})}$ 
 & $\frac{1}{\sqrt{2} \delta_v (x_1+x_2-1)}$ & $\frac{g\, C(x_1,x_2) J}{ 2 y_f \delta_v \delta_{M_1} s_w t_w  c_{2\beta}(x_1+x_2-1)}$
 & $\frac{(x_1+x_2)}{2\sqrt{2}\delta_v x_1x_2}$  & $\frac{g F(x_1,x_2)}{4 \delta_vy_f c_w   }$ 
 & $\frac{(t_\beta-1)E(x_1,x_2)}{2\delta_v t_w^2(t_\beta+1)}$ \\
 &  &  &&&&  \\\hline
  &   &  &&&& \\
$M_1\ll \mu,\,M_2\rightarrow\infty$
 & $\frac{1}{2\sqrt{2} \delta_v }$  
 & $\frac{1}{\sqrt{2} \delta_v }$ & $0$ 
 & $\frac{G(x_1,x_2)}{4\delta_v}$ &  $\frac{g D(x_1,x_2) H}{4 y_f\delta_v}$
 & $\frac{\tilde E(x_1,x_2)}{4\delta_v}$ \\
 & &  &&&& \\\hline
 &    & &&&& \\
$M_1\ll M_2,\,\mu\rightarrow\infty$
 & -
 & \multicolumn{2}{c|}{-}
 & $\frac{G(x_1,x_2)}{4\delta_v}$ & $\frac{g D(x_1,x_2) H}{4 y_f\delta_v}$
 & - \\
  & &  &&&& \\\hline
  &    & &&&& \\
$M_2\ll \mu,\,M_1\rightarrow\infty$ 
  &  $\frac{1}{2\sqrt{2} \delta_v }$ 
  & $\frac{1}{\sqrt{2} \delta_v }$ & $0$  
  & $\frac{E(x_1,x_2)}{4\delta_v}$ & $\frac{g D(x_1,x_2) Z_{f_L}}{2 y_f \delta_v  c_w}$
  & $\frac{E(x_1,x_2)}{4\delta_v}$ \\
  & & &&&&  \\ \hline
  &    & &&&& \\
$M_2\ll M_1,\,\mu\rightarrow\infty$
 & - 
 & \multicolumn{2}{c|}{-}
 & $\frac{E(x_1,x_2)}{4\delta_v}$ & $\frac{g D(x_1,x_2) Z_{f_L}}{2 y_f \delta_v  c_w}$ 
 & - \\
  &  & &&&& \\\hline
Lifting of & isospin & isospin & Yukawa and isospin & isospin & Yukawa and isospin & isospin \\\hline
\end{tabular}
\caption{\label{tab:EnhancementsZ}
As Table \ref{tab:EnhancementsW}, but for $Z\bar f f$ final state. Note that in the third column the longitudinal
 contribution (left) originates from $Z/h$-exchange FSR and VIB,
while the transverse (right) originates from ISR contributions to $Z$ exchange (this implies that these two classes of diagrams are gauge-independent separately, 
which we checked explicitly). Notations are defined as in table \ref{tab:EnhancementsW}, and $y_f=y_{\bar f}\equiv m_f/v_{EW}$ for $f=q,\nu,\ell$.
In addition, we define the $Z$-couplings $Z_{f_L}=q_fs_w^2-t_{3f}$, $Z_{f_R}=q_fs_w^2$,
and $E\equiv \sqrt{x_1^2+x_2^2}/(x_1x_2)$, 
$\tilde E\equiv \sqrt{4(Y_{f_L}+Y_{f_R})^2(x_1-x_2)^2+(x_1+x_2)^2}/(\sqrt{2}x_1x_2)$, 
$F\equiv C(x_1,x_2)\times \sqrt{(2Z_{f_L}(x_1+x_2)-4Z_{f_R})^2+(2Z_{f_R}(x_1+x_2)-4Z_{f_L})^2}/(x_1x_2)$,
$G\equiv \sqrt{\frac{(Y_{f_L}+Y_{f_R})^2}{(Y_{f_L}^2+Y_{f_R}^2)}(x_1-x_2)^2+(x_1+x_2)^2}/(\sqrt{2}x_1x_2)$,
$H\equiv 4\sqrt{Y_{f_R}^4Z_{f_R}^2+Y_{f_L}^4Z_{f_L}^2}/(Y_{f_L}^2+Y_{f_R}^2)$,
$J\equiv 2\sqrt{(Z_{f_L}^2+Z_{f_R}^2)}/c_w^2$.
}
\end{center}
\end{sidewaystable}

\begin{sidewaystable}[ph!]
\begin{center}
\[ \chi\chi\to h\bar f f \]
\begin{tabular}{|c|c|cc|cc|cc|}
\hline
Mass Hierarchy & \multicolumn{3}{c|}{$s$-channel}      & \multicolumn{4}{c|}{$t+u$-channel} \\ \hline
               & $A$ mediator & \multicolumn{2}{c|}{$Z$ mediator} & \multicolumn{2}{c|}{$(I)$} & \multicolumn{2}{c|}{$(II)$} \\
               & $h\bar f_L f_R$ & $h\bar f_L f_R$ & $h\bar f_L f_L$       & $h\bar f_L f_R$ & $h\bar f_L f_L$ & $h\bar f_L f_R$  & $h\bar f_L f_L$  \\ 
               & $h\bar f_R f_L$ & $h\bar f_R f_L$ & $h\bar f_R f_R$       & $h\bar f_R f_L$ & $h\bar f_R f_R$ & $h\bar f_R f_L$  & $h\bar f_R f_R$ \\ \hline
 &  & && & &&\\
$\mu\ll M_2,\,M_1\rightarrow\infty$ 
 & $\frac{1}{2\sqrt{2}\delta_v}$
 & $\frac{1}{\sqrt{2} \delta_v (x_1+x_2-1)}$ & $\frac{g^2 J}{ 2 y_f  (x_1+x_2-2)(x_1+x_2-1)}$
 & $\frac{x_1+x_2}{2\sqrt{2} \delta_v x_1x_2}$ & $\frac{g^2 c_{2\beta} J}{8 y_f  x_1x_2}$
 & $\frac{E(x_1,x_2)}{2\delta_v}$ 
 & $0$\\
 & & & &&&&\\\hline
  &   & &&&&&\\
$\mu\ll M_1,\,M_2\rightarrow\infty$
 & $\frac{1}{2\sqrt{2}\delta_v}$
 & $\frac{1}{\sqrt{2} \delta_v (x_1+x_2-1)}$ & $\frac{g^2(x_1+x_2) J}{ 4 y_f  (x_1+x_2-2)(x_1+x_2-1)}$
 & $\frac{x_1+x_2}{2\sqrt{2} \delta_v x_1x_2}$ & $\frac{g^2 c_{2\beta} J}{8 y_f  x_1x_2}$
 & $\frac{\tilde E(x_1,x_2)}{2 \delta_v}$ 
 & $0$ \\
 &  &  & &&&&\\\hline
  &   &  && &&&\\
$M_1\ll \mu,\,M_2\rightarrow\infty$
 & $\frac{1}{2\sqrt{2}\delta_v}$
 & $\frac{1}{\sqrt{2} \delta_v}$ & $0$
 & $\frac{G(x_1,x_2)}{2\delta_v }$ & $\frac{g^2c_{2\beta}H}{4y_f c_w^2 x_1x_2}$ 
 & $\frac{\tilde E(x_1,x_2)}{2 \delta_v}$  
 & $\frac{g^2\delta_\mu t_w^2 t_\beta (x_1+x_2)L}{4 \sqrt{2} y_f x_1 x_2}$ \\
 & &  & &&&&\\\hline
 &    & & &&&&\\
$M_1\ll M_2,\,\mu\rightarrow\infty$
 & -
 & \multicolumn{2}{c|}{-}
 & $\frac{G(x_1,x_2)}{2\delta_v }$ & $\frac{g^2c_{2\beta}H}{4y_f c_w^2 x_1x_2 }$
 & \multicolumn{2}{c|}{-} \\
  & &  & &&&&\\\hline
  &    & & &&&&\\
$M_2\ll \mu,\,M_1\rightarrow\infty$ 
  & $\frac{1}{2\sqrt{2}\delta_v}$
  & $\frac{1}{\sqrt{2} \delta_v}$ & $0$ 
  & $\frac{E(x_1,x_2)}{4\delta_v}$ & $\frac{g^2c_{2\beta}Z_{f_L}}{4y_f c_w^2 x_1x_2}$
  &  $\frac{E(x_1,x_2)}{4\delta_v}$ 
  & $\frac{g^2\delta_\mu t_\beta (x_1+x_2)}{8 y_f x_1 x_2}$ \\
  & & & &&&& \\ \hline
  &    & &&&&& \\
$M_2\ll M_1,\,\mu\rightarrow\infty$
 & -
 & \multicolumn{2}{c|}{-}
  & $\frac{E(x_1,x_2)}{4\delta_v}$ & $\frac{g^2c_{2\beta}Z_{f_L}}{4y_f c_w^2 x_1x_2}$
 & \multicolumn{2}{c|}{-} \\
  &  & & &&&& \\\hline
Lifting of & isospin & isospin & Yukawa & isospin & Yukawa & isospin & Yukawa \\\hline
\end{tabular}
\caption{\label{tab:EnhancementsHiggs}
As Table \ref{tab:EnhancementsZ}, but for $h\bar f f$ final state for $f=u,c,t,\nu$.
For $f=d,s,b,\ell$ one needs to replace $t_\beta\to 1/t_\beta$ in the last column. 
For the $s$-channel diagrams with electroweak-scale mediator, only those with a $Z$ boson contribute to the
Yukawa or isospin lifting.
The amplitude ratios are normalized as $R_{LL/RR}/\sqrt{(x_1-1)(x_2-1)}$ and $R_{LR/RL}/\sqrt{x_1+x_2-1}$, respectively.
Notations are defined as in the previous tables, in addition $L^2\equiv 4(Y_{f_L}+Y_{f_R})^2+64(Y_{f_L}^2+Y_{f_R}^2)^2$.
}
\end{center}
\end{sidewaystable}

\begin{sidewaystable}
\begin{center}
\[ \chi\chi\to \bar f f \]
\begin{tabular}{|c|c|c|c|c|}
\hline
Mass Hierarchy & \multicolumn{2}{c|}{$s$-channel} & \multicolumn{2}{c|}{$t+u$-channel} \\ \hline
               & $A$ mediator & $Z$ mediator    & $(I)$ & $(II)$\\ \hline
& & & & \\
$\mu\ll M_2,\,M_1\rightarrow\infty$ 
 &  $4g^2y_f\delta_A^2\delta_{M_2}\delta_v(1+s_{2\beta})/t_\beta$ 
 & $\frac12 g^2y_f\delta_{M_2}\delta_vc_{2\beta}$ 
 & $y_f^3 \delta_v/s_\beta^2 $ 
 & $\frac12 y_f g^2 \delta_v \delta_{M_2} ( 1+t_\beta )/t_\beta$ \\
  & & & &\\\hline
 & & & &\\
$\mu\ll M_1,\,M_2\rightarrow\infty$ 
 & $4g^2y_f\delta_A^2\delta_{M_1}\delta_vt_w^2(1+s_{2\beta})/t_\beta$ 
 & $\frac12 g^2y_ft_w^2\delta_{M_1}\delta_vc_{2\beta}$ 
 & $y_f^3 \delta_v/s_\beta^2 $ 
 & $\frac12 y_f g^2 \delta_v \delta_{M_1} t_w^2 ( 1+t_\beta )/t_\beta $ \\
  & & & &\\\hline
  & & & &\\
$M_1\ll \mu,\,M_2\rightarrow\infty$ 
 & $8g^2y_ft_w^2\delta_A^2\delta_\mu \delta_v /t_\beta$ 
 & $g^2y_ft_w^2\delta_\mu^2\delta_v  c_{2\beta}$ 
 & $2 g^2 y_f t_w^2 \delta_v (Y_{f_L}^2+Y_{f_R}^2)$ 
 & $g^2 y_f t_w^2 \delta_v \delta_\mu /t_\beta$ \\
 & & & &\\\hline
 & & & &\\
$M_1\ll M_2,\,\mu\rightarrow\infty$ 
 & - 
 & - 
 & $2 g^2 y_f t_w^2 \delta_v (Y_{f_L}^2+Y_{f_R}^2)$ 
 & - \\
  & & & &\\\hline
 & & & &\\
$M_2\ll \mu,\,M_1\rightarrow\infty$ 
 & $8g^2y_f \delta_A^2\delta_\mu \delta_v/ t_\beta$ 
 & $g^2y_f \delta_\mu^2\delta_v  c_{2\beta}$ 
 & $ g^2 y_f \delta_v $ 
 & $ 2 g^2 y_f \delta_v \delta_\mu / t_\beta$ \\
  & & & &\\\hline
 & & & &\\
$M_2\ll M_1,\,\mu\rightarrow\infty$ 
 & - 
 & - 
 & $ g^2 y_f \delta_v $ 
 & - \\
 & & & &\\\hline
\end{tabular}
\caption{\label{tab:2to2}
Value of the two-to-two matrix elements $\sqrt{|\overline{{\cal M}|^2}_{2\to 2}}$, relative to which
the previous results are given, for $f=u,c,t,\nu$.
For $f=d,s,b,\ell$ one needs to replace $t_\beta\to 1/t_\beta$ and $s_\beta\leftrightarrow c_\beta$.
The matrix element squared has to be multiplied by a color factor $N_c=3$ for quarks. 
Note that all amplitudes involve (at least) one power of $y_f$ and $\delta_v$, that corresponds to Yukawa- and isospin suppression, respectively.}
\end{center}
\end{sidewaystable}

\clearpage

In order to assess the parametric enhancement of 3-body over 2-body processes, it is sufficient to 
consider the amplitude ratios just presented, and we will continue with a more detailed discussion 
of the various lifting mechanisms at the level of individual diagrams in the following 
subsection \ref{app:diagrams}. 
Before doing so, let us briefly remark that the corresponding {\it cross section} ratio for 
$\chi\chi\to B \bar F f$,  normalized to the one for $\chi\chi\to \bar f f$, is obtained by 
\be
\label{svratio}
  \frac{1}{\sigma v_{2\to 2}}d(\sigma v)_{2\to 3} 
= \frac{1}{4\pi^2} \frac{1}{\sqrt{1-m_f^2/m_\chi^2}} \sum_{\lambda, h} \left|R^{\lambda,h}\right|^2 \, dx_1dx_2
\ee
where $x_i=E_i/m_\chi$ are the dimension-less fermion energies of the 3-body final state. Using the results 
from Tables \ref{tab:EnhancementsW} -- \ref{tab:EnhancementsHiggs}, one can thus obtain the contribution to
this ratio from each of the gauge-invariant subsets of diagrams separately.
In the limit of massless final state particles, the integration ranges are $0<x_1<1$ and $1-x_1<x_2<1$,
implying that some of these integrations become logarithmically divergent. This is an expected artefact
of the expansion in $\delta_v$ and, in practice, the corresponding infrared divergent contributions are cut off 
by the non-zero mass of the vector boson. Throughout this work, we assume that the resulting logarithmic 
enhancement ${\cal O}(\frac{\alpha}{\pi}\ln^2(E_B/\sqrt{s}))$ can be treated perturbatively down to the 
infrared cutoff $E_B\sim m_B\sim g v_{EW}$. This imposes an upper limit
on the neutralino mass of roughly $m_\chi \ll {\cal O}(g v_{EW} e^{\pi /g}) \sim {\cal O}(10)$\,TeV. 
If one is interested in higher masses, it would be interesting
to apply the resummation methods discussed e.g.~in 
Refs.~\cite{Bauer:2014ula,Ovanesyan:2014fwa,Baumgart:2014vma}.
On the other hand, we stress that the logarithmic sensitivity to $\ln^2(g \delta_v)$ does not spoil the power counting arguments related to
lifting of isospin suppression factors, since the latter is described by powers $\delta_v^n$ of $\delta_v$. In our numerical results,
we fully take into account the masses of all annihilation products.

\subsection{Suppression lifting from individual diagrams}
\label{app:diagrams}

It is rather illustrative to reflect the results of the  previous subsection  at the level of individual 
diagrams. In Table \ref{fig:lifting}, displayed for clarity already in the main text (see Section 
\ref{sec:lifting}), we therefore organize all relevant amplitudes in a large table, with the four rows 
corresponding to the four gauge-invariant subsets. For each type of diagram, and assuming a Bino-  or 
Wino-like neutralino, we furthermore explicitly indicate the scaling with the  gauge coupling $g$, the 
Yukawa coupling $y_f$, and the vev $v_{EW}$ (we comment on the Higgsino-like case below). 
Let us start our discussion with the first 
column, which contains the diagrams contributing to the 2-body process
$\chi\chi\to\bar ff$. As expected, all these amplitudes scale as $\propto g^2y_fv_{EW}$, but the origin
differs:
\begin{description}
\item[$t$-channel $I$:] The factor $y_f v_{EW}$ enters either via the chirality flip of one of the final-state 
fermions, or via a $L/R$ mixing insertion of the sfermion (for brevity, we show only one representative 
diagram in Table \ref{fig:lifting} for each of these cases).
\item[$t$-channel $II$:] The factor $v_{EW}$ enters via the gaugino/Higgsino mixing insertion on one of 
the initial lines, and the Yukawa suppression enters via the Higgsino-sfermion-fermion coupling.
\item[$s$-channel EW:] The $s$-channel with electroweak-scale mediator corresponds to the $Z$-exchange 
diagram mentioned earlier. In the $s$-wave limit, and from the perspective of the unbroken theory, this 
diagram is represented by the exchange of the pseudoscalar Goldstone boson $G^0$. The factor $v_{EW}$ 
arises from the gaugino/Higgsino mixing, and the Yukawa coupling from the Yukawa interaction $G^0\bar f f$.
\item[$s$-channel $M_A$:] This case is similar to the previous one, except that the mediator is replaced 
by the (physical) heavy pseudoscalar Higgs $A$.
\end{description}

Let us now turn our discussion to the remaining columns of Table \ref{fig:lifting}, which contain 
all relevant 
3-body processes.  Here, the second column shows representative Feynman diagrams that lead to a 
lifting of both isospin and Yukawa suppression, while the third and fourth column show diagrams that 
lift only one of them, respectively:

\begin{description}
\item[Lifting of Yukawa and isospin suppression:] Both suppression factors can be lifted only for two of 
the gauge-invariant sets of diagrams ($t$-$I$ and $s$-EW). In the former case, a transverse $Z_T$ or 
$W_T$ is emitted from either fermion line in the final state, from the sfermion line, or from the initial lines 
(this last case cannot occur in the Bino-like case). We remark that FSR can only lift the helicity suppression 
if the virtual fermion is strongly off-shell, i.e.~not for soft and collinear photons 
(which are sometimes {\it defined} as FSR, see footnote \ref{FSRdef}).
In the $s$-channel case, the diagrams can be thought of as an annihilation 
$\chi\chi\to WW^*$,  with subsequent decay of $W^*$ 
(see Section \ref{sec_threshold} for a discussion of such off-shell internal states). It is 
impossible to lift both suppression factors for the other two classes: for $t$-$II$, this would require a
gaugino-Higgsino-$W/Z$ vertex, which is absent for $v_{EW}\to 0$. The same applies for $s$-$M_A$, 
noting in addition that the $A\bar f f$ coupling requires also the presence of a Yukawa coupling.

\item[Lifting of {\it only} isospin suppression:] The isospin suppression can be lifted for all four subsets, by 
replacing the insertion of $v_{EW}$ within the 2-body amplitude by the emission of a Higgs boson or a 
Goldstone boson, respectively.
Note that for the set $t$-$I$ this amounts to replacing the fermion mass insertion by a 
fermion-fermion-Higgs/Goldstone coupling (or replacing the sfermion $L/R$ mixing insertion by a 
sfermion-sfermion-Higgs/Goldstone coupling, respectively). For all other sets one replaces the 
gaugino-Higgsino mixing insertion in the initial line by a gaugino-Higgsino-Higgs/Goldstone vertex. 
For the $s$-channel, the diagrams can also be thought of as an annihilation into a pair
of scalars, 
 with subsequent decay of one of them.
This mechanism of suppression lifting is very general, and appears for all gauge invariant
subsets of diagrams as well as for all final states (involving $W/Z$ or a Higgs boson).
We expect it to be relevant especially for heavy neutralino masses.

\item[Lifting of {\it only} Yukawa suppression:] This case is in some sense the most difficult to realize. 
The reason is that it requires a Higgs (or Goldstone) boson in the final state, and therefore only 
diagrams where the final-state boson does not couple directly to the final-state fermions can potentially 
contribute in the limit $y_f\to 0$. We identified three such processes, shown in the last column in 
Table \ref{fig:lifting}: For $t$-$I$, the Higgs (or charged Goldstone boson; note that there is no 
sfermion-sfermion-$G^0$ vertex for $y_f\to 0$) can be emitted from the sfermion line in the $t$-channel, 
i.e.~via VIB. The corresponding vertex is derived from a four-scalar sfermion-sfermion-Higgs-Higgs 
interaction, involving the full Higgs doublets. This coupling leads to the required vertices at 
${\cal O}(v_{EW})$, and scales with $g^2$ for $y_f\to 0$ within the MSSM (see Refs.~\cite{Garny:2011cj} 
and \cite{Luo:2013bua} for a discussion within a toy model for the Goldstone- and Higgs-emission, 
respectively). In addition, for $t$-$II$, the Higgs can be emitted via ISR (second row, 
last column of Table \ref{fig:lifting}). While this contribution lifts Yukawa suppression, it is 
suppressed compared to the 2-body process for a large mass hierarchy between gaugino and Higgsino
mass parameters; we nevertheless kept this contribution, because the former effect can easily 
compensate for the latter. Finally, for the $s$-EW case, the Higgs can be emitted from the $s$-channel 
mediator via a Goldstone-Higgs-$Z$ coupling (third row, last column in Table \ref{fig:lifting}). Note 
that this mechanism is distinct from the one discussed in \cite{Kumar:2016mrq}, and that the toy-model 
discussed there cannot be realized within the MSSM. To the best of our knowledge, both the $t$-channel 
ISR and the $s$-channel Higgstrahlung processes that we identified within the MSSM have not been 
discussed before.
\end{description}

One can understand the diagrams that lift Yukawa or isospin suppression as
shown in Table \ref{fig:lifting} based on basic properties of the unbroken MSSM Lagrangian,
as well as the symmetry requirement $J_{CP}=0_-$ of the $s$-wave initial state.
For example, mixing insertions $\propto gv_{EW}$ of the neutralino line can turn a Bino into a Higgsino, 
but not into a Wino. In addition, the Higgsino coupling to fermion/sfermion pairs is proportional to the 
Yukawa coupling, while the corresponding coupling for Bino- and Wino-like neutralinos involves a gauge 
coupling and is therefore generally much less suppressed (except for the top quark). One slightly more 
involved example is the diagram in the last column of the first row. For final states involving a longitudinal 
$W_L$, 
the corresponding sfermion vertex derives from the interaction term 
$\propto g^2(\tilde f_L^\dag H)(H^\dag \tilde f_L)$
 present for sfermion fields that transform as doublet under $SU(2)_L$. 
After inserting the decomposition 
$H=(G^+,(v_{EW}+h+iG^0/\sqrt{2}))$ of the SM-like Higgs doublet one easily verifies that at 
linear order in $v_{EW}$ one obtains a sfermion coupling to $G^\pm$ and $h$, but not to $G^0$, which 
explains why no longitudinal $Z_L$ boson can be produced in this case. The Higgs final state also 
receives a further contribution from the interaction term $\propto H^\dag H \tilde f^\dag\tilde f$,
which exists for all (left {\it and} right) sfermion fields. Furthermore, for the $s$-channel processes of the 
type $\chi\chi\to h B^*\to h\bar f f$ that give a non-zero contribution in the $s$-wave limit, 
the mediator $B$ is a pseudoscalar or transverse vector (i.e. $G^0, A^0$, $Z_T$), while 
for $\chi\chi\to G^0 B^*\to G^0\bar f f$, $B$ is a scalar (i.e. $h,H^0$). This is consistent with the 
odd $CP$ parity of the initial state. 

Note that the above arguments  are only valid when expanding around the unbroken
theory, and representing longitudinal degrees of freedom by Goldstone bosons. In fact, within the broken 
theory, analogous arguments would be hampered by large
cancellations that occur among individual diagrams, and that make the power counting less transparent. 
Nevertheless, we carefully cross checked that all these arguments
can indeed be reproduced when using the full matrix elements within the broken theory, 
and expanding the sum of all diagrams within a gauge invariant subset for heavy neutralino mass.

While the discussion above assumed a gaugino-like neutralino, the case of a Higgsino-like neutralino is 
very similar. For the third and fourth row in Table \ref{fig:lifting}, in particular, nothing changes except that 
the incoming neutralino is now a  Higgsino in the limit $v_{EW}\to 0$, and the insertion 
$\propto gv_{EW}$ denotes mixing with either a Bino or Wino (in addition, both $Z_T$ and $W_T$ ISR
is possible, while only $W_T$ ISR is possible for Wino-like neutralinos). The same applies to
the second line, after interchanging the label of $g$ and $y_f$ on the vertices involving a sfermion in all
diagrams in the first and second column (this does not affect the overall scaling of the amplitude), while 
the diagram in the last column would receive an additional $y_f^2$ suppression.
For the first row, the two neutralino-sfermion-fermion
vertices scale with $y_f$ instead of $g$ in all diagrams. Thus, this class is additionally suppressed by a 
factor $y_f^2$ compared to the other subsets. Nevertheless, for completeness, we kept this case 
because the 3-body processes can lift the additional suppression factors $y_f v_{EW}$ of the 2-body 
amplitude in the same way as for a gaugino-like neutralino.

\smallskip

In summary,  we confirmed the general symmetry arguments outlined in 
Section \ref{sec:liftingModelIndep} for the MSSM and explicitly identified
the contributions to the 3-body amplitudes that realize the suppression lifting, focussing on final states 
containing (tranverse or longitudinal) gauge bosons as well as the SM-like Higgs boson.
By expanding the full amplitudes in various limits that correspond to Bino-, Wino- or Higgsino-like 
neutralino,
respectively, we find that (almost) all of the possibilities allowed by symmetries are realized.
The cases for which we did {\it not} find a contribution within
the MSSM are marked by a `-' in  Table \ref{tab:liftingMSSM}. For processes involving $W$ bosons and 
purely right-handed fermions an additional suppression arises
that can be traced back to the chiral structure of the $SU(2)_L$ interaction. For processes involving 
$Z_L$ (represented by $G^0$) or $A$, and fermions of equal chirality, on the other hand,
lifting of Yukawa suppression would require that the amplitude does not contain Yukawa interaction 
vertices. In addition, vertices such as sfermion-sfermion-$G^0/A$ are
absent for $y_f\to 0$ (as required by $CP$-invariance), such that a $t$-channel process analogous to 
the one in the first row, last column of   Fig.\,\ref{fig:lifting} does not exist. For the $s$-channel,
the symmetries of the initial state would require a CP-even mediator if the $G^0$ or $A$ was emitted via 
ISR. Within the MSSM, only the Higgs bosons are available. However, their coupling to fermions 
necessarily
involve a Yukawa coupling, such that Yukawa suppression cannot be lifted in this specific process. 
Similarly, one can convince oneself that the $s$-channel VIB process (3rd row, 4th column of 
Fig.~\ref{fig:lifting}) as well as the remaining $t$-channel process (2nd row, 4th column) cannot occur 
when replacing $h\to G^0,A$.

\section{Numerical implementation}
\label{app:num}

For each Feynman diagram, we have implemented the full analytical expressions for the helicity 
amplitudes in \ds\ \cite{ds6}. We numerically sum over these contributions to obtain the
total amplitude for a given helicity configuration, 
$\mathcal{M}^{(h, \lambda)}_{\chi \chi \rightarrow  \bar{F} f X}$, as introduced in 
Appendix~\ref{app:ExpIntro}. Differential and partial cross sections are computed
according to Eq.~(\ref{svddiff}), by numerically integrating over the energies of the final 
state particles; for consistency checks, this can be done for any pair of energies and in any 
specified order. In order to improve convergence and accuracy of the numerical integrations, 
we use taylored integration routines that make use of the known locations of kinematic 
 resonances \cite{Calore:2013sqa}. 

For the {\it total cross sections}, we have explicitly implemented  the NWA approximations 
contained in Eqs.~(\ref{eq:subtraction_total}--\ref{eq:subtraction_total_last}). 
We have extensively checked our code, and hence also the prescription of subtracting
the NWA contribution detailed above, by comparing the total cross 
section defined in Eq.~(\ref{svtot}), on a channel-by-channel basis and for various SUSY 
models, with numerical results obtained with {\sf CalcHEP} \cite{Belyaev:2012qa}\footnote{We compared 
our implementation of 3-body cross sections based on \ds\ 5.1 with 
{\sf CalcHEP} 3.4 \cite{Belyaev:2012qa}. 
In particular, we adapted the ewsbMSSM implementation of {\sf CalcHEP} to  compute 
the spectrum from a given set of pMSSM input parameters at scale $Q=M_Z$ (except for $M_A$ which 
is the
pole mass) using SoftSusy 3.4 \cite{Allanach:2001kg}. The Susy les Houches output file written by
SoftSusy is then used as input for \ds {} via the {\it slha} interface.
In order to be able to directly compare the output it is necessary to adapt various routines in order
to match the conventions. Apart from making sure that all SM input parameters agree (we used 
$m_b=4.92$\,GeV, 
$\sin(\theta_W)=0.47162$, $\Gamma_W=2.07$\,GeV, $\Gamma_t=2.0$\,GeV), we made the following 
changes for the purpose of cross checking: 
For {\sf CalcHEP}, we switched off the running bottom mass ({\tt dMbOn=0}) and used unitary gauge (for 
the comparison on a diagram-by-diagram basis;
only the sum is gauge-independent). For \ds\ , the Yukawa couplings are by default read in from the 
blocks YU, YE and YD in the {\it slha} file. For the purpose of comparison, it is convenient to fix the 
Yukawa couplings at $y_i=m_i/v$, especially for the top. Therefore, we commented out the 
corresponding lines in {\tt dsfromslha.f}. Additionally, in {\tt su/dssuconst\_yukawa\_running.f}, we 
commented out the running Yukawas, such that the default Yukawa couplings, which are simply related 
to the (on-shell) masses, are used.
In addition, the call to {\tt dshigwid()} was commented out in {\tt dsfomslha.f} in order to avoid a rescaling 
of Higgs couplings that takes certain NLO corrections into
account. For the purpose of comparison, it is more convenient to have tree-level couplings. In addition, 
we then set the Higgs widths to a common value in both programs.
Finally, we set the first and second generation quark masses to zero and the CKM mixing matrix to unity 
in order to match the conventions of the ewsbMSSM model
implemented in {\sf CalcHEP}. We verified that the conventions agree by comparing also the 2-body 
cross sections for all channels allowed at $s$-wave, for which we find
perfect agreement after the changes described above.}.
For all models, and all annihilation channels, we find remarkable agreement. We also 
checked agreement for individual classes of diagrams ($s/tu$-channel, ISR/FSR/VIB) as 
classified in Section \ref{sec:neutralino-ann}. Let us stress that in terms of computation time the 
implementation via helicity amplitudes, together with the taylored integration routines, is less
expensive compared to the evaluation of squared matrix elements via Monte Carlo integration
as implemented in {\sf CalcHEP}. This is especially significant for the 3-body processes
to which a large number of diagrams contribute, and for which the difference in computation times
amounts to several orders of magnitude in the specific kinematic limit we are interested in here.

For the {\it yields of stable particles}, we have implemented the procedure described 
in Section \ref{sec:NWAdiff}, using unpolarized yields for decaying particles given that these are the 
only ones that are currently available in  \ds{} \cite{Gondolo:2004sc}. 
As discussed, as long as the total yields (i.e.~summed over all channels) are concerned, our 
prescription still 
captures any double counting. We note that  extending our implementation to fully 
polarized yields will be straight-forward for future work, given the results provided in 
Section \ref{sec:NWAdiff} and the helicity amplitudes reported in Appendix \ref{app:HelicityAmplitudes}.

Let us mention a few of the extensive numerical checks that we performed to test the yield 
implementation. We considered, in  particular, models for which the 3-body annihilation is dominated by 
an almost on-shell
intermediate resonance. In this case, the subtraction procedure described in Sec.\,\ref{sec:NWAdiff} is
expected to lead to a large cancellation between the full 3-body contribution and the
NWA term. We explicitly verified this cancellation for all yields of stable particles, and over the full energy 
range. The cancellation amounts to several orders of magnitude in specific cases, and therefore provides 
a robust check of the implementation. In addition, we also verified that the yields obtained from all of the 
models contained in our MSSM scan results pass
a number of checks (e.g.~yields within an expected range at $E>m_\chi/2$ and $E>m_\chi/10$).
Finally, we also considered 3-body final states that contain directly one or more stable
particles (such as e.g. $\chi\chi\to W e \nu$). In this case, we verified that the neutrino and positron
spectra match the analytical result discussed in App. \ref{app:corr} for specific models for which
this final state is dominantly produced by an intermediate $W$ resonance.

\section{Spin correlations of decaying resonances}
\label{app:corr}

In Section \ref{sec:doublecount}, we discussed how to subtract double counting due to on-shell intermediate states
(`resonances') contributing to 3-body annihilation processes. If the resonance carries a spin,
the spectrum of final state particles depends on how much the various helicity states of the
resonance contribute. In Section \ref{sec:doublecount} we argued that for annihilation of Majorana fermions
in the $s$-wave limit, CP and angular momentum conservation uniquely determine the helicity of
all possible intermediate states that can contribute to the 3-body processes considered here.
Here we present a formal derivation of this result, based on a description that would in principle allow
us to treat also more general cases. 

In full generality, several helicity states of the resonance contribute to the amplitude,
and can also interfere with each other when taking the absolute value squared. As a starting point we
consider the example $\chi\chi\to HW \to Hf\bar F$. We are interested in the contribution from the on-shell
intermediate $W$ boson. The full matrix element squared can then be
written in the form
\begin{equation}
 \overline{|{\cal M}_{res}|}^2 \equiv \sum_{s1,s2} \left|\sum_\lambda{\cal M}_{2\to 2}^\mu \epsilon_\mu^{*\lambda}\epsilon_\nu^{\lambda}({\cal M}^\nu_{s1,s2})_{1\to 2}\right|^2 \;,
\end{equation}
where we indicated explicitly the summation over the final-state spins of the fermions, and the polarization states of the internal $W$.
To extract the on-shell contribution in the narrow-width limit we assume that the momentum $q_\mu$ of the $W$ is (almost) on-shell, $q^2\simeq M_W^2$. This implies that the kinematics of the $H$ and $W$ momenta is identical to the 2-body annihilation.
The first term inside the square contains the helicity amplitude for the 2-body part,
\begin{equation}
 {\cal M}_{2\to 2}^\lambda \equiv {\cal M}_{2\to 2}^\mu \epsilon_\mu^{*\lambda}\;.
\end{equation}
For concreteness we can take the momentum of the $W$ to be along the z-axis, $q_\mu=(E_q,0,0,|q|)$,
where $E_q=\sqrt{|q|^2+M_W^2}$ and $|q|$ is determined by the neutralino, W and Higgs mass via the 2-body
kinematics (identical to $|p_3|$, see (\ref{p1p3}) below).
The decay $W\to f\bar F$ gives (fermion momenta $p_1$ and $p_2$)
\begin{equation}
 ({\cal M}^\nu_{s1,s2})_{1\to 2} = \bar u_{s_1}(p_1) (gP_L\gamma^\nu) v_{s_2}(p_2)  
\end{equation}
Inserting these into the resonant matrix element, and writing out the square gives after some renaming of indices
\begin{eqnarray}\label{Mres}
 \overline{|{\cal M}_{res}|}^2 &=& \sum_{\lambda,\lambda'} {\cal M}_{2\to 2}^{*\lambda'}{\cal M}_{2\to 2}^\lambda \epsilon_\mu^{*\lambda'}\epsilon_\nu^{\lambda} \sum_{s_1,s_2} ({\cal M}_{s_1s_2}^{*\mu})_{1\to 2} ({\cal M}_{s_1s_2}^\nu)_{1\to 2} \nn \\
&=& \sum_{\lambda,\lambda'} {\cal M}_{2\to 2}^{*\lambda'}{\cal M}_{2\to 2}^\lambda \times {\cal D}_{\lambda\lambda'}
\end{eqnarray}
where 
\begin{eqnarray}
{\cal D}_{\lambda\lambda'} &\equiv&\epsilon_\mu^{*\lambda'}\epsilon_\nu^{\lambda} \sum_{s_1,s_2} ({\cal M}_{s_1s_2}^{*\mu})_{1\to 2} ({\cal M}_{d,s_1s_2}^\nu)_{1\to 2}\\
&=& \epsilon_\mu^{*\lambda'}\epsilon_\nu^{\lambda}\, g^2\, {\rm tr}((\slashed{p_2}-m_F)\gamma_\mu P_R(\slashed{p_1}+m_f)P_L\gamma_\nu ) \\
&=& \epsilon_\mu^{*\lambda'}\epsilon_\nu^{\lambda}\, 2g^2\, (p_{2}^{\mu}p_{1}^{\nu}  -g^{\mu\nu}p_1\cdot p_2+p_{2}^{\nu}p_{1}^{\mu}+i\epsilon^{\mu\nu\kappa\rho}p_{2\kappa} p_{1\rho})
\end{eqnarray}
The decorrelation-approximation (i.e. the replacement of the matrix element by the product of two matrix elements
with independent summation over spin states, (\ref{decorr})) would be obtained by replacing
\begin{equation}\label{Ddecorr}
  {\cal D}_{\lambda\lambda'} \to {\cal D}_{\lambda\lambda'}^{\rm decorrelated} \equiv \frac13 \delta_{\lambda\lambda'} \sum_{\lambda''}{\cal D}_{\lambda''\lambda''}
\end{equation}
Instead, one can also try to use the full matrix ${\cal D}_{\lambda\lambda'}$.
To compute it, one can use the helicity basis
\begin{equation}
 \epsilon^{\pm}_\mu = (0,1,\pm i,0)/\sqrt{2},\quad  \epsilon^0_\mu=(|q|,0,0,E_q)/M_W \,,
\end{equation}
which fulfill $\epsilon_\mu^{*\lambda'}\epsilon^{\mu\lambda}=-\delta_{\lambda\lambda'}$ as well as $\sum_\lambda \epsilon_\mu^{*\lambda}\epsilon_\nu^{\lambda}=-g_{\mu\nu}+q_\mu q_\nu/M_W^2$.

A convenient frame to evaluate it is the rest frame of the $W$, obtained by boosting along the $z$ direction. In this frame $ \epsilon^0_\mu=(0,0,0,1)$. The momenta of the fermions can be parameterized
by the angle w.r.t to the z-axis (which is singled out as the polarization axis of the $W$),
\begin{equation}
 p_1=(E_{p_1}^*,0,|p_1^*| \sin\theta, |p_1^*|\cos\theta),\quad p_2=(M_W-E_{p_1}^*,0,-|p_1^*| \sin\theta, -|p_1^*|\cos\theta)
\end{equation}
where $|p_1^*|$ and $E_{p_1}^*=\sqrt{|p_1^*|^2+m_f^2}$ are the momentum and energy of $f$ in the $W$ rest frame, see (\ref{p1p3}). Inserting this and evaluating the trace yields an explicit expression for ${\cal D}_{\lambda\lambda'}={\cal D}_{\lambda\lambda'}(\theta)$
in terms of $\theta$.
Using that the polarization vectors have zero temporal component in the basis we are working in,
and that $\vec p_2=-\vec p_1$,
\begin{eqnarray}
{\cal D}_{\lambda\lambda'}(\theta)
&=& 2g^2\, (-2 (\vec\epsilon_{\lambda'}\vec p_1)^*(\vec\epsilon_{\lambda}\vec p_1)+\delta_{\lambda\lambda'}p_1\cdot p_2+iM_W ({\vec\epsilon_{\lambda'}}^{\,*} \times \vec\epsilon_\lambda ) \cdot \vec p_1 )
\end{eqnarray}
Now one can use $p_1\cdot p_2=(M_W^2-m_f^2-m_F^2)/2$ and
\begin{eqnarray}
 \vec\epsilon_{0}\vec p_1 = |p_1^*|\cos\theta,\quad  \vec\epsilon_{\pm}\vec p_1 = \pm i\frac{|p_1^*|}{\sqrt{2}}\sin\theta\\
 \vec\epsilon_+^*\times\vec\epsilon_+=+i\vec\epsilon_0,\quad
 \vec\epsilon_-^*\times\vec\epsilon_-=-i\vec\epsilon_0,  \quad
 \vec\epsilon_0^*\times\vec\epsilon_\pm=\mp i\vec\epsilon_\pm,\quad  
 \vec\epsilon_\pm^*\times\vec\epsilon_0=\mp i\vec\epsilon_\mp\quad
\end{eqnarray}
The result is  
\begin{eqnarray}\label{DlamlamWdecay}
 {\cal D}_{00} &=& 2g^2(p_1\cdot p_2 -2 |p_1^*|^2 \cos^2\theta) \\
 {\cal D}_{\pm\pm} &=& 2g^2(p_1\cdot p_2 - |p_1^*|^2 \sin^2\theta \mp M_W |p_1^*|\cos\theta) \\
 {\cal D}_{\pm\mp} &=& 2g^2|p_1^*|^2\sin^2\theta \\
 {\cal D}_{\pm 0} &=& -i\sqrt{2}g^2 |p_1^*| \sin\theta (M_W \mp 2|p_1^*|\cos\theta) = {\cal D}_{0 \pm}^* \;.
\end{eqnarray}
One can check that the average over the diagonal contributions corresponds to the usual unpolarized decay matrix element,
\begin{equation}
 \overline{|{\cal M}_{1\to 2}|^2} = \frac13 \sum_\lambda {\cal D}_{\lambda\lambda} = \frac13 g^2(6p_1\cdot p_2-4|p_1^*|^2)
 = g^2(M_W^2-m_f^2-m_F^2-\frac43 |p_1^*|^2)
\end{equation}

To obtain the diff. cross section, we use the representation of the phase space in the form
\begin{equation}
d(\sigma v_{Hf\bar F}) = d\Phi \frac{\overline{|{\cal M}|}^2}{4M_\chi^2} = \frac{1}{(2\pi)^5}\frac{1}{16(2M_\chi)^2}\overline{|{\cal M}|}^2 |p_1^*| |p_3| dm_{12} d\Omega_1^* d\Omega_3
\end{equation}
where $p_3$ is the Higgs momentum, and $m_{12}^2=q^2$ the resonant momentum. Now we can do an approximation where we
replace
\begin{equation}
 \overline{|{\cal M}|}^2 \to \overline{|{\cal M}_{res}|}^2 \frac{\pi}{M_W\Gamma_W} \delta(q^2-M_W^2)
\end{equation}
but keep the fully correlated matrix element $\overline{|{\cal M}_{res}|}^2$.
By integrating over $dm_{12}=dm_{12}^2/(2m_{12})=dq^2/(2M_W)$, and doing the trivial Higgs angle $d\Omega_3$  and $d\phi_1^*$
integrals, one obtains the diff. cross section w.r.t to the angle $\theta$ of the fermion $f$ and the polarization axis
of the W boson in the back-to-back system,
\begin{equation}
d(\sigma v_{Hf\bar F})^{NWA} = \frac{2}{(2\pi)^3}\frac{1}{16(2M_\chi)^2}\overline{|{\cal M}_{\rm res}|}^2 \frac{\pi}{M_W\Gamma_W} \frac{1}{2M_W} |p_1^*| |p_3| d\cos\theta
\end{equation}
where
\begin{eqnarray}\label{p1p3}
  |p_1^*| &=& \frac{[(M_W^2-(m_f+m_F)^2)(M_W^2-(m_f-m_F)^2)]^{1/2}}{2M_W} \\
  |p_3| &=& \frac{[((2M_\chi)^2-(m_W+m_H)^2)((2M_\chi)^2-(m_W-m_H)^2)]^{1/2}}{4M_\chi} 
\end{eqnarray}
One can rewrite this expression, using that the two-to-two and W decay rate are given by
\begin{eqnarray}
  \sigma v_{HW} &=& \frac{1}{8\pi}\frac{|p_3|}{(2M_\chi)^2} \sum_\lambda |{\cal M}_{2\to 2}^\lambda|^2\\
  \Gamma_{W\to f\bar F} &=& \frac{1}{8\pi}\frac{|p_1^*|}{M_W^2}  \overline{|{\cal M}_{1\to 2}|}^2 
\end{eqnarray}
where ${\cal M}_{2\to 2}^\lambda$ is the helicity amplitude and $\overline{|{\cal M}_{1\to 2}|}^2$ 
is the usual summed/averaged decay matrix element. Expressed in terms of the matrix introduced above,  $\overline{|{\cal M}_{1\to 2}|}^2= \frac13 \sum_\lambda{\cal D}_{\lambda\lambda}$. The dependence on the angle cancels in this sum.
Then, 
\begin{eqnarray}
\frac{d(\sigma v_{Hf\bar F})^{NWA}}{d\cos\theta} &=& \sigma v_{HW}\, \frac{\Gamma_{W\to f\bar F}}{\Gamma_W} \times \frac{\overline{|{\cal M}_{\rm res}|}^2 }{2(\sum_\lambda |{\cal M}_{2\to 2}^\lambda|^2)(\frac13 \sum_\lambda {\cal D}_{\lambda\lambda})}  \\
&=&  \sigma v_{HW}\,BR(W\to f\bar F)\times {\cal F}_{\chi\chi\to HW\to Hf\bar F}(\theta)
\end{eqnarray}
where we have defined the function ${\cal F}$ which characterizes the angular dependence.
Using also (\ref{Mres}) for ${\cal M}_{\rm res}$, one can write it as
\begin{equation}
{\cal F}_{\chi\chi\to HW\to Hf\bar F}(\theta) = \frac{\sum_{\lambda,\lambda'} {\cal M}_{2\to 2}^{*\lambda'}{\cal M}_{2\to 2}^\lambda \times {\cal D}_{\lambda\lambda'}(\theta)}{2(\sum_\lambda |{\cal M}_{2\to 2}^\lambda|^2)(\frac13 \sum_\lambda {\cal D}_{\lambda\lambda})}   \;.
\end{equation}
%
If one would replace the matrix ${\cal D}_{\lambda\lambda'}$ by the decorrelated approximation (\ref{Ddecorr}),
the last term becomes constant
\begin{equation}
\left. {\cal F}_{\chi\chi\to HW\to Hf\bar F}(\theta) \right|_{{\cal D}\to{\cal D}^{\rm decorrelated}} = \frac12 \;. 
\end{equation}
Integrating over the angle $d\cos\theta$ (which yields a factor $2$), one then recovers the familiar relation for the NWA of the total cross section.
However, in general the matrix ${\cal D}_{\lambda\lambda'}$ differs from the decorrelated approximation, and has a
non-trivial angular dependence as well as off-diagonal entries.

For Majorana DM annihilation into a scalar and a vector, only the longitudinal polarization
contributes to the $s$-wave, i.e. ${\cal M}_{2\to 2}^{\lambda=\pm 1}\to 0$ for $v\to 0$.
Using the explicit expression for ${\cal D}_{\lambda\lambda'}$, this imples that
\begin{equation}
{\cal F}_{\chi\chi\to HW\to Hf\bar F}^{s-{\rm wave}}(\theta)  
= \frac{{\cal D}_{00}(\theta)}{2\,\frac13 \sum_\lambda {\cal D}_{\lambda\lambda}}
=   \frac{M_W^2-m_f^2-m_F^2 -4 |p_1^*|^2 \cos^2\theta}{2(M_W^2-m_f^2-m_F^2-\frac43 |p_1^*|^2)}
\approx \frac{3\sin^2\theta}{4}\;,
\end{equation}
where the last expression applies for massless fermions. This corresponds to the decay spectrum of a longitudinally polarized $W$ boson.
The integral of this expression
over $d\cos\theta$ coincides with the decorrelated case. Therefore, the result
for the total cross section in the NWA is nevertheless accurate, with error governed by $\Gamma_W/M_W$, as expected.

Instead of the angle $\theta$ one can use the energy $E_f$ of the fermion in the rest frame of the annihilating particles,
\begin{equation}
 E_f = \gamma\left(\sqrt{|p_1^*|^2+m_f^2}+|p_1^*|\beta\cos\theta\right)\,, \qquad dE_f=\gamma\beta|p_1^*| d\cos\theta
\end{equation}
where $\beta=|p_3|/\sqrt{|p_3|^2+M_W^2}$ and $\gamma=(1-\beta^2)^{-1/2}$.
This finally yields the fermion spectrum in the narrow-width limit,
\begin{equation}
\left(\frac{d\sigma v_{Hf\bar F}}{dE_f}\right)^{NWA} = \frac{\sigma v_{HW}\, BR(W\to f\bar F)}{\gamma\beta|p_1^*|} \times{\cal F}_{\chi\chi\to HW\to Hf\bar F}(\theta)\Big|_{\cos\theta=\frac{E_f-\gamma\sqrt{|p_1^*|^2+m_f^2}}{|p_1^*|\beta\gamma}}
\end{equation}

This procedure can be generalized to other 3-body final states in a straightforward way.
For example, for $\chi\chi\to Wf\bar F$, the contribution from the $W$ resonance is
\begin{equation}
\left(\frac{d\sigma v_{W^+f\bar F}}{dE_f}\right)^{NWA}\Big|_{R=W} = \frac{\sigma v_{WW}\, BR(W^-\to f\bar F)}{\gamma\beta|p_1^*|} \times   {\cal F}_{\chi\chi\to WW\to W^+f\bar F}(\theta)\Big|_{\cos\theta=\frac{E_f-\gamma\sqrt{|p_1^*|^2+m_f^2}}{|p_1^*|\beta\gamma}}
\end{equation}
where
\begin{equation}
{\cal F}_{\chi\chi\to WW\to W^+f\bar F}(\theta) =  \frac{\sum_{\lambda,\lambda',\lambda_3} {\cal M}_{2\to 2}^{*\lambda_3\lambda'}{\cal M}_{2\to2}^{\lambda_3\lambda} \times {\cal D}_{\lambda\lambda'}(\theta)}{2(\sum_{\lambda,\lambda_3} |{\cal M}_{2\to2}^{\lambda_3\lambda}|^2)(\frac13 \sum_\lambda {\cal D}_{\lambda\lambda})} \;,
\end{equation}
and ${\cal M}_{2\to2}^{\lambda_3\lambda}=\epsilon_\mu^{\lambda_3}(p_3)\epsilon_\mu^{\lambda}(q){\cal M}_{2\to2}^{\mu\nu}$
is the helicity amlitude for the $\chi\chi\to WW$ annihilation process. In comparison to before, we have to sum in addition
over the polarizations of the $W^+$ that appears in the 3-body final state. The matrix ${\cal D}_{\lambda\lambda'}(\theta)$
is the same as before. 

For $s$-wave annihilation the pair of vector bosons is in a state with $S=L=1$, $J=0$, and $L_z=S_z=0$, when choosing the
$z$-axis along the momentum of one of the final state particles. The possible spin projections $m_1$ and $m_2$ of the vector bosons are then determined by the Clebsch-Gordon coefficients
for coupling two spin-1 states ($S_1=S_2=1$) to a total spin $S=1$ state with $m\equiv S_z=0$,
\begin{equation}
 \langle S_1 S_2;m_1 m_2|S_1 S_2;S m\rangle = \left\{\begin{array}{ll}
    \sqrt{1/2} & m_1=1, m_2=-1 \\
    0 & m_1=0, m_2=0 \\
    -\sqrt{1/2} & m_1=1, m_2=-1 \\
  \end{array}\right.
\end{equation}
Since the spatial momenta of the vectors are opposite, this means they can only be in equal helicity states, and additionally have to be transverse, more precisely 
\begin{equation}
{\cal M}_p^{\lambda\lambda'}\propto \langle 1\, 1;\lambda\, (-\lambda')|1\, 1;1\, 0\rangle \propto {\rm diag}(1, 0, -1)\,,
\end{equation}
which implies
\begin{eqnarray}\label{eq:Wpol}
 {\cal F}^{s-{\rm wave}}_{\chi\chi\to WW\to W^+f\bar F}(\theta)
&=& \frac{M_W^2-m_f^2-m_F^2 -2|p_1^*|^2\sin^2\theta}{
2(M_W^2-m_f^2-m_F^2-\frac43 |p_1^*|^2) }  \\
&\approx& \frac34 \left( 1- \frac12 \sin^2\theta \right) \;. \nonumber
\end{eqnarray}
This corresponds to the decay spectrum of transversely polasized $W$ bosons, and the last line applies to massless fermions.
It is straightforward to derive the corresponding matrix ${\cal D}_{\lambda\lambda'}$ for $Z$ decay and to generalize the procedure to a fermionic (top) resonance. One finds similarly that due to CP and angular momentum conservation only a definite helicity state can contribute,
which then determines the decay spectrum.


\bibliographystyle{JHEP}
\bibliography{IBSU2}

\end{document}